\documentclass[prd,preprint,tightenlines,floatfix,showpacs,preprintnumbers,nofootinbib,eqsecnum]{revtex4}
 \usepackage[dvips,final]{graphicx}
  \usepackage{amssymb}
   \usepackage{amsmath}
    \usepackage{epsfig}
     \usepackage{bm}% bold math
      \usepackage{pifont}

\def\muF{\relax\ifmmode\mu_\text{F}^2\else{$\mu_\text{F}^2${ }}\fi}
\def\muR{\relax\ifmmode\mu_\text{R}^2\else{$\mu_\text{R}^2${ }}\fi}
\def\muO{\relax\ifmmode{\mu_{0}^{2}}\else{$\mu_{0}^{2}${ }}\fi}
\def\Mev{\relax\ifmmode{\text{MeV}}\else{MeV{ }}\fi}
\def\x{\overline{x} \,}
\def\y{\overline{y} \,}

\def\MS{$\overline{\text{MS}\vphantom{^1}}${ }}
\def\BLM{$\overline{\text{BLM}\vphantom{^1}}${ }}
\def\Li{\relax\ifmmode{\textbf{Li}_{2}}\else{Li$_2${ }}\fi}
\def\Im{\relax{\textbf{Im}{}}}

\newcommand{\va}[1]{\langle{#1}\rangle}
\newcommand{\gev}[1]{\relax\ifmmode{\text{GeV}^{#1}}\else{GeV$^{#1}${ }}\fi}
\newcommand{\req}[1]{(\ref{#1})}
\def\asb{\relax\ifmmode \bar{\alpha}_s\else{$ \bar{\alpha}_s${ }}\fi}
\def\as{\relax\ifmmode \alpha_s\else{$ \alpha_s${ }}\fi}
\def\acal{\relax\ifmmode{\cal A}\else{${\cal A}${ }}\fi}
\newcommand\convo[1]{\mathop{\otimes}\limits_{#1}}
\begin{document}
\thispagestyle{empty}
%\date{\today}
\preprint{\hbox{IRB-TH-02/04, MIT/CTP-3478, RUB-TPII-02/04}}
\vspace*{-10mm}

\title{Pion form factor in QCD:
       From nonlocal condensates to NLO analytic perturbation
       theory\\}

\author{A.~P.~Bakulev}
 \email{bakulev@thsun1.jinr.ru}
  \affiliation{Bogoliubov Laboratory of Theoretical Physics, JINR,
   141980 Dubna, Russia}

\author{K.~Passek-Kumeri\v{c}ki}
 \email{passek@thphys.irb.hr}
  \affiliation{
   Theoretical Physics Division,
   Rudjer-Bo\v{s}kovi\'{c} Institute,
   P.O.\ Box 180, HR-10002 Zagreb, Croatia}

\author{W.~Schroers}
 \email{wolfram.schroers@feldtheorie.de}
  \affiliation{Center for Theoretical Physics,
  Laboratory for Nuclear Science and Department of Physics,
  Massachusetts Institute of Technology, Cambridge Massachusetts
  02139}

\author{N.~G.~Stefanis}
 \email{stefanis@tp2.ruhr-uni-bochum.de}
  \affiliation{Institut f\"{u}r Theoretische Physik II,
   Ruhr-Universit\"{a}t Bochum,
   D-44780 Bochum, Germany}

%\cleardoublepage
\begin{abstract}
We present an investigation of the pion's electromagnetic form
factor $F_{\pi}(Q^2)$ in the spacelike region utilizing two new
ingredients:
(i) a double-humped, endpoint-suppressed pion distribution
amplitude derived before via QCD sum rules with nonlocal
condensates---found to comply at the $1\sigma$ level with
the CLEO data on the $\pi\gamma$ transition---and
(ii) analytic perturbation theory at the level of parton
amplitudes for hadronic reactions.
The computation of $F_{\pi}(Q^2)$ within this approach is performed
at NLO of QCD perturbation theory (standard and analytic), including
the evolution of the pion distribution amplitude at the same order.
We consider the NLO corrections to the form factor in the \MS scheme
with various renormalization scale settings and also in the
$\alpha_V$-scheme.
We find that using standard perturbation theory, the size of the NLO
corrections is quite sensitive to the adopted renormalization scheme
and scale setting.
The main results of our analysis are the following:
(i) Replacing the QCD coupling and its powers by their analytic images,
both dependencies are diminished and the predictions for the pion form
factor are quasi scheme and scale-setting independent.
(ii) The magnitude of the factorized pion form factor, calculated with
the aforementioned pion distribution amplitude, is only slightly larger
than the result obtained with the asymptotic one in all considered
schemes.
(iii) Including the soft pion form factor via local duality and
ensuring the Ward identity at $Q^2=0$, we present predictions that are
in remarkably good agreement with the existing experimental data both
in trend and magnitude.

\end{abstract}
%Keywords: Renormalization group evolution
%          Perturbative calculations
%          Summation of perturbation theory
%          BLM method of commensurate scale setting
%          Hadron wave functions
%          Electromagnetic form factors
\maketitle

%\tableofcontents

%%%%%%%%%%%%%%%%%%%%%%%%%%%%%%%%%%%%%%%%%%%%%%%%%%%%%%%%%%%%%%%%%%%%%%%
%%%%%%%%%%%%%%%%%%%%%%%%%%%%%%%%%%%%%%%%%%%%%%%%%%%%%%%%%%%%%%%%%%%%%%%
\cleardoublepage
\section{Introduction}
\label{intro}
%%%%%%%%%%%%%%%%%%%%%%%%%%%%%%%%%%%%%%%%%%%%%%%%%%%%%%%%%%%%%%%%%%%%%%%
It is the purpose of this paper to review and discuss questions
relating to the calculation of the electromagnetic pion form factor
with an improved pion distribution amplitude (DA), derived from
QCD sum rules with nonlocal condensates \cite{BMS01}, and to use QCD
Analytic Perturbation Theory (APT)
\cite{SS99,Shi00,DVS00,DVS0012,Shi01} beyond the leading order (LO).
Before going into the details of this framework, let us expose, in
general terms, what these two ingredients mean for the analysis and
also make some introductory remarks.

Hadronic form factors are typical examples of hard-scattering
processes within QCD \cite{CZ77,CZS77,ER80,ER80tmf,LB79,LB80} and
clearly the first level of knowledge necessary to understand the
structure of \emph{intact} hadrons in terms of quarks and gluons.
Such processes have been much explored both because of their
physical relevance, as being accessible to experiments, and
because they allow to assess nonperturbative features of QCD (for
reviews, see, for instance,
\cite{CZ84,BL89,MuellerAH97,Ste99,GPV01}).
In the following, the discussion is centered around the pion's
electromagnetic form factor.
At a more theoretical level, ``hard'' means that at least some part
of the process amplitude, recast in terms of quarks collinear to
hadrons (in an appropriate Lorentz frame), should become amenable to
perturbation theory via factorization theorems on account of the
hard-momentum scale of the process, say, $Q^2$, that should suppress
factorized infrared (IR) subprocesses, thus ensuring short-distance
dominance.
Under these circumstances one can safely evaluate logarithmic scaling
violations by means of perturbative QCD (pQCD) and the
renormalization-group equation.
When no hard momenta flow on the side of the initial (incoming) or the
final (outgoing) hadron, factorization fails and a
renormalization-group analysis cannot be made, so that in order to
calculate the non-factorizable part of the pion form factor, one
has to resort to phenomenological models (prime examples of which
are \cite{IL84,JK86,IL89,JK93}), or employ theoretical concepts
like the (local) quark-hadron duality \cite{SVZ} and their descendants
\cite{NR82,Rad95,BRS00}.

Despite dedicated efforts in the last two decades, exclusive processes
have failed to deliver a complete quantitative understanding within QCD
for a variety of reasons, among others:
\begin{itemize}
\item Limited knowledge of higher-order perturbative and power-law
      behaved (e.g., higher-twist) corrections to the amplitudes.
\item Presence of singularities (of endpoint, mass, soft, collinear,
      or pinch origin) that may spoil factorization in some kinematic
      regions.
\item Insignificant knowledge of hadron distribution amplitudes
      owing to the lack of a reliable non-perturbative approach.
\item Non-factorizing contributions that are not calculable within
      pQCD and hence introduce a strong model dependence.
\end{itemize}

While it may still be not possible to clarify all these theoretical
issues conclusively, we believe that significant progress has quite
recently been achieved in understanding the pion structure both from
the theoretical side---perturbatively
\cite{MNP99a,MNP01a,SSK99,SSK00,MMP02} and nonperturbatively
\cite{BMS01,Kho99,BKM00,BiKho02,BMSdur03}---as well as from the
experimental side \cite{CLEO98,JLAB00,Blok02} and associated
data-processing techniques \cite{SY99,BMS02,BMS03}, a progress that
could bring to a cleaner comparison between data and various
theoretical QCD predictions
\cite{RR96,JKR96,KR96,MuR97,BJPR98,Kho99,DKV01,BMS02,BMS03,BMSefr03,%
Ag04,HWW04}.\footnote{For theoretical predictions on $F_\pi(Q^2)$ in
the timelike region, see, for instance \cite{BRS00,GP95}.}
Moreover, a program to compute the electromagnetic and transition
form factors of mesons on the lattice has been launched by two
collaborations, \cite{LHPC03a,LHPC03b} and \cite{RBC03}, that may
provide valuable insights when it is completed.
This situation prompts an in-depth review and update of these issues,
in an effort to consolidate previous calculations of the pion's
electromagnetic form factor and narrow down theoretical uncertainties.

We will focus our present discussion on two main issues:
\begin{enumerate}
  \item[(i)] How QCD perturbation theory can be safely used to make
predictions in the low-momentum regime where conventional power series
expansions in the QCD coupling break down and nonperturbative effects
dominate.
Such an extension is based on recent works on ``analytization'' of the
running strong coupling $\alpha_{\rm s}(Q^2)$
\cite{DMW96,Gru97,SS97,Shi98,GGK98,SS99,Shi00,Shi01} (see also
\cite{Nes03} for a slightly different approach) and their
generalization to the partonic level
of hadron amplitudes,
like the electromagnetic and the pion-photon
transition form factor, or the Drell--Yan process, beyond the level
of a single scheme scale \cite{SSK99,SSK00,KS01,Ste02}.
In contradiction to the usual assumption of singular growth of
$\alpha_{\rm s}(Q^2)$ in the IR domain, the QCD coupling in this
scheme has an IR-fixed point, with the unphysical Landau pole being
completely absent.
In conventional perturbative approaches, a choice of the
renormalization scale in the region of a few $\Lambda_{\rm QCD}$,
as required, for instance, by the Brodsky--Lepage--Mackenzie (BLM)
scale-fixing procedure \cite{BLM83}, would induce singularities
thus prohibiting the perturbative calculation of hadronic
observables.
Using APT, these singularities are avoided by construction,
i.e., without introducing ad hoc IR regulators, e.g., an
effective gluon mass \cite{BL89,JiAm90}, and therefore the
validity of the perturbative expansion (in mass-independent
renormalization schemes) is not jeopardized by IR-renormalon
power-law ambiguities.
In addition, APT provides a better stability against higher-loop
corrections and a weaker renormalization-scheme dependence
than standard QCD perturbative expansion---see \cite{SSK00} and
Sec.\ \ref{sect:PFFNLO:Num}d.
  \item[(ii)] How to improve the nonperturbative input by employing
a pion DA which incorporates the nonperturbative features of the QCD
vacuum in terms of a nonlocal quark condensate
\cite{MR86,MR89,MR92,BR91,BM95}.
This accounts for the possibility that vacuum quarks can flow
with a nonzero average
virtuality $\lambda_{q}^{2}$, in an attempt to connect dynamic
properties of the pion, like its electromagnetic form factor,
directly with the QCD vacuum structure (we refer to \cite{BMSdur03}
for further details).
Within this scheme, the pion DA (termed BMS\ \cite{BMS01} in the
following) turns out to be double-humped with strongly suppressed
endpoints ($x=0,1$), the latter feature being related to the
nonlocality parameter $\lambda_{q}^{2}$.
It has been advocated, for example, in \cite{SSK99,SSK00} (see also
\cite{Ste95,Ste99}), that a suppression of the endpoint region
(which is essentially nonperturbative) as strong as possible is a
prerequisite for the self-consistent application of QCD perturbation
theory within a factorization scheme.
\end{enumerate}

In a recent series of papers \cite{BMS01,BMS02,BMS03}, two of us
together with S.~V.~Mikhailov have conducted an analysis of the CLEO
data \cite{CLEO98} on the pion-photon transition using attributes from
QCD light-cone sum rules \cite{Kho99,SY99}, NLO
Efremov--Radyushkin--Brodsky--Lepage (ERBL)
\cite{ER80,ER80tmf,LB79,LB80} evolution \cite{Mul94,Mul95}, and
detailed estimates of uncertainties owing to higher-twist contributions
and NNLO\ perturbative corrections \cite{MMP02}.
These works confirmed the gross features of the previous
Schmedding--Yakovlev (SY) analysis \cite{SY99}; notably, both the
Chernyak--Zhitnitsky (CZ) \cite{CZ84} pion DA as well as the asymptotic
one are incompatible with the CLEO data \cite{CLEO98} at the $4\sigma$-
and $3\sigma$-level, respectively, whereas the aforementioned BMS\ pion
DA, which incorporates the vacuum nonlocality, is within the $1\sigma$
error ellipse.
Moreover, this approach revealed the possibility of using the CLEO
experimental data to estimate the value of the QCD vacuum correlation
length $\lambda_{q}^{-1}$.
Indeed, it turns out that the extracted value
$\lambda_{q}^{2} \simeq 0.4$~GeV${}^2$ is consistent with those
obtained before using QCD sum rules \cite{BI82,OPiv88,KPS02} and also
with numerical simulations on the lattice \cite{DDM99,BM02}.
In addition, it was shown \cite{BMS03} that the value of the inverse
moment
\hbox{$\langle x^{-1}\rangle_{\pi}(\mu^2)
=\int^1_0 \varphi_\pi(x;\mu^2){x}^{-1}dx$}
of the pion DA, calculated by means of an \emph{independent} QCD sum
rule, is compatible with that extracted from the CLEO data.
These findings give us confidence to use the BMS\ pion DA (including
also the range of its intrinsic theoretical uncertainties) in order to
derive predictions for the electromagnetic pion form factor within the
factorization scheme of QCD at NLO, presenting also results which
include the non-factorizing soft contribution \cite{NR82,BRS00} to
compare with available experimental data.

The structure of the paper is as follows.
In Sec.\ \ref{sect:Factor} we shall recall the QCD factorization of the
pion's electromagnetic form factor.
Sec.\ \ref{sect:PionDA} deals with the basics of the pion distribution
amplitude and its derivation from QCD sum rules with nonlocal
condensates.
The perturbative results for the pion form factor at NLO order, on the
basis of the results given in \cite{MNP99a}, are summarized in Sec.\
\ref{sect:PFFNLO:AR}, whereas issues related to the setting of the
renormalization scheme and scale are discussed in Sec.\
\ref{sect:RenSc}.
The important topic of the non-power series expansion of the pion form
factor in the context of Analytic Perturbation Theory is considered in
Sec.\ \ref{sect:APT}.
Our numerical analysis and the comparison of our results with available
experimental data is presented in Sec.\ \ref{sect:PFFNLO:Num}.
Finally, in Sec.\ \ref{sect:concl} we give a summary of the results
and draw our conclusions.
Important technical details of the analysis are supplied in five
appendixes.

%%%%%%%%%%%%%%%%%%%%%%%%%%%%%%%%%%%%%%%%%%%%%%%%%%%%%%%%%%%%%%%%%%%%%%%
%%%%%%%%%%%%%%%%%%%%%%%%%%%%%%%%%%%%%%%%%%%%%%%%%%%%%%%%%%%%%%%%%%%%%%%
\section{QCD factorization applied to the pion form factor}
\label{sect:Factor}
%%%%%%%%%%%%%%%%%%%%%%%%%%%%%%%%%%%%%%%%%%%%%%%%%%%%%%%%%%%%%%%%%%%%%%%
The outstanding virtue of factorization is that a hadronic process can
be dissected in such a way as to isolate a partonic part accessible to
pQCD.
Provided that the partonic subprocesses are free of IR-singularities,
then at large momentum transfer $F_{\pi}(Q^2)/f_{\pi}^2 \sim 1/Q^2$,
modulo logarithmic corrections due to renormalization.
Hence, the amplitude for the electromagnetic pion form factor is
short-distance dominated and can be expressed in terms of its
constituent quarks collinear to the pion with the errors of this
replacement being suppressed by powers of $1/Q$.
Even more, one can rigorously dissect the QCD amplitude into a
coefficient function, that contains the hard quark-gluon interactions,
and two matrix elements corresponding to the initial and final pion
states of the leading twist operator with the quantum numbers of the
pion according to the Operator Product Expansion (OPE).
In this way, one establishes that the coefficient function will
scale asymptotically according to its dimensionality modulo anomalous
dimensions controlled by the renormalization group equation.

\begin{figure}[h]
 \centerline{\includegraphics[width=\textwidth]{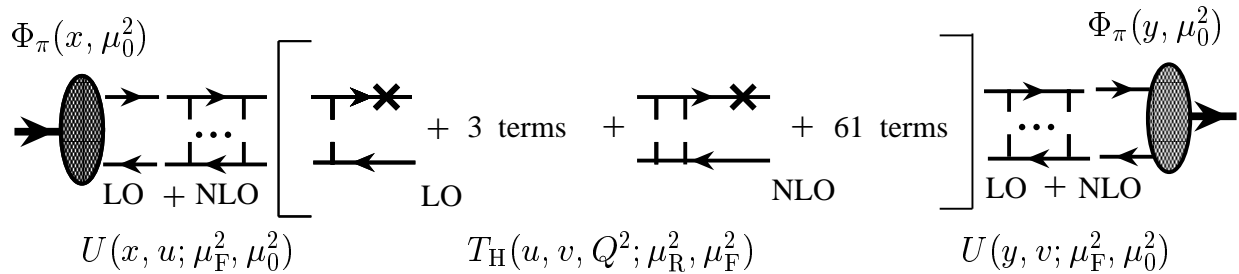}}
  \caption[fig:pidata]
        {\footnotesize
         Illustration of the structure of the factorized pion form
         factor within pQCD at NLO of the hard scattering
         amplitude and the evolution effect of the pion DA.
         Hard gluons are indicated by broken lines, whereas the
         external off-shell photon is denoted by a cross.
         \label{fig:feynmanpi}}
\end{figure}

The pion's electromagnetic form factor is defined by the matrix element
\begin{equation}
 \va{\pi^{+}(P^{\prime})| J_{\mu}(0) | \pi^{+}(P)}
 =  {\left( P + P^{\prime}\right)}_{\mu} F_{\pi}(Q^{2}) \, ,
\label{eq:pionvertex}
\end{equation}
%%Eq (2.1) Definition of pion form factor as vertex function
where $J_{\mu}$ is the electromagnetic current expressed in terms of
quark fields, $(P^{\prime} - P)^2=q^2\equiv -Q^2$ is the photon
virtuality, i.e., the large momentum transfer injected into the pion,
and $F_\pi$ is normalized to $F_\pi(0)=1$.
Based on the above considerations, the pion form factor can be
generically written in the form
\cite{ER80,ER80tmf,LB79,LB80}
\begin{eqnarray}
 \label{eq:pff}
 F_{\pi}(Q^{2})
 = F_{\pi}^\text{Fact}(Q^{2})
  +  F_{\pi}^\text{non-Fact}(Q^{2}) \,,
\end{eqnarray}
%%Eq (2.2) Total pion form factor: factorizable + nonfactorizable parts
where
$F_{\pi}^\text{Fact}(Q^{2})$
is the factorized part within pQCD and
$F_{\pi}^\text{non-Fact}(Q^{2})$
is the non-factorizable part---usually being referred to as the
``soft contribution'' \cite{NR82}---that contains subleading
power-behaved (e.g., twist-4 and higher-twist) contributions
originating from nonperturbative effects.
It is important to understand that Eq.\ (\ref{eq:pff}) becomes
increasingly unreliable as $Q^2\to 0$, owing to the breakdown of
perturbation theory at such low momentum scales.
Hence, we expect the real form factor to be different from the RHS
of this equation at low $Q^2$.
We shall show in Sec.\ \ref{sect:expdata} how to remedy this problem.
The leading-twist factorizable contribution can be expressed as a
convolution in the form
\begin{eqnarray}
 F_{\pi}^\text{Fact}(Q^{2}; \muR)
=
   \Phi_{\pi}^{*}(x,\muF)\convo{x}
     T_\text{H}(x,y,Q^{2};\muF,\muR)\convo{y}
      \Phi_{\pi}(y,\muF)\, ,
 \label{eq:pff-Fact}
\end{eqnarray}
%%Eq (2.3) Convolution form of factorizable pion form factor
where $\otimes$ denotes the usual convolution symbol
($A(z)\convo{z}B(z) \equiv \int_0^1 dz A(z) B(z)$)
over the longitudinal momentum fraction variable $x$ ($y$) and
$\mu_\text{F}$ represents the factorization scale at which the
separation between the long- (small transverse momentum) and
short-distance (large transverse momentum) dynamics takes place,
with $\mu_\text{R}$ standing for the renormalization (coupling
constant) scale.
A graphic illustration of the factorized pion form factor in terms of
Feynman diagrams is given in Fig.\ \ref{fig:feynmanpi}.

%%%%%%%%%%%%%%%%%%%%%%%%%%%%%%%%%%%%%%%%%%%%%%%%%%%%%%%%%%%%%%%%%%%%%%%
%                              FIGURE 1                               %
%%%%%%%%%%%%%%%%%%%%%%%%%%%%%%%%%%%%%%%%%%%%%%%%%%%%%%%%%%%%%%%%%%%%%%%
%%%%%%%%%%%%%%%%%%%%%%%%%%%%%%%%%%%%%%%%%%%%%%%%%%%%%%%%%%%%%%%%%%%%%%%

Here, $T_\text{H}(x,y,Q^{2};\muF,\muR)$ is the hard-scattering
amplitude, describing short-distance interactions at the parton level,
i.e., it is the amplitude for a collinear valence quark-antiquark pair
with total momentum $P$ struck by a virtual photon with momentum $q$
to end up again in a configuration of a parallel valence
quark-antiquark pair with momentum $P'=P+q$ and can be calculated
perturbatively in the form of an expansion in the QCD coupling, the
latter to be evaluated at the reference scale of
renormalization $\muR$:
\begin{eqnarray}
   T_\text{H}(x,y,Q^2;\muF,\muR) =
       \alpha_s(\muR)\,  T_\text{H}^{(0)}(x,y,Q^2)
         + \frac{\alpha_s^2(\muR)}{4 \pi} \,
             T_\text{H}^{(1)}(x, y, Q^2;\muF, \muR)
                + \ldots
 \label{eq:TH}
\end{eqnarray}
%%Eq (2.4) Hard-scattering amplitude as power series in pQCD
The explicit results for the hard-scattering amplitude in LO and
NLO accuracy are supplied in Appendix \ref{app:HSAnlo}.
All transverse momenta below the factorization scale that would cause
divergences associated with the propagation of partons over long
distances have been  absorbed into the pion DAs, which have the
correct long-distance behavior, as dictated by nonperturbative QCD.

Because the QCD perturbation series expansion in the strong
coupling is only asymptotic, this calculation bears an intrinsic error
owing to its truncation that is of the order of the first neglected
term $\sim C/Q^p$, with $C$ and $p$ being, in general, dependent on
the particular observable in question---here the pion form factor.
Lacking all-order results for the perturbative coefficients, one has to
resort to fixed-order, renormalization and factorization
scheme-dependent contributions to $F_{\pi}^\text{Fact}(Q^{2},
\muR)$ that do not exceed beyond the NLO \cite{MNP99a}.
The truncation of this series expansion at any finite order introduces
a residual dependence of the corresponding fixed-order or partly
resummed hard-scattering amplitude and, consequently, also of the
finite-order prediction for $F_{\pi}^{\text{Fact}}$, on the
renormalization scheme adopted and on the renormalization scale
$\mu_\text{R}$ chosen.
In order that the perturbative prediction comes as close as possible to
the physical form factor, measured in experiments---which is exactly
independent of the renormalization (or any other unphysical scheme)
scale---the best perturbative expansion would be the one which
minimizes the error owing to the disregarded higher-order corrections.
This can be accomplished, for instance, by trading the conventional
power-series perturbative expansion in favor of a non-power series
expansion in terms of an analytic strong running coupling, performing
the calculations in the framework of Analytic Perturbation Theory to be
discussed in Sec.\ \ref{sect:APT}.
Here it suffices to state that in this framework the QCD running
coupling has an IR-fixed point and hence avoids eo\ ipso
IR-renormalon ambiguities allowing to adopt a BLM\ scale setting
procedure.

By convoluting the finite-order result for the hard-scattering
amplitude, expressed in the form of Eq.\ \req{eq:TH}, with the
distribution amplitude \req{eq:PhipVP} truncated at the same order in
$\alpha_s$, an additional residual dependence on the factorization
scheme and the factorization scale $\mu_\text{F}$ appears.
We show in Appendix \ref{app:Fact} how to get rid of the factorization
scale dependence at fixed order of perturbation theory (NLO) by proving
that non-cancelling terms in
$F_\pi^\text{Fact}(Q^2;\muR)$
are of order $\alpha_s^3$.
For an alternative way of handling the $\muF$ dependence, we refer the
reader to \cite{MNP01a,MNP01b}.
For practical purposes, the preferable form of the convolution equation
for $F_{\pi}^{\text{Fact}}(Q^2)$ is  given by adopting the so-called
``default'' choice, i.e., setting in Eqs.\ (\ref{eq:pff-Fact}),
(\ref{eq:TH}) $\muF=Q^2$.
Note, however, that the same choice of scale in different schemes
yields also to different results for finite-order approximants for
the pion form factor \cite{CS83}.
Problems connected with heavy-quark mass thresholds in the $\beta$
function are given below particular attention both in the
hard-scattering part and in the evolution part.

Another crucial question is whether the factorizable pQCD\
contribution to the pion form factor is actually sufficient to
describe the available experimental data, or if one has to take
into account the soft part as well.
It has been advocated in
\cite{IL84,IL89,NR82,BR91,JK93,AMN95,JKR96,BRS00} that at
momentum-transfer values probed experimentally so far, this latter
contribution, though power suppressed because it behaves like $1/Q^4$
for large $Q^2$, dominates and mimics rather well the observed
$1/Q^2$ behavior.
To account for this effect, we will include the soft contribution
\cite{BRS00} (discussed in Sec.\ \ref{sect:NFPFF}) into our form-factor
prediction when comparing with the data, albeit the poor quality of the
latter at higher $Q^2$ makes it impossible to draw any definite
conclusions about the transition from one regime to the other.
Therefore, our main purpose in this paper is to calculate the
factorizable contribution as accurately as possible.
The calculational ingredients will be to
\begin{itemize}
\item use as a nonperturbative input a set of pion DAs
      $\varphi_\pi(x,\mu_0^2)$, derived in \cite{BMS01} from QCD sum
      rules with nonlocal condensates, with the optimum one, termed
      BMS\ model, standing out;
\item evolve $\varphi_\pi(x,\mu_0^2)$ by employing a kernel and
      corresponding anomalous dimensions up to NLO
      \cite{KMR86,Mul94,Mul95} both within the standard and the
      analytic perturbation theory;
\item employ a hard-scattering amplitude
      $T_\text{H}(x,y,Q^2;\muF, \muR)$
      up to NLO order \cite{FGOC81,DR81,Sar82,RK85,KMR86,BT87,MNP99a},
      using both standard power and also non-power series expansions;
\item take into account the soft (non-factorizable) contribution,
      $F_{\pi}^\text{non-Fact}(Q^{2})$, on the basis of the Local
      Duality (LD) approach when comparing the theoretical predictions
      with the experimental data.
\end{itemize}

%%%%%%%%%%%%%%%%%%%%%%%%%%%%%%%%%%%%%%%%%%%%%%%%%%%%%%%%%%%%%%%%%%%%%%%
%%%%%%%%%%%%%%%%%%%%%%%%%%%%%%%%%%%%%%%%%%%%%%%%%%%%%%%%%%%%%%%%%%%%%%%
\section{Pion Distribution Amplitude}
\label{sect:PionDA}
%%%%%%%%%%%%%%%%%%%%%%%%%%%%%%%%%%%%%%%%%%%%%%%%%%%%%%%%%%%%%%%%%%%%%%%
\subsection{Nonperturbative input}
\label{sect:NPInput}
%%%%%%%%%%%%%%%%%%%%%%%%%%%%%%%%%%%%%%%%%%%%%%%%%%%%%%%%%%%%%%%%%%%%%%%
Turning our attention now to the pion distribution amplitude, we
note that $\Phi_{\pi}(x,\muF)$ specifies in a process- and
frame-independent way\footnote{Provided the same factorization scheme
is used for all considered processes \cite{Ste95,Ste99}.}
the longitudinal-momentum $xP$ distribution of the valence quark
(and antiquark which carries a fraction $\bar{x}=1-x$) in the pion with
momentum $P$.
At the twist-2 level it is defined by the following universal operator
matrix element (see, e.g., \cite{CZ84} for a review)
\begin{eqnarray}
 \label{eq:pi-DA-ME}
 \va{0\mid\bar{d}(z)\gamma^{\mu}\gamma_5\,
 {\cal C}(z,0) u(0)\mid\pi(P)}
  \Big|_{z^2=0}
 &=& i P^{\mu}
%%Eq (3.1) Dimensionful Pion distribution amplitude
      \int^1_0 dx e^{ix(zP)}\
      \Phi_{\pi}\left(x,\muO \sim z^{-2}\right)\ ;\\
 \int_0^1 \Phi_{\pi}(x,\muO)\, dx
  &=& f_{\pi}\, ,
 \label{eq:pi-DA-fpi}
\end{eqnarray}
%%Eq (3.2) Normalization of pion DA to f_pi
with $f_{\pi} = 130.7 \pm 0.4$~MeV \cite{PDG2002} being the pion decay
constant defined by
\begin{equation}
\langle 0|\bar{d}(0)\gamma_{\mu} \gamma_{5} u(0)|\pi^{+}(P) \rangle
 =
ip_{\mu} f_{\pi}
\label{eq:fpi}
\end{equation}
%%Eq (3.3) Definition of f_pi
and where
\begin{equation}
  {\cal C}(0,z)
  = {\cal P}
  \exp\!\left[-ig_s\!\!\int_0^z t^{a} A_\mu^{a}(y)dy^\mu\right]
\label{eq:pexponent}
\end{equation}
%%Eq (3.4) Path-ordered exponential (connector)
is the Fock--Schwinger phase factor
(coined the color ``connector'' in \cite{Ste84}), path-ordered along
the straight line connecting the points $0$ and $z$, to preserve gauge
invariance.
The scale $\muO$, called the normalization scale of the pion DA, is
related to the ultraviolet (UV) regularization of the quark-field
operators on the light cone whose product becomes singular for $z^2=0$.

Although the distribution amplitude is intrinsically a nonperturbative
quantity, its evolution is governed by pQCD (a detailed discussion is
relegated to Appendix \ref{app:DAnlo}) and can be expressed in the form
\begin{eqnarray}
 \Phi_{\pi}(x, \muF)
  = U(x,s; \muF,\muO)\convo{s}\Phi_{\pi}(s, \muO)\,,
 \label{eq:PhipVP}
\end{eqnarray}
%%Eq (3.5) Evolution of dimensionful pion distribution amplitude
where $\Phi_{\pi}(s, \muO)$ is a nonperturbative input determined at
some low-energy normalization point $\muO$ (where the local operators
in Eq.\ (\ref{eq:pi-DA-ME}) are renormalized)---which is of the order
of 1 GeV${}^2$---while $U(x,s; \muF, \muO)$ is the operator to evolve
that DA from the scale $\muO$ to the scale $\muF$ and is calculable in
QCD perturbation theory.
In the asymptotic limit, the shape of the pion DA is completely fixed
by pQCD with the nonperturbative input being solely contained in
$f_{\pi}$.

Neglecting the ${\mathbf k}_{\perp}$ dependence of the hard-scattering
amplitude at large $Q^2$,\footnote{This actually means that for all
initial ${\mathbf k}^2_{\perp i}\ll Q^2$ and analogously for all final
${\mathbf l}^2_{\perp i}\ll Q^2$,
radiative corrections sense only single quark and gluon lines.}
it is convenient to introduce a dimensionless pion DA,
$\varphi_\pi(x)$, normalized to 1,
\begin{eqnarray}
 \Phi_{\pi}(x,\muO) = f_{\pi}\, \varphi_\pi(x,\muO)
  \label{eq:NormDA}
\end{eqnarray}
%%Eq (3.6) Normalization of dimensionless pion DA to unity
that can be defined as the probability amplitude for finding two
partons with longitudinal momentum fractions $x$ and $\bar{x}$
``smeared'' over transverse momenta
${\mathbf k}_{\perp}^{2} \leq {\mu^2}$, i.e.,
\begin{equation}
 \varphi_\pi(x,\mu^2)
=
 \int_{0}^{{\mathbf k}_{\perp}^{2} \leq {\mu^2}}
 \left[d{}^{2} {\mathbf k}_{\perp} \right]
 \psi\left( x, {\mathbf k}_{\perp} \right) \, ,
 \label{eq:DAoverWF}
\end{equation}
%%Eq (3.7) Pion DA as integral over transverse momenta of pion WF
where $ \left[ d^{2}{\mathbf k}_{\perp} \right]$ is an appropriate
integration measure over transverse momenta \cite{LB80}, helicity
labels have been suppressed, and a logarithmic pre-factor due to
quark self-energy and photon-vertex corrections has been absorbed
for the sake of simplicity into the definition of the pion wave
function.

%%%%%%%%%%%%%%%%%%%%%%%%%%%%%%%%%%%%%%%%%%%%%%%%%%%%%%%%%%%%%%%%%%%%%%%
%                              FIGURE 2                               %
%%%%%%%%%%%%%%%%%%%%%%%%%%%%%%%%%%%%%%%%%%%%%%%%%%%%%%%%%%%%%%%%%%%%%%%
\begin{figure}[t]
 \centerline{\includegraphics[width=0.45\textwidth]{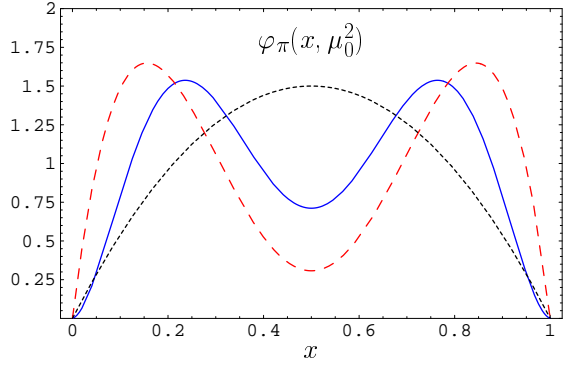}}
  \caption{\footnotesize Comparison of selected pion DAs denoted by
   obvious acronyms:
   $\varphi_\text{as}$ (dotted line),
   $\varphi_\text{CZ}$ (dashed line) \protect\cite{CZ84}, and
   $\varphi_\text{BMS}$ (solid line) \protect\cite{BMS01}, defined by
   Eq.\ (\ref{eq:phi024mu0}) in conjunction with (\ref{eq:DAcan}).
   All DAs are normalized at the same scale $\muO\approx1$ GeV$^2$.
   \label{fig:pion_das}}
\end{figure}
%%%%%%%%%%%%%%%%%%%%%%%%%%%%%%%%%%%%%%%%%%%%%%%%%%%%%%%%%%%%%%%%%%%%%%%
The nonperturbative input, alias the pion DA at the initial
normalization scale $\muO$, $\varphi_\pi(x,\mu_0^2)$, will be
expressed as an expansion over Gegenbauer polynomials which form an
eigenfunctions decomposition, having recourse to a convenient
representation which separates the $x$ and $\mu^2$ dependence (a
detailed exposition can be found in \cite{Ste99}).
Then, the pion DA at the initial scale $\muO$ reads
\begin{eqnarray}
 \varphi_\pi(x,\mu_0^2)
  = 6 x (1-x)
     \left[ 1
          + a_2(\muO) \, C_2^{3/2}(2 x -1)
          + a_4(\muO) \, C_4^{3/2}(2 x -1)
          + \ldots
     \right]\, ,
\label{eq:phi024mu0}
\end{eqnarray}
%%Eq (3.8) Pion DA in terms of Gegenbauer polynomials
with all nonperturbative information being encapsulated in the
coefficients $a_n$.
These coefficients will be taken from a QCD sum-rules calculation
employing nonlocal condensates \cite{BMS01,BMSdur03}, and we refer the
reader to these works for more details.
Here we only use the results obtained there.
We found at $\muO = 1.35$~GeV${}^2$ and for a quark virtuality of
$\lambda_{q}^{2}=0.4$~GeV${}^2$:
\begin{eqnarray}
\begin{array}{lll}
 a_0  =    1 & \quad\quad\quad
 a_2 =     0.19 & \quad\quad\quad
 a_4  =    -0.13 \\
 a_6  =    5 \times 10^{-3} & \quad\quad\quad
 a_8 =     4 \times 10^{-3} & \quad\quad\quad
 a_{10} =  4 \times 10^{-3}  \, .
\end{array}
\label{eq:bmscoef}
\end{eqnarray}
%%Eq (3.9) Gegenbauer coefficients for BMS pion DA
One appreciates that all Gegenbauer coefficients with $n>4$ are close
to zero and can therefore be neglected.
Hence, to model the pion DA, it is sufficient to keep only the first
two coefficients, which we display below in comparison with those for
the asymptotic DA and the CZ~\cite{CZ84} model after 2-loop evolution
to the reference scale $\muO=1$~GeV${}^2$, i.e.,
\begin{eqnarray}
  \begin{array}{rllll}
   \varphi_\text{as}: &  a_n=0 \:, n \geq 2 &  &  &
             \quad \muO = \muF,
              \\
   \varphi_\text{BMS}:& a_n=0 \:, n > 4 & \quad a_2=0.20 &
            \quad a_4=-0.14 &
             \quad \muO = 1~\text{GeV}^2 \quad %\protect\cite{BMS01}\,.
              \\
   \varphi_\text{CZ}: &  a_n=0 \:, n > 2 & \quad a_2=0.56 & &
            \quad \muO = 1~\text{GeV}^2 \quad % \protect\cite{BMS02}
      \,.
  \end{array}
\label{eq:DAcan}
\end{eqnarray}
%%Eq (3.10) Gegenbauer coefficients for as, BMS, CZ pion DAs
%%%%%%%%%%%%%%%%%%%%%%%%%%%%%%%%%%%%%%%%%%%%%%%%%%%%%%%%%%%%%%%%%%%%%%%
%                              FIGURE 3                               %
%%%%%%%%%%%%%%%%%%%%%%%%%%%%%%%%%%%%%%%%%%%%%%%%%%%%%%%%%%%%%%%%%%%%%%%
\begin{figure}[b]
 \centerline{\includegraphics[width=0.9\textwidth]{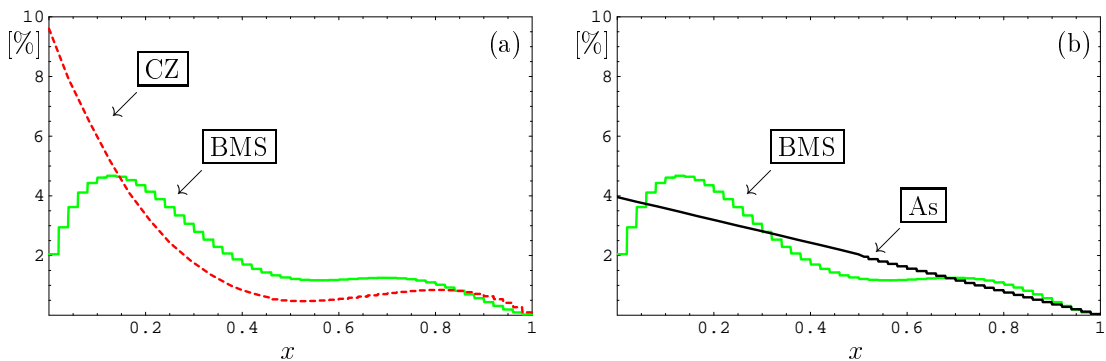}}
   \caption{\footnotesize
   Percentage distribution (see text) of the first inverse moment in
   $x$ of the BMS\ model DA \cite{BMS01} in comparison with the CZ one
   \protect\cite{CZ84} (a) and the asymptotic DA (b).
   \label{fig:hyst_comp}}
\end{figure}
%%%%%%%%%%%%%%%%%%%%%%%%%%%%%%%%%%%%%%%%%%%%%%%%%%%%%%%%%%%%%%%%%%%%%%%
The shapes of these DAs are displayed in Fig.\ \ref{fig:pion_das}.

At this point some important remarks and observations are in order.
\begin{itemize}
\item The BMS\ pion DA, though doubly-peaked, has its endpoints
$x\to 0$ and $x\to 1$ strongly suppressed due to the nonlocality
parameter $\lambda_{q}^{2}$.
Hence, fears frequently expressed in the literature that
double-humped pion DAs should be avoided because they
may emphasize the endpoint region, where the use of perturbation theory
is unjustified, are unfounded.
\item The BMS\ pion DA approaches asymptotically $\varphi_\text{as}$
in the endpoints from \textit{below}, whereas $\varphi_\text{CZ}$
approaches the asymptotic limit from \textit{above}, which means
that the endpoint behavior of the latter is dangerous until very
large values of $Q^2$.
It is well-known \cite{MuellerAH97,SSK99,Ste99} that in the endpoint
region $x\to1$ the spectator quark in the hard process, carrying the
small longitudinal momentum fraction $\bar{x}$, can ``wait'' for a long
time until it exchanges a soft gluon with the struck quark to fit
again into the final pion wave function.
As a result, a strong Sudakov suppression \cite{LS92} is needed in that
case in order to justify the use of perturbation theory.
In contrast, the endpoint behavior of the BMS\ DA is not controversial
because, though doubly peaked, it does not emphasize the endpoint
regions.
Even more, as Fig.\ \ref{fig:hyst_comp} shows by plotting the first
inverse moment $\langle x^{-1} \rangle_\pi$, calculated as
$\int_{x}^{x+0.02} \varphi_\pi(x) x^{-1} dx$ and normalized to $100\%$
($y$-axis), the BMS\ DA receives in this region even less contributions
than the asymptotic DA, as we explained above.
\item By the same token, the Sudakov suppression of the endpoint region
of the BMS\ DA is less crucial compared to endpoint-concentrated DAs.
The implementation of Sudakov corrections using the analytic
factorization scheme was considered in technical detail in \cite{SSK99}
for the case of the asymptotic pion DA.
Such an analysis for the BMS\ DA is more involved and will be conducted
in a future publication.
\item The deep reason for the failure of the CZ DA was provided in
\cite{MR89,MR92,Rad97}.
The condensate terms in the CZ sum rules are strongly peaked at the
endpoints $x\to 0$ and $x\to 1$, the reason being that the vacuum quark
distribution in the longitudinal momentum fractions is approximated by
a $\delta$-function $\delta(x)$ and its derivatives.
For that reason, the condensate terms, i.e., the nonperturbative
contributions to the sum rule, force the pion DA to be
endpoint-concentrated, with the perturbative loop contribution
proportional to $x(1-x)$ being insufficient to compensate these
two sharp peaks at $x=0$ and $x=1$.
Allowing for a smooth distribution in the longitudinal momentum for the
vacuum quarks, i.e., using nonlocal condensates in the QCD sum rules
(as done in the derivation of the BMS\ pion DA), the endpoint regions
of the extracted DA are suppressed, despite the fact that its shape
is doubly peaked.
%%%%%%%%%%%%%%%%%%%%%%%%%%%%%%%%%%%%%%%%%%%%%%%%%%%%%%%%%%%%%%%%%%%%%%%
%                              FIGURE 4                               %
%%%%%%%%%%%%%%%%%%%%%%%%%%%%%%%%%%%%%%%%%%%%%%%%%%%%%%%%%%%%%%%%%%%%%%%
\begin{figure}[bh]
 \centerline{\includegraphics[width=0.5\textwidth]{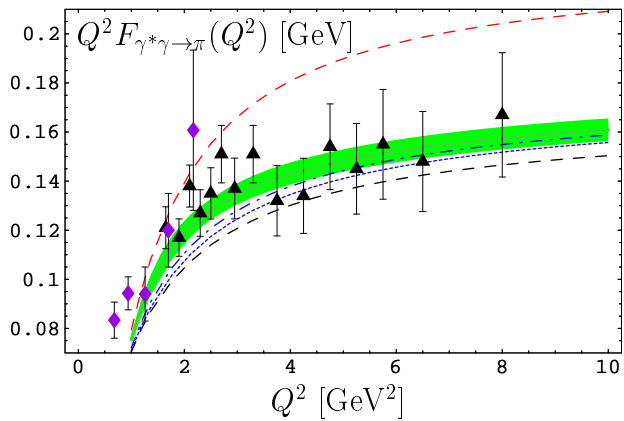}}
  \caption{\footnotesize
   Light-cone sum-rule predictions for
   $Q^2F_{\gamma^*\gamma\to\pi}(Q^2)$
   in comparison with the CELLO (diamonds, \protect\cite{CELLO91})
   and the CLEO (triangles, \protect\cite{CLEO98}) experimental data
   evaluated with the twist-4 parameter value
   $\delta_\text{Tw-4}^2=0.19$~GeV$^2$ \protect\cite{BMS02,BMS03}.
   The predictions shown correspond to the following pion DA models:
   $\varphi_\text{CZ}$ (upper dashed line) \protect\cite{CZ84},
   BMS-``bunch'' (shaded strip) \protect\cite{BMS01},
   two instanton-based models, viz., \cite{PPRWG99} (dotted line)
   and \cite{PR01} (dashed-dotted line), and the asymptotic pion DA
   $\varphi_\text{as}$ (lower dashed line) \protect\cite{ER80,LB79}.
   A recent transverse lattice result \protect{\cite{Dal02}} is very
   close to the dash-dotted line, but starts to be closer to the center
   of the BMS\ strip for $Q^2\geq6$~GeV$^2$ .}
 \label{fig:Formfactor}
\end{figure}
%%%%%%%%%%%%%%%%%%%%%%%%%%%%%%%%%%%%%%%%%%%%%%%%%%%%%%%%%%%%%%%%%%%%%%%
\item The endpoint behavior of the pion DA is the route cause why the
pion-photon transition form factor---which in LO is purely
electromagnetic---calculated with the CZ pion DA was found \cite{KR96}
to overshoot the CLEO data.
More recently, the analysis of the CLEO data using light-cone sum rules
\cite{Kho99,SY99,BMS02,BMS03} has excluded the CZ pion DA at the
4$\sigma$ confidence level, while the BMS\ DA was found to be inside
the 1$\sigma$ error ellipse (for $\lambda_{q}^{2} = 4$~GeV${}^2$),
whereas even the asymptotic DA  was also excluded by the CLEO data at
the 3$\sigma$ level.
These quoted findings are reflected in the behavior of the predictions
for the pion-photon transition form factor displayed in Fig.\
\ref{fig:Formfactor}, which is based on the corrected version of
\cite{BMS01} (the displayed strip is therefore slightly different from
that in \cite{BMSefr03}).
\end{itemize}

To make these statements more transparent, let us define the DA profile
deviation factor
\begin{equation}
  \Delta_{\varphi}
   \equiv
    \frac{\langle x^{-1} \rangle^{\rm \varphi}_{\pi}}
         {\langle x^{-1} \rangle^{\rm As}_{\pi}}
    = 1 + a_2 + a_4 + \ldots
\label{eq:Delta}
\end{equation}
%%Eq (3.11) Profile deviation factor from asymptotic pion DA
which quantifies the deviation of a model DA from the asymptotic one
and supply its value in Table \ref{tab:profilefactor} for
several pion DAs suggested in the literature in comparison
with the constraints from the experimental data and theoretical
calculations.
Reading this Table in conjunction with Fig.\ \ref{fig:Formfactor},
one comes to the conclusion that the BMS\ ``bunch'' provides the best
agreement with the CLEO and CELLO experimental constraints, being also
in compliance with various theoretical constraints and lattice
calculations.
%%%%%%%%%%%%%%%%%%%%%%%%%%%%%%%%%%%%%%%%%%%%%%%%%%%%%%%%%%%%%%%%%%%%%%%
%                                TABLE 1                              %
%%%%%%%%%%%%%%%%%%%%%%%%%%%%%%%%%%%%%%%%%%%%%%%%%%%%%%%%%%%%%%%%%%%%%%%
\begin{table}[h]
\caption{Estimates for the Gegenbauer coefficients and the DA profile
deviation factor $\Delta_{\varphi}$ up to polynomial order 4 for the
asymptotic, the BMS, and the CZ DAs compared with constraints derived
from light-cone sum rules (LCSR), QCD sum rules with nonlocal
condensates (NL QCD SR) for the DA and the inverse moment $\va{x^{-1}}$,
\protect\footnote{Note that the uncertainties on the Gegenbauer
coefficients $a_2$ and $a_4$ are correlated.
Here, the rectangular limits of the fiducial ellipses extracted from
the NL QCD SRs \cite{BMS01} and from the CLEO data in
\protect\cite{BMS02,BMS03} are shown.} and by analyzing the CLEO data.
Also shown are the corresponding entries for instanton-based models
(ADT, PPRWG, PR) and those associated with a transverse-lattice
calculation---labelled Lattice.
All values displayed are normalized at the scale $\mu^2=1.35$~GeV$^2$,
corresponding to the scale of NL QCD SRs.
\label{tab:profilefactor}}
\begin{ruledtabular}
\begin{tabular}{cccccc}
DA models/methods
     & $a_2$
        & $a_4$
           & $a_2+a_4$
              & $a_2-a_4$
                 & $\Delta_{\varphi}$
             \\ \hline \hline
As   & 0
        & 0
           & $0$
              & $0$
                 & $1$
             \\
BMS  & $0.19$
        & $-0.13$
           & $0.06$
              & $0.32$
                 & $1.06$
              \\
CZ   & $0.52$
        & $0$
           & $0.52$
              & $0.52$
                 & $1.52$
               \\
PPRWG~\cite{PPRWG99}
     & $0.042$
        & $0.006$
           & $0.05$
              & $0.04$
                 & $1.05$
              \\
PR~\cite{Pra01}
     & $0.09$
        & $-0.02$
           & $0.07$
              & $0.10$
                 & $1.07$
              \\
ADT~\cite{ADT00}
     & $0.05$
        & $-0.04$
           & $0.01$
              & $0.09$
                 & $1.01$
              \\
Lattice~\cite{Dal02}
     & $0.08$
        & $0.02$
           & $0.10$
              & $0.06$
                 & $1.10$
              \\
LCSRs \protect{\cite{BiKho02}}
     & $[0.07, 0.37]$
        & ---
           & ---
              & ---
                 & $1.22 \pm 0.15$
               \\
NL QCD SRs for DA \protect{\cite{BMS01}}
     & $[0.13, 0.25]^a$
        & $[-0.04, -0.21]^a$
           & $[+0.02, +0.09]$
               & $[+0.18, +0.46]$
                 & $1.06 \pm 0.04$
              \\
NL QCD SR for $\va{x^{-1}}$ \protect{\cite{BMS01}}
     & ---
        & ---
           & $[+0.00, +0.20]$
               & ---
                 & $1.10 \pm 0.10$
              \\
CLEO $1\sigma$-limits \protect{\cite{BMS03}}
     & $[0.15, 0.43]^a$
        & $[-0.60, -0.04]^a$
           & $[-0.21, +0.15]$
               & $[+0.21, +1.00]$
                  & $0.97 \pm 0.18$
              \\
CLEO $2\sigma$-limits \protect{\cite{BMS03}}
     & $[0.11, 0.47]^a$
        & $[-0.71, +0.07]^a$
           & $[-0.31, +0.25]$
               & $[+0.07,+1.14]$
                  & $0.97 \pm 0.28$
              \\
CLEO $3\sigma$-limits \protect{\cite{BMS03}}
     & $[0.07, 0.51]^a$
        & $[-0.82, +0.19]^a$
           & $[-0.41, +0.35]$
               & $[-0.07, +1.28]$
                  & $0.97 \pm 0.38$
\end{tabular}
\end{ruledtabular}
\end{table}
%%%%%%%%%%%%%%%%%%%%%%%%%%%%%%%%%%%%%%%%%%%%%%%%%%%%%%%%%%%%%%%%%%%%%%%

%%%%%%%%%%%%%%%%%%%%%%%%%%%%%%%%%%%%%%%%%%%%%%%%%%%%%%%%%%%%%%%%%%%%%%%
%%%%%%%%%%%%%%%%%%%%%%%%%%%%%%%%%%%%%%%%%%%%%%%%%%%%%%%%%%%%%%%%%%%%%%%
\subsection{Perturbative NLO evolution}
 \label{sect:NLOEvo}
%%%%%%%%%%%%%%%%%%%%%%%%%%%%%%%%%%%%%%%%%%%%%%%%%%%%%%%%%%%%%%%%%%%%%%%
Let us now discuss how the pion DA evolves at NLO using first standard
perturbation theory to be followed by analogous considerations within
APT.
The evolution of the distribution amplitude \req{eq:phi024mu0}
proceeds along the lines outlined in Appendix \ref{app:DAnlo}.
Taking into account only the first two Gegenbauer coefficients
and LO evolution, one obtains
\begin{eqnarray}
 \varphi_{\pi}^{\text{LO}}(x,\muF)
  = 6 x (1-x)
     \left[ 1
          + a_2^{\text{D,LO}}(\muF) \, C_2^{3/2}(2 x -1)
          + a_4^{\text{D,LO}}(\muF) \, C_4^{3/2}(2 x -1)
     \right]\,,
\label{eq:phi024LO}
\end{eqnarray}
%%Eq (3.12) Evolution of pion DA as Gegenbauer expansion up to
%%          polynomial order 2
where $a_2^{\text{D,LO}}(\muF)$ and $a_4^{\text{D,LO}}(\muF)$
are given by \req{eq:anLO} taking recourse to \req{eq:EnLO},
and D denotes ``diagonal'', while ND below stands for ``nondiagonal''.
%%%%%%%%%%%%%%%%%%%%%%%%%%%%%%%%%%%%%%%%%%%%%%%%%%%%%%%%%%%%%%%%%%%%%%%
%                              FIGURE 5                               %
%%%%%%%%%%%%%%%%%%%%%%%%%%%%%%%%%%%%%%%%%%%%%%%%%%%%%%%%%%%%%%%%%%%%%%%
\begin{figure}[ht]
 \centerline{\includegraphics[width=0.9\textwidth]{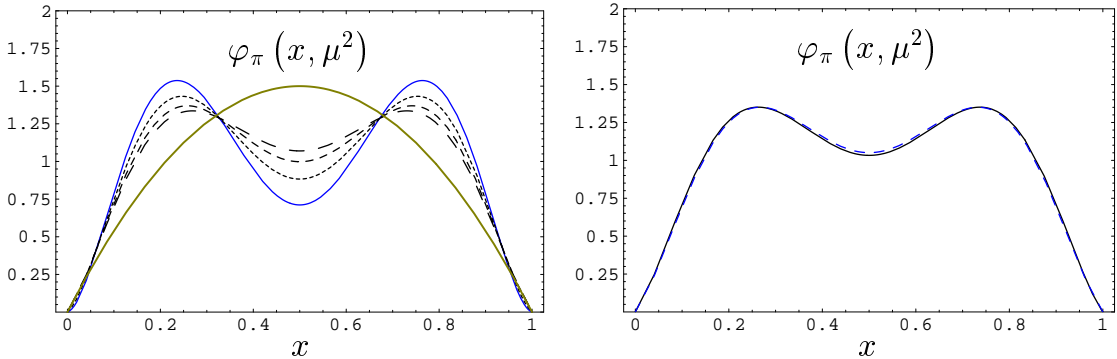}}
  \caption{\footnotesize
    The left panel shows the effect of the LO diagonal part of the
    evolution equation, Eq.\ (\ref{eq:phi024LO}), on the BMS\ DA.
    The convex solid line denotes the asymptotic profile of the pion
    DA, the other solid one stands for
    $\varphi_\text{BMS}^\text{LO}(x)$ at 1 GeV$^2$,
    while the  broken lines represent $\varphi_\text{BMS}^\text{LO}(x)$
    at 4, 20, and 100 GeV$^2$ (with the larger scale corresponding also
    to the larger value of the DA at the middle point).
    Right panel:
    Comparison of $\varphi_\text{BMS}^\text{LO}$
    (Eq.\ (\ref{eq:phi024LO})---solid line)
    and $\varphi_\text{BMS}^\text{NLO}$
    (Eq.\ (\ref{eq:phi024NLOall})---dashed line) at 20 GeV$^2$ to
    illustrate the effect of NLO evolution.
    \label{fig:bms_evo}}
\end{figure}
%%%%%%%%%%%%%%%%%%%%%%%%%%%%%%%%%%%%%%%%%%%%%%%%%%%%%%%%%%%%%%%%%%%%%%%
On the other hand, the solution of the NLO evolution equation takes the
form
\begin{subequations}
\label{eq:phi024NLOall}
\begin{eqnarray}
 \varphi_{\pi}^{\text{NLO}}(x,\muF)
  = \varphi_{\pi}^\text{D,NLO}(x,\muF)
  + \frac{\alpha_s(\muF)}{4\pi}\,
     \varphi_{\pi}^\text{ND,NLO}(x,\muF)\,,
 \label{eq:phi024NLO}
\end{eqnarray}
%%Eq (3.13a) General solution of NLO evolution equation for pion DA
where
\begin{eqnarray}
 \varphi_{\pi}^\text{D,NLO}(x,\muF)
  = 6 x (1-x)
     \left[1
        + \sum_{n=2, 4}a_n^\text{D,NLO}(\muF) \, C_n^{3/2}(2x-1)
     \right]
\label{eq:phi024NLOdia}
\end{eqnarray}
%%Eq (3.13b) Diagonal part of NLO pion DA
and
\begin{eqnarray}
 \varphi_{\pi}^\text{ND,NLO}(x,\muF)
  &=& 6 x (1-x)
       {\sum_{n\geq 2}}'
        a_n^\text{ND,NLO}(\muF) \, C_n^{3/2}(2 x -1)\,.
 \label{eq:phi024NLOnondia}
\end{eqnarray}
%%Eq (3.13c) Nondiagonal part of NLO pion DA
\end{subequations}
The coefficients
$a_n^\text{D,NLO}(\muF)$ and
$a_n^\text{ND,NLO}(\muF)$
are given in \req{eq:anNLO0} and \req{eq:anNLO1}, respectively, by
employing (\ref{eq:EnNLO}), while ${\sum}'$ denotes the sum over
even indices only.
Note that, although the input DA, $\varphi_\pi(x,\muO)$, was
parameterized by only two Gegenbauer coefficients $a_2$ and $a_4$,
higher harmonics also appear due to the nondiagonal nature of the NLO
evolution.\footnote{Since $a_n^\text{ND,NLO}(\muF)$ decreases with
$n$, for the purpose of numerical calculations, we use an approximate
form of $\varphi_\text{ND}^{\text{NLO}}(x,\muF)$
in which we neglect $a_n^\text{ND,NLO}(\muF)$ for $n>100$.}
The effect of the inclusion of the LO diagonal part of the evolution
kernel is important, as one sees from the left part of Fig.\
\ref{fig:bms_evo}, which shows this effect for the BMS\ pion DA.

On the other hand, from the right part of Fig. \ref{fig:bms_evo},
we deduce that the NLO nondiagonal evolution is rather small.
We note that in the above computation the exact two-loop expression
for $\alpha_s$ \cite{Mag98} in the \MS-scheme
($\Lambda_\text{QCD}=410$ MeV, $N_f=3$) was employed, cf.\
(\ref{eq:alphaexact}), in which matching at the heavy-flavor
thresholds $M_4=1.3$ GeV and $M_5=4.3$ GeV (with $M_1=M_2=M_3=0$)
has been employed \cite{SM94}.
A discussion of the relation of this exact solution to the usual
approximation, promoted by the Particle Data Group~\cite{PDG2002},
has recently been given in \cite{BMS03} (see also Appendix C).

%%%%%%%%%%%%%%%%%%%%%%%%%%%%%%%%%%%%%%%%%%%%%%%%%%%%%%%%%%%%%%%%%%%%%%%
%%%%%%%%%%%%%%%%%%%%%%%%%%%%%%%%%%%%%%%%%%%%%%%%%%%%%%%%%%%%%%%%%%%%%%%
\section{Pion form factor at NLO: Analytic results}
\label{sect:PFFNLO:AR}
%%%%%%%%%%%%%%%%%%%%%%%%%%%%%%%%%%%%%%%%%%%%%%%%%%%%%%%%%%%%%%%%%%%%%%%
The NLO results for the hard-scattering amplitude $T_\text{H}$ are
summarized in Appendix \ref{app:HSAnlo}.
Setting in \req{eq:TH} $\muF=Q^2$ and taking into account the NLO
evolution of the pion DA $\varphi$ via \req{eq:phi024NLOall}, we
obtain from (\ref{eq:pff-Fact})
\begin{eqnarray}
 F_{\pi}^\text{Fact}(Q^2; \muR)
  = F_{\pi}^\text{LO}(Q^2;\muR)
  + F_{\pi}^\text{NLO}(Q^2;\muR) \, ,
 \label{eq:Q2pff}
\end{eqnarray}
%%Eq (4.1) Factorized pion FF as sum of LO and NLO terms
where the LO term is given by
\begin{eqnarray}
 F_{\pi}^\text{LO}(Q^2;\muR)
  &=& \alpha_s(\muR)\, {\cal F}_{\pi}^\text{LO}(Q^2)\\
  Q^2 {\cal F}_{\pi}^\text{LO}(Q^2)
  &\equiv& 8\,\pi\,f_{\pi}^2\,
     \left[1 + a_2^\text{D,NLO}(Q^2) + a_4^\text{D,NLO}(Q^2)\right]^2
 \label{eq:Q2pffLO}
\end{eqnarray}
%%Eq (4.2) LO pion FF in terms of calligraphic FF
%%Eq (4.3) LO calligraphic pion FF with NLO evolution of pion DA
and the calligraphic designations denote quantities with their
$\alpha_s$-dependence pulled out.
In order to make a distinction between the contributions stemming from
the diagonal and the nondiagonal parts of the NLO evolution equation of
the pion DA, we express the NLO correction to the form factor in the
form
\begin{eqnarray}
 F_{\pi}^\text{NLO}(Q^2;\muR)
 = \frac{\alpha_s^2(\muR)}{\pi}\,
   \left[{\cal F}_\pi^\text{D,NLO}(Q^2;\muR)
    + {\cal F}_\pi^\text{ND,NLO}(Q^2;N_\text{Max}=\infty)
   \right]
   \label{eq:Q2pffNLO}
\end{eqnarray}
%%Eq (4.4) NLO pion FF in terms of calligraphic pion FF
and write the diagonal contribution
\begin{eqnarray}
 {\cal F}_\pi^\text{D,NLO}(Q^2;\muR)
 &\equiv& b_0\, {\cal F}_{\pi}^{(1,\beta)}(Q^2;\muR)
         + C_\text{F} {\cal F}_{\pi}^{(1,\text{F})}(Q^2)
         + C_\text{G} {\cal F}_{\pi}^{(1,\text{G})}(Q^2)
 \label{eq:Q2pffNLODia}
\end{eqnarray}
%%Eq (4.5) NLO Diagonal part of calligraphic pion FF
as a color decomposition (in correspondence with (\ref{eq:THNLOpff}))
in terms of
\begin{subequations}
\begin{eqnarray}
 Q^2 {\cal F}_{\pi}^{(1,\beta)}(Q^2;\muR)
& = & 2\,\pi\,f_{\pi}^2\,
      \left[\displaystyle
            \frac{5}{3}
          + \frac{3 + \displaystyle (43/6) a_2^\text{D,NLO}(Q^2)
                    + (136/15) a_4^\text{D,NLO}(Q^2)}
                 {1 + a_2^\text{D,NLO}(Q^2) + a_4^\text{D,NLO}(Q^2)}
          -  \ln \frac{Q^2}{\muR}
     \right]\nonumber\\
  & & \times
       \left[
             1 + a_2^\text{D,NLO}(Q^2) + a_4^\text{D,NLO}(Q^2)
       \right]^2 \, ,
\label{eq:Q2pff1beta}
\end{eqnarray}
%%Eq (4.6a) (1, beta) term of NLO part of calligraphic pion FF
\begin{eqnarray}
Q^2 {\cal F}_{\pi}^{(1,\text{F})}(Q^2)
  &=& 2\,\pi\,f_{\pi}^2\,
      \left\{- \frac{71}{6}
            - a_2^\text{D,NLO}(Q^2)
              \left[\frac{497}{36} - \frac{161}{24}\,
              a_2^\text{D,NLO}(Q^2)
                   \right]
      \right.\nonumber\\
  & & \left. \qquad\quad
            -\ a_4^\text{D,NLO}(Q^2)
             \left[\frac{1123}{450}
                 - \frac{9793}{300}\, a_2^\text{D,NLO}(Q^2)
                 - \frac{1387}{50}\, a_4^\text{D,NLO}(Q^2)\right]
      \right\}\, ,
\nonumber\\
\label{eq:Q2pff1F}
\end{eqnarray}
%%Eq (4.6b) (1, F) term of NLO part of calligraphic pion FF
and
\begin{eqnarray}
 Q^2 {\cal F}_{\pi}^{(1,\text{G})}(Q^2)
  &=& 2\,\pi\,f_{\pi}^2\,
      \left\{- 0.67
            + a_2^\text{D,NLO}(Q^2)
             \left[18.70 + 16.35\, a_2^\text{D,NLO}(Q^2)
                   \right]
      \right. \nonumber\\
  & & \left. \qquad\quad
            +\ a_4^\text{D,NLO}(Q^2)
              \left[24.23 + 36.76\, a_2^\text{D,NLO}(Q^2)
                         + 20.26\, a_4^\text{D,NLO}(Q^2)\right]
      \right\}\, ,
\nonumber\\
\label{eq:Q2pff1G}
\end{eqnarray}
\label{eq:Q2pff1}
\end{subequations}
%%Eq (4.6c) (1, G) term of NLO part of calligraphic pion FF
$\!\!$where the superscripts F  and G
refer to the color factors $C_\text{F}$
and $C_\text{G}=C_\text{F}-C_\text{A}/2$,
respectively.
Note that for the matter of calculational convenience, we also display
the sum of these two terms (cf.\ Eq.\ (\ref{eq:TH1C1})):
\begin{eqnarray}
 Q^2 {\cal F}_{\pi}^{(1,\text{FG})}(Q^2)
& = & 2\,\pi\,f_{\pi}^2\,
      \left\{- 15.67
             - a_2^\text{D,NLO}(Q^2)
             \left[21.52 - 6.22\, a_2^\text{D,NLO}(Q^2)
             \right]
      \right.\nonumber\\
  & & \left. -\ a_4^\text{D,NLO}(Q^2)
              \left[7.37 - 37.40\, a_2^\text{D,NLO}(Q^2)
                         - 33.61\, a_4^\text{D,NLO}(Q^2)
              \right]
      \right\}\,.
\nonumber\\
\label{eq:Q2pff1FG}
\end{eqnarray}
%%Eq (4.7) (1, FG) term of NLO part of calligraphic pion FF
On the other hand, the nondiagonal term reads
\begin{eqnarray}
 Q^2 {\cal F}_\pi^\text{ND,NLO}(Q^2;N_\text{Max})
 = 4\pi\,f_{\pi}^2
    \left[1+a_2^\text{D,NLO}(Q^2)+a_4^\text{D,NLO}(Q^2)\right]
     \sum_{k=2}^{N_\text{Max}}{}' a_{k}^\text{ND,NLO}(Q^2)\,,
\label{eq:Q2pffNLOnondia}
\end{eqnarray}
%%Eq (4.8) Nondiagonal part of NLO calligraphic pion FF
where $N_\text{Max}$ denotes the maximal number of Gegenbauer harmonics
taken into account in order to achieve the desired accuracy.

As it was shown in Ref.\ \cite{MNP99a}, the effects of the LO DA
evolution are crucial.
In order to investigate the importance of the NLO DA evolution, we
compare the predictions obtained using the complete NLO results,
given above, with those derived by employing only the LO DA evolution
via \req{eq:phi024LO}.
The corresponding expressions in this latter case follow from those
above by performing the replacements
\begin{eqnarray}
a_n^\text{D,NLO} \to a_n^{\text{D,LO}}
 \qquad \text{and} \qquad
a_n^\text{ND,NLO} \to  0\,,
\label{eq:onlyLOev}
\end{eqnarray}
%%Eq (4.9) LO coefficients by replacement from NLO ones
so that the contribution
${\cal F}_\pi^\text{ND,NLO}(Q^2;\muR)$
is absent.
Introducing the notation
\begin{eqnarray}
  \widetilde{\cal F}_{\pi}^{i}
  \equiv {\cal F}_{\pi}^{i}|_{\text{LO DA evolution}}
  \,,
 \label{eq:FpionlyLOev}
\end{eqnarray}
%%Eq (4.10) Notation for tilded calligraphic pion FF
we analyze the relative importance of the various contributions
(LO, NLO, and Local Duality (LD) part---see Sec.\ \ref{sect:NFPFF})
by defining the following ratios
\begin{eqnarray}
 \label{eq:Ratio}
  R\left(Q^2,N_\text{Max}\right)
   &=& \frac{{\cal F}_{\pi}^\text{LO}\left(Q^2\right) + (\alpha_s/\pi)
             {\cal F}_\pi^\text{ND,NLO}\left(Q^2;N_\text{Max}
                                                \right)}
            {{\cal F}_{\pi}^\text{LO}\left(Q^2\right) + (\alpha_s/\pi)
             {\cal F}_\pi^\text{ND,NLO}
                     \left(Q^2;N_\text{Max}\approx\infty\right)}
  \,;\\
  \label{eq:hat-Ratio}
\hat{R}\left(Q^2,N_\text{Max}\right)
   &=& \frac{F_{\pi}^\text{LD}\left(Q^2\right)
            + \alpha_s {\cal F}_{\pi}^\text{LO}\left(Q^2\right)
            + (\alpha_s^2/\pi)
              {\cal F}_\pi^\text{ND,NLO}\left(Q^2;N_\text{Max}
                                                 \right)}
            {F_{\pi}^\text{LD}\left(Q^2\right)
            + \alpha_s {\cal F}_{\pi}^\text{LO}\left(Q^2\right)
            + (\alpha_s^2/\pi)
              {\cal F}_\pi^\text{ND,NLO}\left(Q^2;N_\text{Max}
   \approx\infty\right)}
  \,;\\
 \label{eq:mod-Ratio}
  R_\text{mod}\left(Q^2\right)
   &=& \frac{\widetilde{\cal F}_{\pi}^\text{LO}\left(Q^2\right)}
            {{\cal F}_{\pi}^\text{LO}\left(Q^2\right)
            + (\alpha_s/\pi)
              {\cal F}_\pi^\text{ND,NLO}\left(Q^2;N_\text{Max}
   \approx\infty\right)}\,;\\
  \label{eq:mod-hat-Ratio}
  \hat{R}_\text{mod}\left(Q^2\right)
   &=& \frac{F_{\pi}^\text{LD}\left(Q^2\right)
          + \alpha_s \widetilde{\cal F}_{\pi}^\text{LO}\left(Q^2\right)}
            {F_{\pi}^\text{LD}\left(Q^2\right)
          + \alpha_s {\cal F}_{\pi}^\text{LO}\left(Q^2\right)
          + (\alpha_s^2/\pi)
            {\cal F}_\pi^\text{ND,NLO}\left(Q^2;N_\text{Max}
   \approx\infty\right)}\,.
\end{eqnarray}
%%Eqs (4.11) (4.12) (4.13) (4.14) Ratios of LO, NLO, and LD parts of
%%                                pion FF
%%%%%%%%%%%%%%%%%%%%%%%%%%%%%%%%%%%%%%%%%%%%%%%%%%%%%%%%%%%%%%%%%%%%%%%
%                              FIGURE 6                               %
%%%%%%%%%%%%%%%%%%%%%%%%%%%%%%%%%%%%%%%%%%%%%%%%%%%%%%%%%%%%%%%%%%%%%%%
\begin{figure}[ht]
 \centerline{\includegraphics[width=0.49\textwidth]{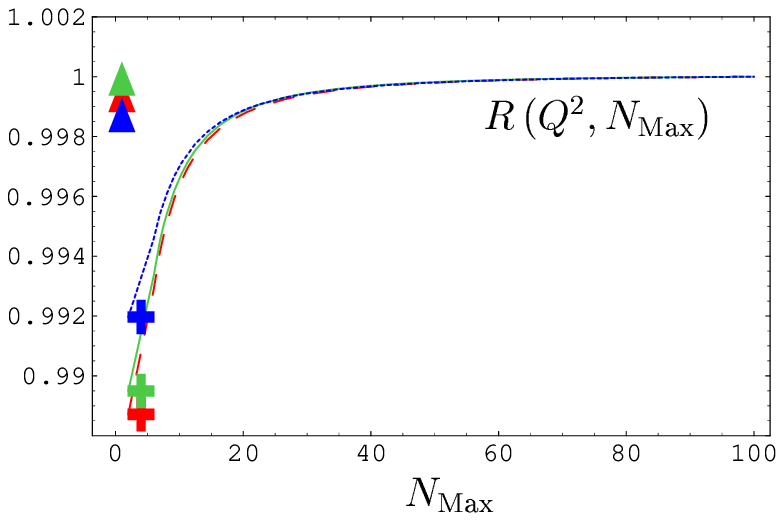}~~~%
  \includegraphics[width=0.49\textwidth]{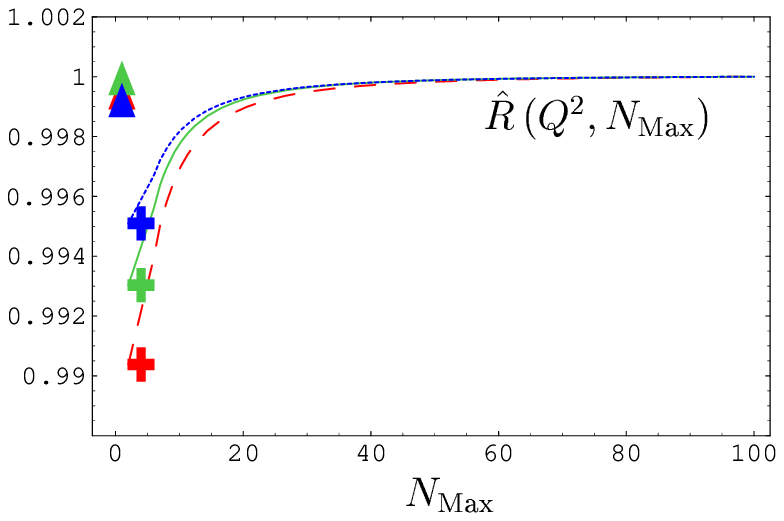}}
    \caption{\footnotesize
     Left: The ratio $R\left(Q^2,N_\text{Max}\right)$, defined in Eq.\
     (\ref{eq:Ratio}), for three different values of $Q^2$ as a
     function of $N_\text{Max}$.
     The blue dotted curve corresponds to $Q^2=2$~GeV$^2$, the
     green solid curve to $Q^2=10$~GeV$^2$, and the red dashed one
     to $Q^2=50$~GeV$^2$.
     Crosses of the same color represent the values of
     $R\left(Q^2,N_\text{Max}=4\right)$, whereas triangles refer to
     $R_\text{mod}\left(Q^2\right)$, Eq.\ (\ref{eq:mod-Ratio}).
     Right: The same designations hold as for the left side, but now
     for the ratio
     $\hat{R}\left(Q^2,N_\text{Max}\right)$
     given by Eq.\ (\ref{eq:hat-Ratio}) and correspondingly for
     $\hat{R}_\text{mod}$---Eq.\ (\ref{eq:mod-hat-Ratio}).
    \label{fig:ratio}}
\end{figure}
%%%%%%%%%%%%%%%%%%%%%%%%%%%%%%%%%%%%%%%%%%%%%%%%%%%%%%%%%%%%%%%%%%%%%%%
\\
\noindent
These ratios are displayed graphically in Fig.\ \ref{fig:ratio} for the
BMS\ DA ($a_2(\muO)=0.2$, $a_4(\muO)=-0.14$)
in the region $N_\text{Max}=4-100$.
We infer from this figure that, adopting in our calculations an
accuracy on the order of 99.5\%, we can safely neglect the non-diagonal
part of the NLO evolution equation and use for the pion form-factor
computations to follow the approximate expression
(omitting the superscript Approx)
\begin{eqnarray}
 F_{\pi}^\text{Fact-Approx}(Q^2; \muR)
  \equiv
  \alpha_s(\muR)\,
  \tilde{\cal F}_{\pi}^\text{LO}(Q^2;\muR)
  + \frac{\alpha_s^2(\muR)}{\pi}\,
  {\cal F}_\pi^\text{D,NLO}(Q^2;\muR)
  \,.
 \label{eq:Q2pffApprox}
\end{eqnarray}
%%Eq (4.15) Approximated form of pion FF at NLO
Actually, the difference between
${\cal F}_\pi^\text{D,NLO}(Q^2;\muR)$
and
$\tilde{\cal F}_\pi^\text{D,NLO}(Q^2;\muR)$
is of the order of $\alpha_s(Q^2)$,
so that it is safe to use everywhere only the LO evolution.
We have verified in our numerical calculations
that the difference is indeed less than 1\%.

%%%%%%%%%%%%%%%%%%%%%%%%%%%%%%%%%%%%%%%%%%%%%%%%%%%%%%%%%%%%%%%%%%%%%%%
%%%%%%%%%%%%%%%%%%%%%%%%%%%%%%%%%%%%%%%%%%%%%%%%%%%%%%%%%%%%%%%%%%%%%%%
\section{Setting the Renormalization Scheme and Scale}
\label{sect:RenSc}
%%%%%%%%%%%%%%%%%%%%%%%%%%%%%%%%%%%%%%%%%%%%%%%%%%%%%%%%%%%%%%%%%%%%%%%
The choice of the expansion parameter represents the major ambiguity in
the interpretation of the pQCD predictions because finite-order
perturbative predictions depend unavoidably on the renormalization
scale and scheme choice.\footnote{Actually, to the order we are
calculating these dependencies, they can be represented by a single
parameter, say, the renormalization scale, because $b_0$ and $b_1$ are
renormalization-scheme invariant.}
If one could optimize the choice of the renormalization scale and
scheme according to some sensible criteria, the size of the
higher-order corrections, as well as the size of the expansion
parameter, i.e., the QCD running coupling, could serve as sensible
indicators for the convergence of the perturbative expansion.
In what follows, we shall consider several scheme and scale-setting
options.

%%%%%%%%%%%%%%%%%%%%%%%%%%%%%%%%%%%%%%%%%%%%%%%%%%%%%%%%%%%%%%%%%%%%%%%
\subsection{$\overline{\text{MS}}$ scheme}
 \label{sect:MS-bar}
%%%%%%%%%%%%%%%%%%%%%%%%%%%%%%%%%%%%%%%%%%%%%%%%%%%%%%%%%%%%%%%%%%%%%%%
The results we have presented in the previous subsection were obtained
in the \MS renormalization (and factorization) scheme.
Let us discuss the choice of the renormalization scale $\mu_\text{R}$
in some more detail.
We see that in our NLO results, Eq.\ (\ref{eq:Q2pffNLO}), this
dependence is contained in the coupling constant $\alpha_s(\muR)$
as well as in the NLO correction ${\cal F}_{\pi}^{(1,\beta)}$.
The latter correction is proportional to the $b_0$ coefficient of the
$\beta$ function and is $N_f$-dependent.
Hence, a natural question arises:
How can we determine the right value of $N_f$ in the form-factor
expression?

We propose here to apply the following procedures.\\
(i) The first one concerns the standard choice $\muR=Q^2$ and
    suggests to shift $\muR$ at the heavy-quark threshold in order to
    ensure the continuity of the form factor according to
\begin{subequations}\label{eq:FF-Threshold}
 \begin{eqnarray}
   F_{\pi}^\text{Fact}(Q^2)
    = \left\{
       \begin{array}{ll}
        F_{\pi}^\text{Fact}(Q^2; \muR=Q^2)\big|_{N_f=3}
         &\text{for}~Q^2\leq M_4^2;\\
        F_{\pi}^\text{Fact}(Q^2;
        \muR=Q^2+\delta\mu^2_4)\big|_{N_f=4}\vphantom{^{\Big|}}
         &\text{for}~\left\{\begin{array}{l}
                             M_4^2\leq Q^2\\
                             Q^2+\delta\mu^2_4\leq M_5^2
                            \end{array}
                     \right.; \\
        F_{\pi}^\text{Fact}(Q^2;
        \muR=Q^2+\delta\mu^2_4
       +\delta\mu^2_5)\big|_{N_f=5}\vphantom{^{\Big|}}
         &\text{for}~M_5^2\leq Q^2+\delta\mu^2_4\,.
       \end{array}\right.
  \end{eqnarray}
As a result, we have to fulfill the following matching conditions
 \begin{eqnarray}
   F_{\pi}^\text{Fact}(M_4^2; M_4^2)\big|_{N_f=3}
   &=& F_{\pi}^\text{Fact}(M_4^2;
   M_4^2+\delta\mu^2_4)\big|_{N_f=4}\,;\\
   F_{\pi}^\text{Fact}(M_5^2-\delta\mu^2_4;
   M_5^2)\big|_{N_f=4}
   &=& F_{\pi}^\text{Fact}(M_5^2-\delta\mu^2_4;
   M_5^2+\delta\mu^2_5)\big|_{N_f=5}.
   \quad
 \end{eqnarray}
\end{subequations}
%%Eq (5.1a) (5.1b) (5.1c) Matching conditions for quark-mass thresholds

\noindent
(ii) The second procedure addresses specifically the BLM\ scale setting
     $\muR=\mu_\text{BLM}^2$.
     In this case, the only problem is the small value of the BLM\
     scale (see Table \ref{t:scales}) due to the fact that the
     $b_0$-term is completely absent and $N_f$-dependent terms do not
     arise.
     Therefore, we propose to implement the BLM\ scale setting only
     above some minimal scale: $\mu_\text{min}$.
     Below this scale, which is in the range of the typical meson
     scales and hence only the light-quark sector ($N_f=3$)
     contributes, we fix $\muR=\mu_\text{min}^2$ and set $N_f=3$
     using the
     ${\cal F}_{\pi}^{(1,\beta)}(Q^2;\mu_\text{min}^2)$-term
     in the form provided by (\ref{eq:BLM-improvement})---more
     explanations will be given shortly.

The truncation of the perturbative series to a finite order introduces
a residual dependence of the results on the scale $\mu_\text{R}$,
while the inclusion of higher-order corrections decreases this
dependence.
Nonetheless, we are still left with an intrinsic theoretical
ambiguity of the perturbative results.
One can try to estimate the uncertainty entailed by this ambiguity
(see, for example, \cite{MNP99a}) or choose the renormalization scale
$\mu_\text{R}$ on the basis of some physical arguments.

The simplest and widely used choice for $\mu_\text{R}$ is to identify
it with the large external scale, i.e., to set
\begin{eqnarray}
\muR = Q^2
\,,
\label{eq:muRQ2}
\end{eqnarray}
%%Eq (5.2) Default choice of renormalization scale
the justification for adopting this choice being mainly a pragmatic
one.
However, physical arguments suggest that a more appropriate scale
should be lower.
Namely, since each external momentum entering an exclusive reaction is
partitioned among many propagators of the underlying hard-scattering
amplitude in the associated Feynman diagrams, the physical scales
that control these processes are related to the average momentum
flowing through the internal quark and gluon lines and are therefore
inevitably softer than the overall momentum transfer.
To treat this problem, several suggestions have been made in the
literature.
According to the so-called Fastest Apparent Convergence (FAC)
procedure \cite{Grunberg80,Grunberg84}, the scale $\mu_\text{R}$ is
determined by the requirement that the NLO coefficient in the
perturbative expansion of the physical quantity in question
vanishes which here means
\begin{eqnarray}
F_{\pi}^{\text{NLO}}(Q^2;\muR=\mu_{\text{FAC}}^2)=0 \, .
\label{eq:FACpff}
\end{eqnarray}
%%Eq (5.3) Definition of FAC scheme
On the other hand, following the Principle of Minimum Sensitivity
(PMS) \cite{Stevenson81a,Stevenson81,Stevenson82,Stevenson84}, one
mimics locally the global independence of the  all-order
expansion by choosing the renormalization scale $\mu_\text{R}$ to
coincide with the stationary point of the truncated perturbative
series.
In our case, this reads
\begin{eqnarray}
\frac {d} {d\muR} \left[ F_{\pi}^{\text{LO}}(Q^2;\muR) +
                         F_{\pi}^{\text{NLO}}(Q^2;\muR)
                  \right]
         \left|_{\displaystyle\muR=\displaystyle\mu_{\text{PMS}}^2}
     \right.= 0 \, .
\label{eq:PMSpff}
\end{eqnarray}
%%Eq (5.4) Definition of PMS scheme

In the Brodsky--Lepage--Mackenzie (BLM) procedure \cite{BLM83},
all vacuum-po\-la\-ri\-zat\-ion effects from the QCD
$\beta$-function (i.e., the effects of quark loops) are absorbed
into the renormalized running coupling by resumming the large
($b_0 \alpha_S$)$^n$ terms giving rise to infrared renormalons.
According to the BLM\ procedure, the renormalization scale best
suited to a particular process at a given order of expansion can
be, in practice, determined by demanding that the terms proportional
to the $\beta$-function should vanish.
This naturally connects to conformal field theory and we refer the
interested reader to \cite{BKM03} for a recent review.
The optimization of the renormalization scale and scheme setting in
exclusive processes by employing the BLM\ scale fixing was elaborated
in \cite{BJPR98} and in references cited therein.
The renormalization scales in the BLM\ method are ``physical'' in the
sense that they reflect the mean virtuality of the gluon propagators
involved in the Feynman diagrams.
According to the BLM\ procedure, the renormalization scale is
determined by the condition
\begin{eqnarray}
{\cal F}_{\pi}^{(1,\beta)}(Q^2;\muR=\mu_{\text{BLM}}^2)=0
     \, .
\label{eq:BLMpff}
\end{eqnarray}
%%Eq (5.5) BLM scale setting condition
For calculational convenience, we express $\mu_\text{R}^2$ in
terms of $Q^2$:
\begin{eqnarray}
\muR = a(Q^2) \; Q^2
\label{eq:muRaQ2}
\end{eqnarray}
%%Eq (5.6) Renormalization scale in terms of Q^2
and proceed to calculate this quantity in the above-mentioned
scale-setting schemes.
Then, the FAC\ procedure leads to
\begin{eqnarray}
\label{eq:muFACpff}
  a_{\text{FAC}}(Q^2)& =& \exp \left[
       - \frac{5}{3}
       - \frac{3 +\displaystyle (43/6) a_2^{\text{D,NLO}}(Q^2) +
                   (136/15) a_4^{\text{D,NLO}}(Q^2)}{%
               1 + a_2^{\text{D,NLO}}(Q^2) + a_4^{\text{D,NLO}}(Q^2)}
          \right. \nonumber \\ & & \left. \qquad
       -  \frac{4}{b_0}\,
           \frac{{\cal F}_{\pi}^{(1,\text{FG})}(Q^2)
       +{\cal F}_\pi^\text{ND,NLO}(Q^2)}
                {{\cal F}_{\pi}^{\text{LO}}(Q^2)}
           \right]\,,
\end{eqnarray}
%%Eq (5.7) Rescaling factor for FAC scheme
which can be related to the PMS\ procedure via
\begin{eqnarray}
  a_{\text{PMS}}(Q^2)& =& e^{-c_1/2} a_{\text{FAC}}(Q^2)
\label{eq:PMSsc}
\end{eqnarray}
%%Eq (5.8) Relation between FAC and PMS schemes
with $c_1\equiv b_1/b_0^2$.
This value corresponds to the stationary point (the maximum) of the
NLO prediction for $F_{\pi}^\text{Fact}$.

On the other hand, for the BLM\ scale one obtains
\begin{subequations}
\begin{eqnarray}
  \mu_{\text{BLM}}^2
  = a_{\text{BLM}}(Q^2) \, Q^2\, ,
 \label{eq:muBLMpff}
\end{eqnarray}
%%Eq (5.9a) BLM renormalization scale in terms of Q^2
where
\begin{eqnarray}
  a_{\text{BLM}}(Q^2) = \exp \left[ \displaystyle
       - \frac{5}{3}
       - \frac{3 +\displaystyle (43/6) a_2^{\text{D,NLO}}(Q^2) +
                   (136/15) a_4^{\text{D,NLO}}(Q^2)}{%
               1 + a_2^{\text{D,NLO}}(Q^2) + a_4^{\text{D,NLO}}(Q^2)}
                      \right]
                 \, .
\label{eq:aBLMpff}
\end{eqnarray}
%%Eq (5.9b) BLM rescaling factor
\label{eq:BLMscPff}
\end{subequations}

%%%%%%%%%%%%%%%%%%%%%%%%%%%%%%%%%%%%%%%%%%%%%%%%%%%%%%%%%%%%%%%%%%%%%%%
%                                TABLE 2                              %
%%%%%%%%%%%%%%%%%%%%%%%%%%%%%%%%%%%%%%%%%%%%%%%%%%%%%%%%%%%%%%%%%%%%%%%
\begin{table}
\caption{Scales $\mu_{\text{PMS}}$, $\mu_{\text{FAC}}$,
 $\mu_{\text{BLM}}$, and $\mu_{V}$
 for the asymptotic, the BMS, and the CZ
 DAs. \label{t:scales}}
\begin{ruledtabular}
\begin{tabular}{cccccc}
DA   & $Q^2/\mu_{\text{FAC}}^2$
        & $Q^2/\mu_{\text{PMS}}^2$
           & $Q^2/\mu_{\text{BLM}}^2$
              & $Q^2/\mu_{V\vphantom{_{|}}}^2$
                 & $Q^2$ \\ \hline \hline
As   & $\displaystyle 18$
        & $\displaystyle 27$
           & $\displaystyle 106$
              & $\displaystyle 20$
                 & any   \\
BMS  & $\displaystyle
       16-20$
        & $\displaystyle 24-29$
           & $\displaystyle 105-117$
              & $\displaystyle 20-22$
                 & $1-50$ GeV$^2$  \\
CZ  & $\displaystyle 146-62$
        & $\displaystyle 217-92$
           & $\displaystyle 475-278$
              & $\displaystyle 90-52$
                 & $1-50$ GeV$^2$
\end{tabular}
\end{ruledtabular}
\end{table}
%%%%%%%%%%%%%%%%%%%%%%%%%%%%%%%%%%%%%%%%%%%%%%%%%%%%%%%%%%%%%%%%%%%%%%%
The values of the scales $\mu_{\text{PMS}}$, $\mu_{\text{FAC}}$,
and $\mu_{\text{BLM}}$ for the asymptotic, the CZ, and the BMS\
DAs, defined in \req{eq:DAcan}, are listed in Table \ref{t:scales}.
One notices that the BLM\ scale is rather low for all considered DAs.
This makes its applicability at experimentally accessible $Q^2$ values
rather questionable.
But it is possible to improve this scale-setting procedure in the
following way.

First of all, let us rewrite the BLM\ prescription in the more
suggestive form
\begin{eqnarray}
 &&\Bigg\{F_\pi(Q^2;\muR),
       {\cal F}_{\pi}^{(1,\beta)}(Q^2;\muR)
       = \frac{1}{4}{\cal F}_{\pi}^\text{LO}(Q^2)
          \ln\left[\frac{\muR}{\mu_{\text{BLM}}^2(Q^2)}
             \right]
 \Bigg\}
 \stackrel{\text{BLM}}\Rightarrow \nonumber\\ &&
\label{eq:BLM-principle} \\
 && \qquad\Big\{F_\pi(Q^2;\mu_{\text{BLM}}^2(Q^2)),
       {\cal F}_{\pi}^{(1,\beta)}(Q^2;\mu_{\text{BLM}}^2(Q^2))
       = 0
 \Big\}\,.
 \nonumber
\end{eqnarray}
%%Eq (5.10) Rewriting the BLM prescription in suggestive form
It becomes evident that when the BLM\ scale yields $\alpha_s$
values close to unity, perturbation theory breaks down.
To avoid this happening, one can, of course, introduce ad hoc a cutoff
for $\alpha_s$, operative, say, above $0.5-0.6$, or one can ``freeze''
$\alpha_s$ at low $Q^2$ scales to some finite value by introducing
an effective gluon mass \cite{JiAm90,BJPR98}.\footnote{Restricting the
value of $\alpha_s$ does not necessarily limit the quark and gluon
virtualities in the Feynman diagrams to values for which perturbation
theory applies.}
Still another possibility is to use the analytic coupling \cite{SS97},
as done in \cite{SSK99,SSK00} (see next section).

In order to protect the BLM\ scale from intruding into the forbidden
nonperturbative soft region, where perturbation theory becomes invalid,
one can make use of a minimum scale, $\mu_\text{min}$, based on the
grounds of QCD factorization theorems and the OPE, as applied for
instance in \cite{Kor89a,Kor89b,GKKS97,Ste98} and also in \cite{BKM00}:
\begin{eqnarray}
\label{eq:mi_min}
 \mu^2_\text{min} &\geq& \muO\,.
\end{eqnarray}
%%Eq (5.11) Minimal scale setting
Here $\muO$ stands for a typical nonperturbative (hadronic) scale in
the range $[0.4-1.5]$~GeV${}^2$ and corresponds roughly to the inverse
distance at which the parton and hadron representations have to match
each other.
Note that the smaller $\mu_\text{min}^2$ is chosen, the deeper the
endpoint region $x\to 1$ can be explored for smaller values of $Q^2$.
It is intuitively clear that the typical parton virtuality in the
(hard) Feynman diagrams---let us call it $\mu_q^2$---should not become
less than its counterpart in the pion bound state: $\mu_{\pi}^2$.
Because the latter is linked to the scale $\muO$, the scale
$\mu_\text{min}^2$ should be limited from below by this scale.
Consequently, we assume that the following hierarchy of
scales---partonic (i.e., perturbative) and hadronic (i.e.,
nonperturbative)---holds:
\begin{equation}
  \lambda_q^2 < \muO \leq \mu_q^2 \leq \mu_\text{R-scheme}^2 \, .
\label{eq:scalehierachy}
\end{equation}
%%Eq (5.12) Hierarchy of intrinsic scales
Then, if
$\mu^2_\text{BLM}<\muO$,
one obtains instead of Eq.\ (\ref{eq:BLM-principle}), the IR
protected version (termed in our analysis \BLM prescription)
\begin{eqnarray}
 \Big\{F_\pi\left[Q^2,\mu_{\text{BLM}}^2(Q^2)\right],
       {\cal F}_{\pi}^{(1,\beta)}\left[Q^2,\mu_{\text{BLM}}^2(Q^2)
                                 \right]
       = 0
 \Big\}
 \stackrel{\text{\BLM}}\Rightarrow&& \nonumber\\
&&\label{eq:BLM-improvement} \\
 \left\{F_\pi(Q^2;\mu_\text{min}^2),
       {\cal F}_{\pi}^{(1,\beta)}(Q^2;\mu_\text{min}^2)
       = \frac{1}{4}{\cal F}_{\pi}^\text{LO}(Q^2)
          \ln\left(\frac{\mu_\text{min}^2}{\mu_{\text{BLM}}^2(Q^2)}
             \right)
 \right\}\,.
 \nonumber
\end{eqnarray}
%%Eq (5.13) Modified BLM prescription and scheme
This modification of the BLM\ scale setting enables us to treat the
problem of the $N_f$-dependence of the $\beta$ function in the term
${\cal F}_{\pi}^{(1,\beta)}(Q^2;\muR)$ without any further
assumptions or modifications.
Because of the fact that the scale $\muR$ is now bounded from below by
(\ref{eq:mi_min}), one is not faced with ambiguities related to the
variation of the number of active flavors $N_f$ due to
heavy-quark thresholds in the $b_0$ coefficient entering
${\cal F}_{\pi}^{(1,\beta)}(Q^2;\muR)$.
According to this, we set $N_f=3$ for $\muR=\mu_\text{min}^2$, whereas
for $\muR=\mu_\text{BLM}^2>\mu_\text{min}^2$ there is no ambiguity by
virtue of ${\cal F}_{\pi}^{(1,\beta)}(Q^2;\mu_\text{BLM}^2)=0$.
Therefore, the bona fide BLM\ scale setting reads
\begin{equation}
\label{eq:BLM-bar}
  \mu_\text{\BLM}
=
  \mbox{max}\left\{\mu_\text{BLM}, \mu_\text{min}\right\}\, ,
\end{equation}
%%Eq (5.14) Scale prescription for BLM_mod
where $\mu_\text{min}$ will be specified later on in connection with
the soft part of the form factor.

%%%%%%%%%%%%%%%%%%%%%%%%%%%%%%%%%%%%%%%%%%%%%%%%%%%%%%%%%%%%%%%%%%%%%%%
\subsection{$\alpha_V$ scheme}
 \label{sect:V-Scheme}
%%%%%%%%%%%%%%%%%%%%%%%%%%%%%%%%%%%%%%%%%%%%%%%%%%%%%%%%%%%%%%%%%%%%%%%
The self-consistency of perturbation theory implies that the difference
in the calculation to order $n$ of the same physical quantity in two
different schemes must be of order \hbox{$n+1$}.
This means that relations among different physical observables must be
independent of the renormalization scale and scheme conventions to any
fixed order of perturbation theory.
In Ref. \cite{BrodskyL95} it was argued that by applying the BLM\
scale-fixing procedure to perturbative predictions of two observables
in, for example, the \MS-scheme, and then algebraically eliminating
$\alpha_{\overline{\text{MS}}}$,
one can link to each other any perturbatively calculable observables
without  scale and scheme ambiguity.
Within this approach, the choice of the BLM\ scale ensures that the
resulting ``commensurate scale relation''  is independent of the choice
of the intermediate renormalization scheme employed.
On these grounds, Brodsky {\em et al.} in \cite{BJPR98} have analyzed
several exclusive hadronic amplitudes in the $\alpha_V$ scheme, in
which the effective coupling $\alpha_V(\mu^2)$ is defined by utilizing
the heavy-quark potential $V(\mu^2)$.
The $\alpha_V$ scheme is a ``natural'', physically motivated scheme,
which by definition, automatically incorporates vacuum polarization
effects due to the fermion-antifermion pairs into the coupling.
The $\mu_V^2$ scale which then appears in the argument of the
$\alpha_V$ coupling reflects the mean virtuality of the exchanged
gluons.
Furthermore, since $\alpha_V$ is an effective running coupling defined
by virtue of a physical quantity, it must be \textit{finite} at low
momenta, and, therefore, an appropriate parameterization of the
low-energy region should, in principle, be included.

The scale-fixed relation between the couplings
$\alpha_{\overline{\text{MS}}}$ and $\alpha_V$
is given by \cite{BJPR98}
\begin{subequations}
\begin{eqnarray}
 \alpha_s(\mu_{\text{BLM}}^2)
 =
 \alpha_V(\mu_V^2) \left[ 1 + \frac{\alpha_V(\mu_V^2)}{4 \pi}
          \, \frac{8 C_\text{A}}{3} + \cdots \right] \,,
 \label{eq:defalphaV}
\end{eqnarray}
%%Eq (5.15a) Strong coupling in V scheme
where
\begin{eqnarray}
 \mu_V^2 = e^{5/3} \, \mu_{\text{BLM}}^2
             \, .
\label{eq:defmuV}
\end{eqnarray}
\label{eq:alphaV}
\end{subequations}
%%Eq (5.15b) Relation of scales between BLM and V schemes
The scales $\mu_V$ associated with selected pion DAs are included in
Table \ref{t:scales}.

Taking into account Eqs.\ (\ref{eq:alphaV}), the NLO prediction for
the pion form factor, given by
Eqs.\ (\ref{eq:Q2pff})--(\ref{eq:Q2pffNLOnondia}), gets modified as
follows
\begin{eqnarray}
 \alpha_s(\muR)
   & \rightarrow & \alpha_V(\mu_V^2)
  \nonumber \\   \label{eq:FpialphaV} \\
  {\cal F}_\pi^\text{D,NLO}(Q^2)
   & \rightarrow &
  {\cal F}_\pi^\text{D,NLO}(Q^2)
  = {\cal F}_{\pi}^{(1,\text{FG})}(Q^2)
  + 2 {\cal F}_{\pi}^{\text{LO}}(Q^2)\,.
          \nonumber
\end{eqnarray}
%%Eq (5.16) NLO calligraphic pion FF in V scheme
We are not going to present predictions in this scheme using the
standard QCD coupling, as this would require the introduction of
exogenous parameters, like an effective gluon mass, that cannot be
fixed within the same approach but have to be taken from elsewhere.
For such an application, we refer the interested reader to the analysis
of \cite{BJPR98}.
The connection of \cite{BJPR98} to the analytic approach, which we
will use below, was discussed in detail in \cite{SSK00}.
Predictions for the pion form factor within the $\alpha_V$ scheme will
be presented below in the context of Analytic Perturbation Theory.

%%%%%%%%%%%%%%%%%%%%%%%%%%%%%%%%%%%%%%%%%%%%%%%%%%%%%%%%%%%%%%%%%%%%%%%
%%%%%%%%%%%%%%%%%%%%%%%%%%%%%%%%%%%%%%%%%%%%%%%%%%%%%%%%%%%%%%%%%%%%%%%
\section{Strong Running Coupling and Non-power Series Expansions}
\label{sect:APT}
%%%%%%%%%%%%%%%%%%%%%%%%%%%%%%%%%%%%%%%%%%%%%%%%%%%%%%%%%%%%%%%%%%%%%%%
\subsection{One-loop case}
\label{sec:one-loopana}
In the one-loop approximation we have a rather simple
renormalization-group (RG) equation for the running coupling constant:
\begin{eqnarray}
\label{eq:RG-Eq_alpha}
 \frac{d\ \alpha_s(\mu^2)}{d\ln\mu^2}
  &=& \beta(\mu^2)\,;\\
 \label{eq:beta_1-loop}
 \beta_\text{1-loop}(\mu^2)
  &=& -b_0\left(\frac{\alpha_s^2(\mu^2)}{4\pi}\right)
\end{eqnarray}
%%Eq (6.1) RGE for running coupling
%%Eq (6.2) Beta function definition in 1-loop
with $b_0$ given in Appendix~\ref{app:HSAnlo}.
The solution of this equation has the form
\begin{eqnarray}
 \alpha_s^{(1)}(Q^2)
  = \frac{4\pi}{b_0\ln(Q^2/\Lambda^2)}\,,
 \label{eq:alphaSS_1Loop}
\end{eqnarray}
%%Eq (6.3) Solution for strong coupling in 1-loop
where $\Lambda\equiv \Lambda_\text{QCD}$ is the QCD scale parameter.
A well-known problem here is the appearance of an IR pole at
$Q^2=\Lambda^2$,
which spoils the analyticity of the QCD running coupling.

In a series of papers \cite{SS97,SS98,SS99,SS01} Shirkov and Solovtsov
introduced an analytic running coupling that avoids by construction
the Landau singularity, thus generalizing earlier attempts by Radyushkin
\cite{Rad82} and Krasnikov and Pivovarov \cite{KP82}.
To this end, they used the spectral representation for the QCD running
coupling $\asb(Q^2)$ (the bar over $\as$ means that the analyticity
property is valid) and expressed it in the form
\begin{eqnarray}
 \asb(Q^2)
  = \frac{1}{\pi}\int_0^{\infty}
     d\sigma\frac{\rho(\sigma)}{\sigma+Q^2-i\epsilon}
 \label{eq:2.1}
\end{eqnarray}
%%Eq (6.4) Spectral representation for strong coupling
without subtractions due to the fact
that the spectral density $\rho(\sigma)$ decreases as $1/\ln^2\sigma$
for large $\sigma$.
The corresponding one-loop spectral density reads
\begin{eqnarray}
 \rho^{(1)}(\sigma)
  = \left(\frac{4\pi}{b_0}\right)
    \frac{\pi}{\ln^2(\sigma/\Lambda^2)+\pi^2}
 \label{eq:2.3}
\end{eqnarray}
%%Eq (6.5) Spectral density for strong coupling
and provides the one-loop singularity-free coupling function
\begin{eqnarray}
 \asb^{(1)}(Q^2/\Lambda^2)
 = \frac{4\pi}{b_0}
    \left[\frac{1}{\ln(Q^2/\Lambda^2)}
        + \frac{\Lambda^2}{\Lambda^2-Q^2}
    \right] \, .
 \label{eq:2.4}
\end{eqnarray}
%%Eq (6.6) Analytic strong coupling in 1-loop approximation
The first term on the RHS expresses the standard UV\ behavior of the
invariant coupling, while the second one compensates the ghost pole
at $Q^2=\Lambda^2$ and has a nonperturbative origin, being
suppressed at $Q^2\to\infty$.

Let us now consider powers of the analytic coupling function.
By performing an analytic continuation of the $k$-th power of the
function (\ref{eq:alphaSS_1Loop}) in the complex $Q^2$-plane, one
determines the corresponding spectral functions
$\rho_k^{(1)}(\sigma)$, ($k=1,2\ldots$):
\begin{eqnarray}
 \label{eq:1loop-ro_k}
  \rho_k^{(1)}(\sigma)
   = \left(\frac{4\pi}{b_0}\right)^{k}
     \Im\left(\frac{1}{\ln(-\sigma/\Lambda^2)}\right)^{k}\,,
\end{eqnarray}
%%Eq (6.7) Spectral density for k power
which in turn determine the analytic image $\acal_{k}^{(1)}(Q^2)$
of $\left[\as^{(1)}(Q^2)\right]^k$, i.e.,
\begin{eqnarray}
 \acal_{k}^{(1)}(Q^2)
  = \frac{1}{\pi}\int_0^{\infty}
     d\sigma\frac{\rho_k^{(1)}(\sigma)}{\sigma+Q^2-i\epsilon}\,.
\label{eq:1loop-A_k}
\end{eqnarray}
%%Eq (6.8) Analytic image of k power as spectral representation
For $k=1, 2, \ldots$, we have
\begin{eqnarray}
 \label{eq:1loop-iter}
  \acal_{k+1}^{(1)}(Q^2)
  = - \left(\frac{4\pi}{k b_0}\right)
   \frac{\partial \acal_{k}^{(1)}(Q^2)}{\partial \ln Q^2}
   \quad \text{and}\quad
  \acal_{1}^{(1)}(Q^2)\equiv \asb^{(1)}(Q^2/\Lambda^2)\,,
\end{eqnarray}
%%Eq (6.9) Analytic image of k power in derivative form
which for $k=1$ reduces to
\begin{eqnarray}
 \acal_{2}^{(1)}(Q^2)
  = \left(\frac{4\pi}{b_0}\right)^2
    \left[\frac{1}{\ln^2(Q^2/\Lambda^2)}
        + \frac{Q^2\Lambda^2}{\left(\Lambda^2-Q^2\right)^2}
    \right]\,.
 \label{eq:1loop-A_2}
\end{eqnarray}
%%Eq (6.10) Analytic image of second power
Notice at this point some key properties of these functions:
\begin{itemize}
  \item each $\acal_{k}^{(1)}(Q^2)$ with $k\geq2$ tends to zero for
        $Q^2\to0$;
  \item each $\acal_{k}^{(1)}(Q^2)$ has exactly $k-1$ zeros for
        $Q^2\in[0,\infty)$;
  \item when $Q^2\to\infty$, each
        $\acal_{k}^{(1)}(Q^2)\Big|_{Q^2\to\infty}\sim 1/\ln^k[Q^2]$
        tends to 0.
\end{itemize}
These properties are universal in the sense that they do not depend on
the loop order.
The functions $\acal_{k}(Q^2)$ are used in the so-called Analytic
Perturbation Theory  \cite{SS97,Shi98,GGK98,SS99,Shi00,Shi01},
where standard perturbative series, for example, for the Adler function
\begin{eqnarray}\label{eq:PT_Adler}
  D^\text{PT}(Q^2)
  &=& N_c\sum_{f}e_f^2
   \left\{
    1
    + \frac{\alpha_s(Q^2)}{\pi}
    + d_1 \left[\frac{\alpha_s(Q^2)}{\pi}\right]^2
    + \ldots
   \right\}
\end{eqnarray}
%%Eq (6.11) Adler function as power series in strong coupling
is recast into a non-power series expansion to obtain
\begin{eqnarray}\label{eq:APT_Adler}
  D^\text{APT}(Q^2)
  &=& N_c\sum_{f}e_f^2
   \left[
    1
    + \frac{\acal_{1}(Q^2)}{\pi}
    + d_1 \frac{{\cal A}_2(Q^2)}{\pi^2}
    + \ldots
   \right]\,.
\end{eqnarray}
%%Eq (6.12) Adler function as non-power series expansion in APT
The one-loop expressions for $\acal_{1}$ and ${\cal A}_2$ are given in
(\ref{eq:1loop-iter}) and (\ref{eq:1loop-A_2}), respectively.

\subsection{Two-loop case}
\label{sec:two-loopana}
In the two-loop case the situation is more complicated.
The corresponding $\beta$-function reads
\begin{eqnarray}
  \beta_\text{2-loop}(\alpha)
   = \frac{-b_0\alpha^2}{4\pi}
      \left(1+\frac{b_1\alpha}{b_0\ 4\pi}\right)
       \label{eq:2.5}
\end{eqnarray}
%%Eq (6.13) Beta function in 2-loops
with the first two beta coefficients given in Appendix
\ref{app:HSAnlo}.
Integrating the RG\ equation (\ref{eq:RG-Eq_alpha}), we obtain the
transcendental equation
\begin{eqnarray}
 L_Q = \frac{4\pi}{\alpha\left(L_Q\right)b_0}
       - c_1\ln\left(c_1+\frac{4\pi}{\alpha\left(L_Q\right)b_0}\right),
  \qquad c_1 = \frac{b_1}{b_0^2}\, ,
  \qquad L_Q\equiv\ln\left(Q^2/\Lambda^2\right)\,.
\label{eq:2.6}
\end{eqnarray}
%%Eq (6.14) Transcendental equation for first 2 beta coefficients
As has been shown in \cite{Mag98}, the two-loop running coupling in
QCD, being the solution of this equation, can be written via the
Lambert $W_{-1}$ function
\begin{eqnarray}
 \label{eq:alphaexact}
  \as^{(2)}\left(Q^2\right)
  = -\frac{4\pi}{b_0 c_1}
      \left[1
          + W_{-1}\left(-\frac{1}{c_1e}
                      \left(\frac{\Lambda^2}{Q^2}
                      \right)^{1/c_1}
             \right)
      \right]^{-1}\,.
\end{eqnarray}
%%Eq (6.15) 2-loop strong coupling in terms of Lambert function
For some more explanations we refer the interested reader to
\cite{BMS02}, Appendix C, Eqs.\ (C15) and (C20) in conjunction with
figure 5.
By performing the analytic continuation of function
(\ref{eq:alphaexact}) in the complex $Q^2$-plane, the spectral function
$\rho^{(2)}(\sigma)$ can be determined \cite{KM01}:
\begin{eqnarray}
 \label{eq:ro}
  \rho^{(2)}(\sigma)
   = \frac{4\pi}{b_0 c_1}
      \Im\left(-\frac{1}{1+W_{1}(z(\sigma))}\right)\,,
\end{eqnarray}
%%Eq (6.16) 2-loop spectral density
where
\begin{eqnarray}
 z(\sigma)
  = \frac{1}{c_{1}e}
     \exp[-\sigma/c_{1}+i(1/c_{1}-1)\pi]\,.
\end{eqnarray}
%%Eq (6.17) Argument of Lambert function
Then, the analytic coupling $\asb^{(2)}(Q^2)$ in the 2-loop
approximation becomes
\begin{eqnarray}
 \asb^{(2)}(Q^2)
  = \frac{1}{\pi}\int_0^{\infty}
     d\sigma\frac{\rho^{(2)}(\sigma)}{\sigma+Q^2-i\epsilon} \, .
 \label{eq:2.1-2loop}
\end{eqnarray}
%%Eq (6.18) Spectral representation of 2-loop strong coupling
However, this expression is too complex to be treated exactly.
For that reason, Shirkov and Solovtsov suggested in \cite{SS99} to use
instead the approximate expression
\begin{eqnarray}
 \label{eq:asb_2-Appro}
 \asb^{(2,\text{approx})}(Q^2)
  = \frac{4\pi}{b_0}\
      \bar{a}_s\left[\ell\left(L_Q ,c_1\right)\right]\,,
  \qquad\qquad\\
  \bar{a}_s(\ell)
  \equiv \frac{1}{\ell}
    + \frac{1}{1-\exp(\ell)}\,,\qquad
  \ell(L_Q,c) \equiv L_Q + c\,\ln\sqrt{L_{Q}^{2}+4\pi^2}
 \label{eq:ell_2-Appro}
\end{eqnarray}
%%Eq (6.19) Approximate form of 2-loop strong coupling
%%Eq (6.20) Definitions of arguments
with the same $L_Q$ as in (\ref{eq:2.6}).
This expression reproduces both the UV two-loop asymptotic behavior
as well as the value at the infrared fixed point $Q^2=0$ rather well.
More specifically, above about $Q^2\geq1\,$GeV${}^2$, it resembles the
exact result with an accuracy in the range of $99\%$ and can be used
for all higher $Q^2$ values.
Note in this context that the one-loop expression $\asb^{(1)}(Q^2)$,
Eq.\ (\ref{eq:alphaSS_1Loop}), can be represented by
\begin{eqnarray}
 \asb^{(1)}(Q^2)
  = \frac{4\pi}{b_0}\
      \bar{a}_s\left(L_Q\right)\,.
\end{eqnarray}
%%Eq (6.21) Approximate form of 1-loop strong coupling
%%%%%%%%%%%%%%%%%%%%%%%%%%%%%%%%%%%%%%%%%%%%%%%%%%%%%%%%%%%%%%%%%%%%%%%
%                             TABLE 3                                 %
%%%%%%%%%%%%%%%%%%%%%%%%%%%%%%%%%%%%%%%%%%%%%%%%%%%%%%%%%%%%%%%%%%%%%%%
\begin{table}[t]
\caption{\label{tab-2}
 Parameters entering Eq.\ (\protect{\ref{eq:asb_2-App-Global}})
 for different values of the QCD scale parameter
 $\Lambda_\text{QCD}^{N_f=3}$.}
\begin{ruledtabular}
 \begin{tabular}{cccc}
    Parameters   & $\Lambda_\text{QCD}^{N_f=3}=350$~MeV
                 & $\Lambda_\text{QCD}^{N_f=3}=400$~MeV
                            & $\Lambda_\text{QCD}^{N_f=3}=450$~MeV
                                                   \\ \hline
 $c_{21}^\text{fit}$
                & $-1.012$ & $-1.015$  & $-1.091$  \\
 $\Lambda_{21}$ & $57$~MeV & $67$~MeV  & $69$~MeV  \\
 $\ln(\Lambda_{21}^2/1~\gev{2})$
                & $-5.738$ & $-5.412$  & $-5.349$
 \end{tabular}
 \end{ruledtabular}
\end{table}
%%%%%%%%%%%%%%%%%%%%%%%%%%%%%%%%%%%%%%%%%%%%%%%%%%%%%%%%%%%%%%%%%%%%%%%
%%%%%%%%%%%%%%%%%%%%%%%%%%%%%%%%%%%%%%%%%%%%%%%%%%%%%%%%%%%%%%%%%%%%%%%
%                               FIGURE 7                              %
%%%%%%%%%%%%%%%%%%%%%%%%%%%%%%%%%%%%%%%%%%%%%%%%%%%%%%%%%%%%%%%%%%%%%%%
\begin{figure}[b]
 \centerline{\includegraphics[width=0.5\textwidth]{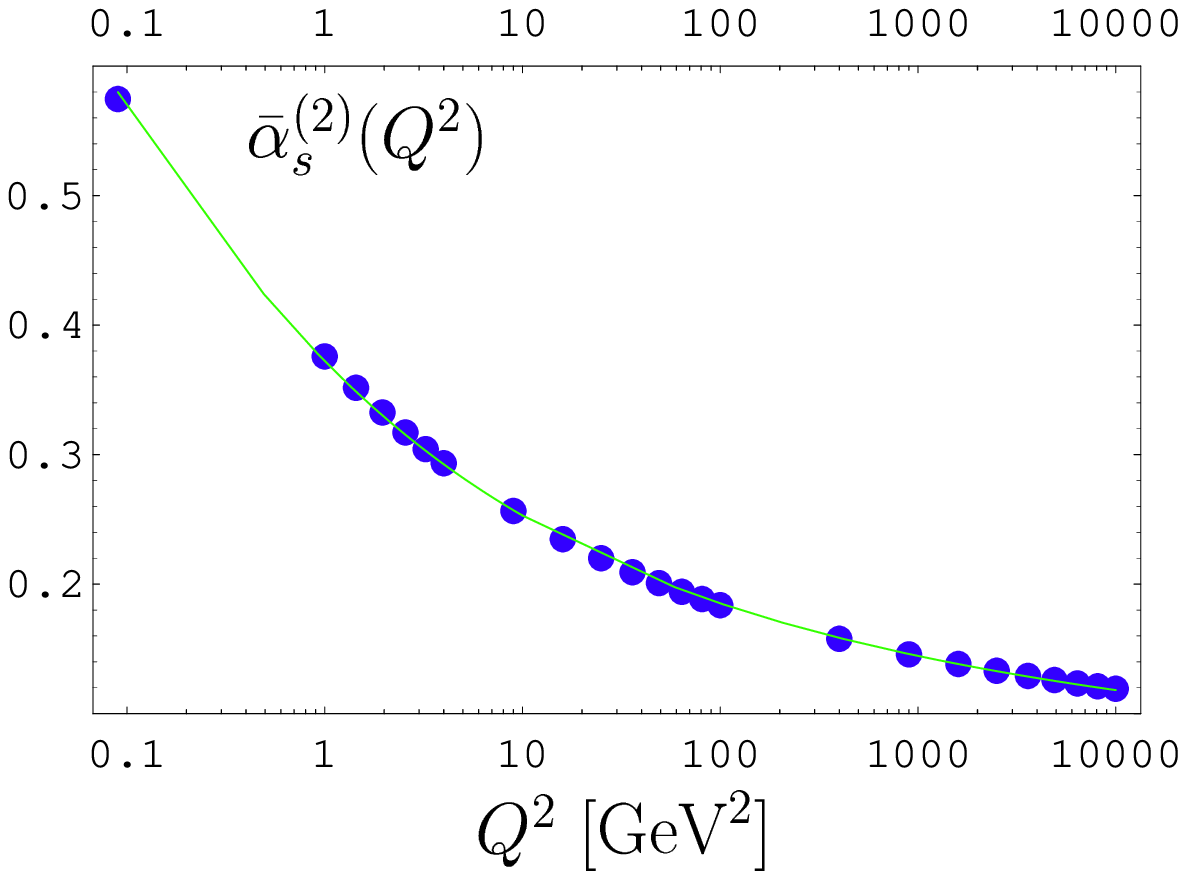}}
  \caption{\footnotesize \footnotesize %
    The solid line shows the approximate expression, given by
    $\asb^{(2,\text{fit})}(Q^2)$,
    Eq.\ (\ref{eq:asb_2-App-Global}), whereas the bullets represent the
    exact values of $\asb^{(2)}(Q^2)$ taking into account heavy-quark
    threshold matching.
    \label{fig-al400}}
\end{figure}
%%%%%%%%%%%%%%%%%%%%%%%%%%%%%%%%%%%%%%%%%%%%%%%%%%%%%%%%%%%%%%%%%%%%%%%
\\
\noindent
The only feature not yet taken into account in the above approximation
is the matching at the quark-flavor thresholds:
$M_4=1.3$~GeV, $M_5=4.3$~GeV, and $M_6=170$~GeV (with $M_1=M_2=M_3=0$).
However, taking into account this matching, the approximate formula
(\ref{eq:asb_2-Appro}) starts to become inaccurate.
As a result of the interpolation procedure, we obtain then in this
(so-called ``global'' fit in the Shirkov--Solovtsov terminology
\cite{DVS00}---abbreviated by the self-explaining label ``fit'')
case another approximation:\footnote{This interpolation is based upon
data contained in \protect\cite{KM01,KM03} and also on unpublished
data provided to us by B.~A.~Magradze.}
\begin{eqnarray}
 \label{eq:asb_2-App-Global}
  \asb^{(2,\text{fit})}(Q^2)
  &=& \frac{4\pi}{b_0(N_f=3)}\,
       \bar{a}_s\left[\ell
                 \left(\ln\frac{Q^2}{\Lambda_{21}^2},c_{21}^\text{fit}
                 \right)
                \right]\,,
\end{eqnarray}
%%Eq (6.22) Fit form of 2-loop strong analytic coupling
with the parameters $c_{21}^\text{fit}$ and $\Lambda_{21}$ listed
in Table \ref{tab-2}.
The quality of this approximation ensures a deviation less than $1\%$
in the whole $Q^2$ interval and is illustrated in Fig. \ref{fig-al400}.

To fix the parameter
$\Lambda_\text{QCD}^{N_f=3}$, we use \cite{PDG2002}
\begin{eqnarray}\label{eq:alpha_s(m_Z))}
 \asb^{(2)}(m_Z^2) = 0.120
\end{eqnarray}
%%Eq (6.23) Value of strong coupling at Z mass
that gives us
\begin{eqnarray}\label{eq:Lambda_400}
 \Lambda_\text{QCD}^{N_f=3} = 400\text{ MeV}\,.
\end{eqnarray}
%%Eq (6.24) Value of Lambda_QCD for 3 flavors

Let us now focus our attention to powers of the analytic coupling
function.
By performing the analytic continuation of the $k$-th power of function
(\ref{eq:alphaexact}) in the complex $Q^2$-plane, one determines the
corresponding spectral functions
$\rho_k^{(2)}(\sigma)$, $k=1,2\ldots$:
\begin{eqnarray}
 \label{eq:ro_k}
  \rho_k^{(2)}(t)
   = \left(\frac{4\pi}{b_0 c_1}\right)^{k}
     \Im\left(-\frac{1}{1+W_{1}(z(t))}\right)^{k}\,,
\end{eqnarray}
%%Eq (6.25) Powers of spectral density for strong coupling at 2-loops
which in turn provide the analytic images
$\acal_{k}^{(2)}(Q^2)$ of $\left[\as^{(2)}(Q^2)\right]^k$; viz.,
\begin{eqnarray}
 \acal_{k}^{(2)}(Q^2)
  = \frac{1}{\pi}\int_0^{\infty}
     d\sigma\frac{\rho_k^{(2)}(\sigma)}{\sigma+Q^2-i\epsilon}\,.
 \label{eq:2loop-A_k}
\end{eqnarray}
%%Eq (6.26) Analytic image for k power in 2-loop order in spectral form
%%%%%%%%%%%%%%%%%%%%%%%%%%%%%%%%%%%%%%%%%%%%%%%%%%%%%%%%%%%%%%%%%%%%%%%
%                               TABLE 4                               %
%%%%%%%%%%%%%%%%%%%%%%%%%%%%%%%%%%%%%%%%%%%%%%%%%%%%%%%%%%%%%%%%%%%%%%%
\begin{table}[b]
\caption{\label{tab-3}
 Parameters entering Eq.\ (\ref{eq:asb_2-App-GlobalA2}) for different
 values of the QCD scale parameter $\Lambda_\text{QCD}^{N_f=3}$.}
\begin{ruledtabular}
 \begin{tabular}{cccc}
  Parameters     & $\Lambda_\text{QCD}^{N_f=3}=350$~MeV
                 & $\Lambda_\text{QCD}^{N_f=3}=400$~MeV
                           & $\Lambda_\text{QCD}^{N_f=3}=450$~MeV
                                                    \\ \hline
 $c_{22}^\text{fit}$
                & $-1.549$  & $-1.544$   & $-1.534$   \\
 $\Lambda_{22}$ & $29$~MeV& $34.5$~MeV& $41$~MeV\\
 $\ln(\Lambda_{22}^2/1~\gev{2})$
                & $-7.088$  & $-6.734$  & $-6.399$
 \end{tabular}
 \end{ruledtabular}
\end{table}
%%%%%%%%%%%%%%%%%%%%%%%%%%%%%%%%%%%%%%%%%%%%%%%%%%%%%%%%%%%%%%%%%%%%%%%
\\ \noindent
These functions obey a more complicated recurrence relation:
($k=1, 2\ldots$)
\begin{eqnarray}
 \label{eq:2loop-iter}
  \frac{\partial \acal_{k}^{(2)}(Q^2)}{\partial \ln Q^2} =
  - k \frac{b_0}{4\pi}
      \left[\acal_{k+1}^{(2)}(Q^2)
          + \frac{b_1}{4\pi b_0}\acal_{k+2}^{(2)}(Q^2)
      \right]
  \,\,\,\text{and}\,\,\,
  \acal_{1}^{(2)}(Q^2)\equiv \asb^{(2)}(Q^2/\Lambda^2)\,.\,\,\,
\end{eqnarray}
%%Eq (6.27) Recurrence relation for k power of analytic image

\begin{figure}[t]
 \centerline{\includegraphics[width=0.45\textwidth]{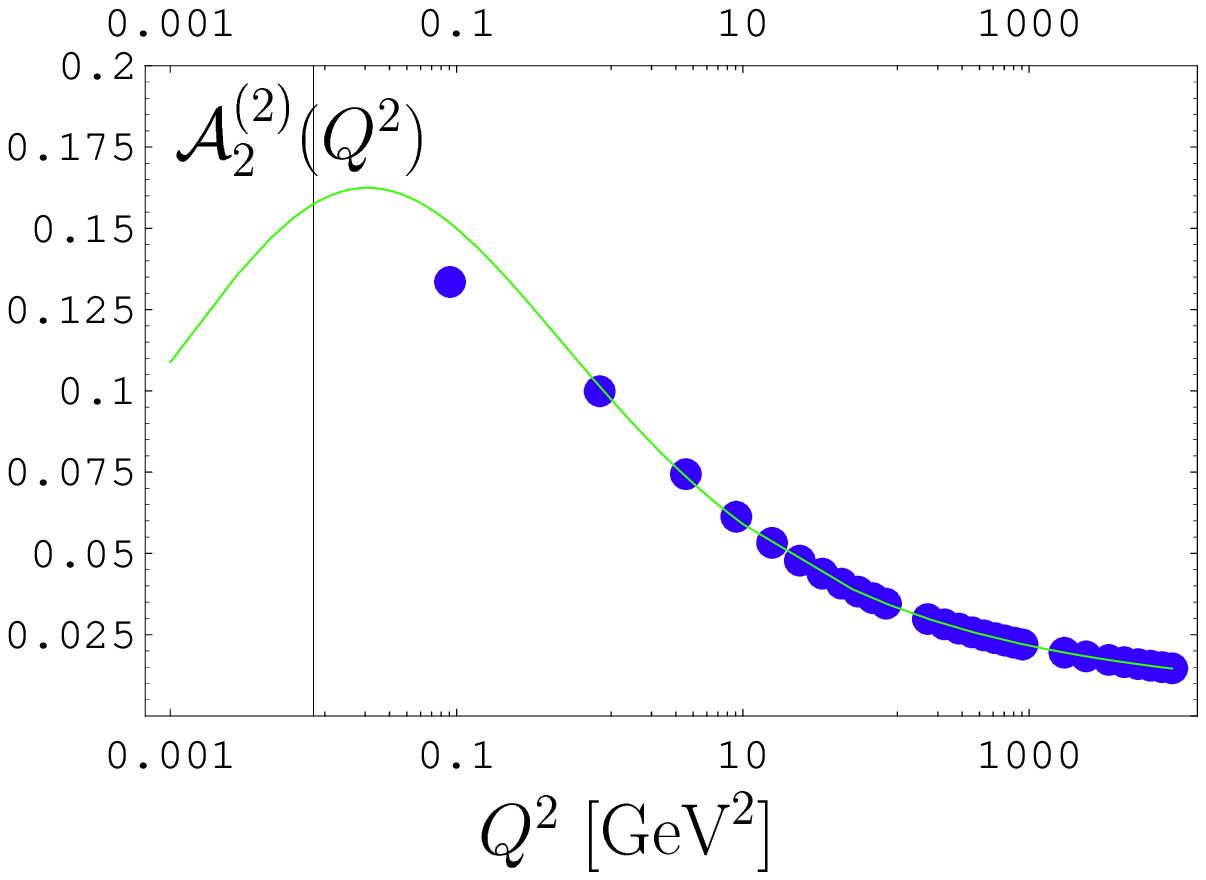}~~~%
            \includegraphics[width=0.4\textwidth]{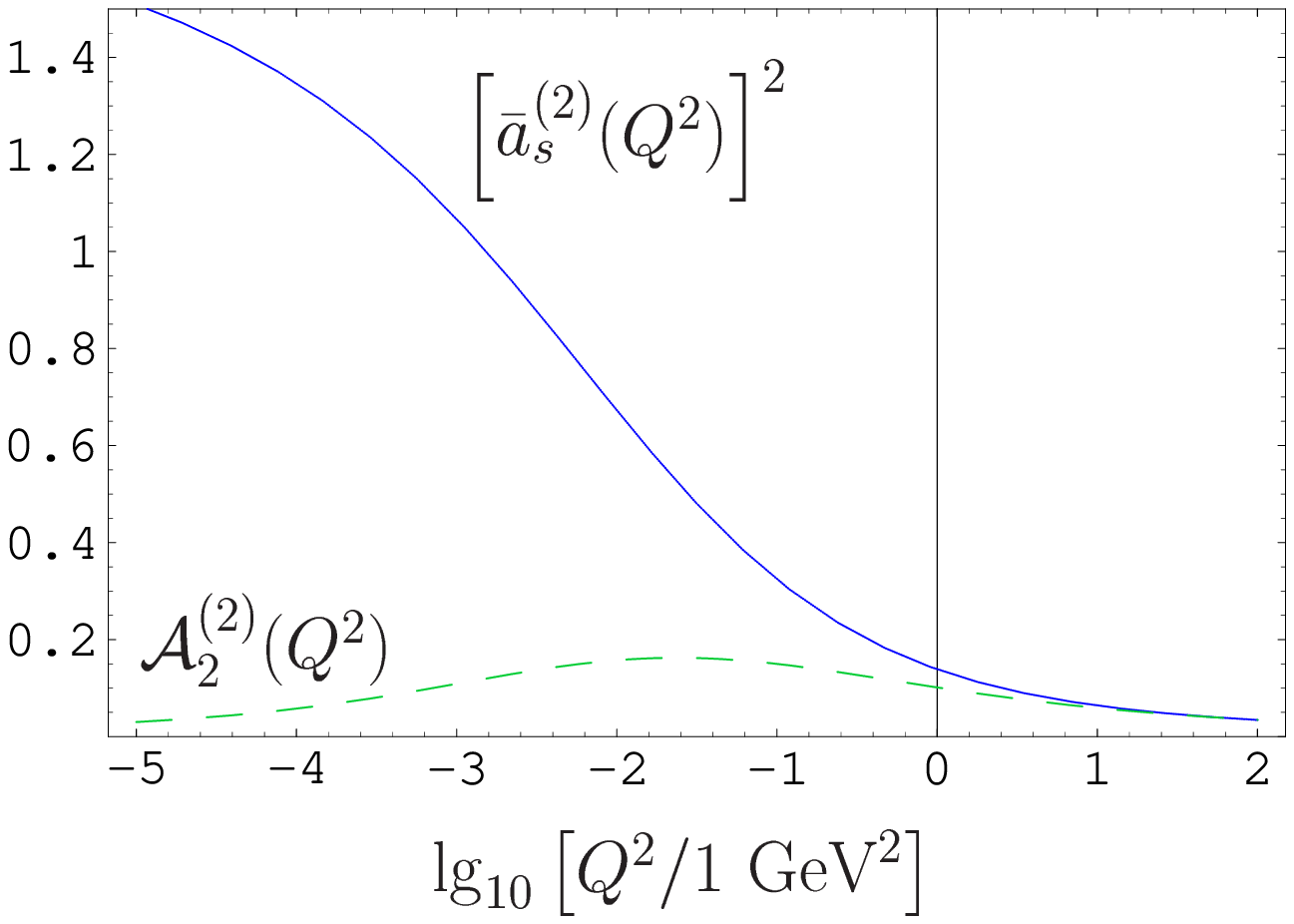}}
   \caption{\footnotesize
    (Left) The solid line represents the approximate expression
    $\acal_{2}^{(2,\text{fit})}(Q^2)$ given by
    Eq.\ (\ref{eq:asb_2-App-GlobalA2}),
    whereas the bullets denote the exact values of
    $\acal_{2}^{(2)}(Q^2)$
    taking into account the heavy-quark threshold matching.
    (Right) Comparison of $\acal_{2}^{(2,\text{fit})}(Q^2)$
    (dashed curve) with $[\asb^{(2,\text{fit})}(Q^2)]^2$ (solid curve).
     Note the modified scale of the abscissa.
    \label{fig-A2400}}
\end{figure}

As a result of the interpolation procedure, we obtain in the
``global'' case the following approximation for $k=2$
\begin{eqnarray}
\label{eq:asb_2-App-GlobalA2}
 \acal_{2}^{(2,\text{fit})}(Q^2)
  &=& \left[\frac{4\pi}{b_0(N_f=3)}\right]^2
      \left\{\frac{1}{L^2} - \frac{\exp(L)}{\left[1 - \exp(L)\right]^2}
      \right\}_{%
  L=\ell\left[\ln\left(
  \frac{Q^2}{\Lambda_{22}^2}\right),c_{22}^\text{fit}\right]} \, ,
\end{eqnarray}
%%Eq (6.28) Approximation of analytic image in global case
with the parameters $c_{22}^\text{fit}$ and $\Lambda_{22}$
being listed in Table \ref{tab-3}.
The quality of the approximation is high
with the deviation restricted to about $1\%$ ($10\%$)
for $Q^2\geq 1$~GeV$^2$ ($Q^2\leq 0.1$~GeV$^2$),
as illustrated in Fig.\ \ref{fig-A2400}.
One sees from this figure that for $Q^2\geq 10$ GeV$^2$ the difference
between the exact and approximate expression starts to be negligible
with the sizable deviation being confined in the region
$Q^2\leq 1$~GeV$^2$.
%%%%%%%%%%%%%%%%%%%%%%%%%%%%%%%%%%%%%%%%%%%%%%%%%%%%%%%%%%%%%%%%%%%%%%%
%                             FIGURE 8                                %
%%%%%%%%%%%%%%%%%%%%%%%%%%%%%%%%%%%%%%%%%%%%%%%%%%%%%%%%%%%%%%%%%%%%%%%
%%%%%%%%%%%%%%%%%%%%%%%%%%%%%%%%%%%%%%%%%%%%%%%%%%%%%%%%%%%%%%%%%%%%%%%

%%%%%%%%%%%%%%%%%%%%%%%%%%%%%%%%%%%%%%%%%%%%%%%%%%%%%%%%%%%%%%%%%%%%%%%
\subsection{Factorization of the pion form factor at NLO under
            analytization}
\label{sec:two-loop-ana-fact}
%%%%%%%%%%%%%%%%%%%%%%%%%%%%%%%%%%%%%%%%%%%%%%%%%%%%%%%%%%%%%%%%%%%%%%%
The analytization procedure of the pion form factor at NLO leads to
ambiguities, first discussed in \cite{SSK00}.
The key question is: according to what analytization prescription are
we replacing the running strong coupling and its powers by their
analytic images?
In fact, it is possible to impose the analytization of the NLO term
of $F_{\pi}^\text{Fact}$ following two different main options:
\begin{itemize}
\item In keeping with our philosophy of the analytization of observables
as a whole \cite{KS01,Ste02}, we may adopt a \emph{Maximally Analytic}
prescription and use in the NLO term of the pion form factor also the
analytic image of $\alpha_s^2$.
This amounts to
\begin{subequations}
\begin{eqnarray}
 \!\!\!\!\!\!\left[F_{\pi}^\text{Fact}(Q^2; \muR)\right]_\text{MaxAn}
 \ =\ \asb^{(2)}(\muR)\, {\cal F}_{\pi}^\text{LO}(Q^2)
   + \frac{1}{\pi}\,
      \acal_{2}^{(2)}(\muR)\,
       {\cal F}_{\pi}^\text{NLO}(Q^2;\muR)\,,
 \label{eq:pffMaxAn}
\end{eqnarray}
%%Eq (6.29a) Imposition of Maximal Analytization on NLO pion FF
which will be evaluated with the aid of Eq.\
(\ref{eq:asb_2-App-GlobalA2}).
\item  Another procedure, we call \emph{Naive Analytic}, replaces the
strong coupling and its powers by the analytic coupling $\asb$ and
its powers $\asb^{2}$ everywhere in the NLO term of
$F_{\pi}^\text{Fact}$.
This is actually the analytization procedure proposed in \cite{SSK00}
and amounts to the following requirement
\begin{eqnarray}
 \!\!\!\!\!\!\!\!\!
\left[F_{\pi}^\text{Fact}(Q^2; \muR)\right]_\text{NaivAn}
\ = \ \asb^{(2)}(\muR)\, {\cal F}_{\pi}^\text{LO}(Q^2)
   + \frac{1}{\pi}\,
      \left[\asb^{(2)}(\muR)\right]^2\,
       {\cal F}_{\pi}^{\text{NLO}}(Q^2;\muR)\,.
 \label{eq:pffNaivAn}
\end{eqnarray}
\end{subequations}
%%Eq (6.29b) Imposition of Naive analytization on NLO pion FF
Note that the naive analytization does not respect nonlinear relations
of the coupling owing to different dispersive images.
\end{itemize}

Anticipating our detailed numerical analysis of the pion form factor
using APT, we define
\begin{eqnarray}
  \Delta F_{\pi}^{\text{an}}(Q^2)
\equiv
  \left[F_{\pi}^\text{Fact}(Q^2)    \right]_\text{MaxAn}
  - \left[F_{\pi}^{\text{Fact}}(Q^2)\right]_{\text{NaivAn}}
   \, ,
 \label{eq:Naiv-Max}
\end{eqnarray}
%%Eq (6.30) Difference between Maximal and Naive Analytization
which provides a quantitative measure for the analytization ambiguity.

%%%%%%%%%%%%%%%%%%%%%%%%%%%%%%%%%%%%%%%%%%%%%%%%%%%%%%%%%%%%%%%%%%%%%%%
%%%%%%%%%%%%%%%%%%%%%%%%%%%%%%%%%%%%%%%%%%%%%%%%%%%%%%%%%%%%%%%%%%%%%%%
\section{Pion Form Factor at NLO: Numerical Analysis and Comparison
         with Experimental Data}
\label{sect:PFFNLO:Num}
%%%%%%%%%%%%%%%%%%%%%%%%%%%%%%%%%%%%%%%%%%%%%%%%%%%%%%%%%%%%%%%%%%%%%%%
In this section we would like to present our predictions for the pion
form factor utilizing the BMS\ pion DA and pQCD at the level of NLO
accuracy.
First, we consider the standard perturbative approach with different
scale settings within the \MS scheme and
continue then with a detailed discussion of the pion form factor as a
non-power series expansion of the QCD analytic coupling.
To this end, we employ the analytization procedures discussed before
to obtain $Q^2F_{\pi}^{\text{Fact}}$ in
the \MS scheme, with different scale settings, and also in the
$\alpha_V$ scheme.
To confront our theoretical predictions with the experimental data in
the last subsection, we will include the soft non-factorizable
contribution, modelled on the basis of local duality.
To join properly the hard and soft contributions, local duality
and the Ward identity at $Q^2=0$ will be employed in order to ensure
that each of these contributions is evaluated in its own region of
validity, according to the factorization of the parton and hadron
representations.
A comparison of these predictions with the corresponding ones obtained
with the asymptotic pion DA will be included.

%%%%%%%%%%%%%%%%%%%%%%%%%%%%%%%%%%%%%%%%%%%%%%%%%%%%%%%%%%%%%%%%%%%%%%%
\subsection{Standard perturbative approach}
\label{sect:SHSA}
%%%%%%%%%%%%%%%%%%%%%%%%%%%%%%%%%%%%%%%%%%%%%%%%%%%%%%%%%%%%%%%%%%%%%%%
%%%%%%%%%%%%%%%%%%%%%%%%%%%%%%%%%%%%%%%%%%%%%%%%%%%%%%%%%%%%%%%%%%%%%%%
%                              FIGURE 9                               %
%%%%%%%%%%%%%%%%%%%%%%%%%%%%%%%%%%%%%%%%%%%%%%%%%%%%%%%%%%%%%%%%%%%%%%%
\begin{figure}[b]
\vspace*{0mm}
\centerline{\includegraphics[width=0.9\textwidth]{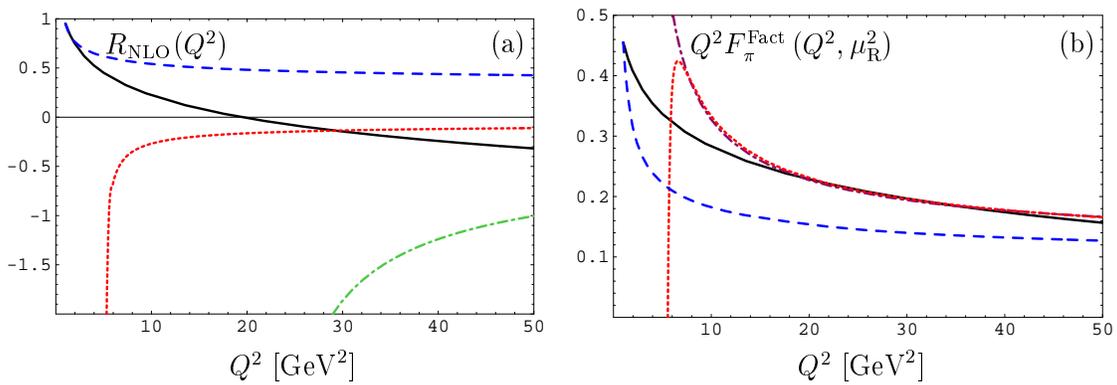}}
\vspace*{0mm}
\caption{\footnotesize
  The ratio $R_\text{NLO}(Q^2)$, (a), and the NLO results for
  $Q^2 F_{\pi}^{\text{Fact}}$, (b), in the
  $\overline{\strut \text{MS}}$ scheme with various renormalization
  scale settings.
  The dashed line corresponds to $\muR=Q^2$, the solid line to
  the $\overline{\strut \text{BLM}}$ scale setting with
  $\mu_\text{min}^2=\muO=1$~GeV${}^2$,
  while the dash-dotted one denotes the result obtained with the
  BLM\ (a) and FAC\ (b) scale settings.
  The analogous result for the PMS\ scale setting is shown
  as a dotted line.
  Note that in both panels the BMS\ DA has been employed and
  that for the FAC\ and PMS\ scale settings $N_f$ has been fixed to 3.
 \label{f:asczPff}}
\end{figure}
%%%%%%%%%%%%%%%%%%%%%%%%%%%%%%%%%%%%%%%%%%%%%%%%%%%%%%%%%%%%%%%%%%%%%%%
As outlined in Sec.\ \ref{sect:RenSc}, the NLO prediction for the pion
form factor, as any other finite-order prediction, contains a
theoretical uncertainty stemming from its dependence on the
renormalization scale $\mu_\text{R}$ and the scheme used.
This dependence is, however, reduced in comparison with the LO
prediction due to the inclusion of the NLO correction.
To quantify these statements, we plot in Fig.\ \ref{f:asczPff}(a) the
ratio $R_\text{NLO}(Q^2) = F_\pi^\text{NLO}(Q^2)/F_\pi^\text{LO}(Q^2)$
and in Fig.\ \ref{f:asczPff}(b) the result for the factorized form
factor at NLO, using the BMS\ DA in the \MS scheme with different
scale settings.
The main observation from these figures is the strong sensitivity of
$R_\text{NLO}(Q^2)$ and the moderate dependence of
$F_\pi^\text{Fact}(Q^2)$ on the scale-setting procedure
adopted---especially at $Q^2$ values accessible to present
experiments.

Let us discuss these figures in a systematic way.
\begin{itemize}
\item For $\mu_\text{R}^2=Q^2$, the ratio $R_\text{NLO}(Q^2)$ is
positive, large (on the order of about $50\%$) and decreases very
slowly, while $\alpha_s$ is small ($\sim 0.3$).
As a result, the LO contribution is about twice as the NLO one and the
form factor is small.
\item Using the FAC\ scale setting, the whole NLO contribution
vanishes, so that also the ratio is zero.
In this case, the form factor is rather moderate down to momenta of
the order of $10$~GeV$^2$, where the QCD effective coupling becomes
of order unity.
\item Applying the PMS\ scale setting, the NLO contribution is
negative with $R_\text{NLO}(Q^2)$ being small and also negative
down to a critical value of $Q^2\simeq 6$~GeV$^2$ (see Table
\ref{t:scales}), where the absolute value of the NLO contribution
becomes equal to the LO one and the form factor becomes zero.
For this scale setting, already at $Q^2\simeq 6$~GeV$^2$, the QCD
effective coupling starts ``feeling'' the Landau singularity and
becomes excessively large, while above $10$~GeV$^2$ the form factor
is rather moderate.
\item Adopting the BLM\ procedure, the results are quite similar to
those obtained with the PMS\ scale setting with respect to the ratio
$R_\text{NLO}(Q^2)$, whereas the form factor now is negative and
very large below $50$~GeV$^2$ (lying outside the range of
Fig.\ \ref{f:asczPff}(b)) because the corresponding NLO correction
is again negative and even larger.
The reason for this behavior is that in this scheme the typical
parton virtualities in the Feynman diagrams are much lower than the
external scale $Q^2$ (see Table \ref{t:scales}) giving rise to a
large value of the QCD effective coupling.
\item The \BLM scale setting has two distinct regimes, characterized
by the fact that the ratio $R_\text{NLO}(Q^2)$ changes its sign
around $20$~GeV$^2$: in the regime below this momentum value, the
result for the form factor resembles that found with the
$\mu_\text{R}^2=Q^2$ scale setting, though its fall-off with $Q^2$ is
not that steep. On the other hand, above $20$~GeV$^2$, the form factor
almost coincides with the one calculated with the PMS\ scale setting.
\end{itemize}
A further complication: it is not clear how to implement quark-mass
thresholds when using the FAC\ and PMS\ scale settings.
Therefore, the predictions shown have been obtained by fixing $N_f=3$.
This is because both scales depend on $\beta_0$ and this induces
discontinuities in the form factor at the quark-mass thresholds.
For that reason, we refrain from using the FAC\ and PMS\ schemes in
our further considerations.
To summarize, all scale settings can be safely used above about
$20$~GeV$^2$, while at smaller $Q^2$ values, the PMS\ and FAC\
settings become unphysical, whereas the \BLM and $\mu_\text{R}^2=Q^2$
scale-setting procedures can further be used at values of $Q^2$
exhausting the validity domain of pQCD.
On the other hand, the BLM scale setting remains inapplicable up to
scales of the order of $50$~GeV$^2$ (see Fig.\ \ref{f:asczPff}a).
As already explained before, no predictions in the $\alpha_V$ scheme
have been shown because this would require the introduction of
exogenous IR regulators.

%%%%%%%%%%%%%%%%%%%%%%%%%%%%%%%%%%%%%%%%%%%%%%%%%%%%%%%%%%%%%%%%%%%%%%%
\subsection{Use of non-power series expansions}
 \label{sect:NPE:APT}
\begin{figure}[t]
 \centerline{\includegraphics[width=0.9\textwidth]{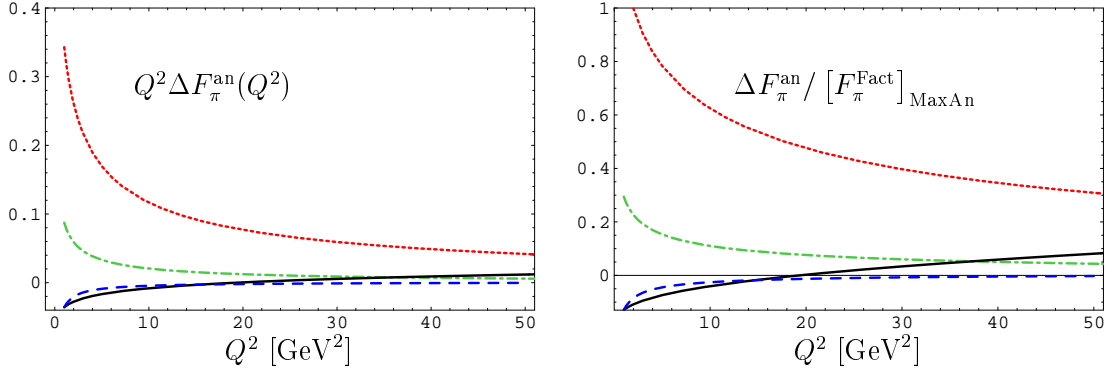}}
   \caption{\footnotesize The NLO analytization ambiguity
     $Q^2\Delta F_{\pi}^{\text{an}}(Q^2)$ (left)
     and the ratio
     $\Delta F_{\pi}^{\text{an}}/
     \left[F_{\pi}^{\text{Fact}}\right]_\text{MaxAn}$
     of the NLO analytization ambiguity relative to the factorized pion
     form factor, computed with the ``Maximally Analytic'' procedure,
     (right) within the $\overline{\strut \text{MS}}$ scheme
     with various scale settings:
     $\mu_\text{R}^2=Q^2$ (dashed line), BLM\ (dotted line),
     $\overline{\strut \text{BLM}}$ (solid line),
     and the $\alpha_V$-scheme (dash-dotted line).
     The curves shown correspond to the BMS\ DA.
     \label{fig:23+24}}
\end{figure}

%%%%%%%%%%%%%%%%%%%%%%%%%%%%%%%%%%%%%%%%%%%%%%%%%%%%%%%%%%%%%%%%%%%%%%%
We turn now to the results obtained in APT.
To exploit the effect of the analytization ambiguity on the factorized
pion form factor, according to (\ref{eq:Naiv-Max}),
we plot in Fig.\ \ref{fig:23+24} (left panel)
%%%%%%%%%%%%%%%%%%%%%%%%%%%%%%%%%%%%%%%%%%%%%%%%%%%%%%%%%%%%%%%%%%%%%%%
%                              FIGURE 10                              %
%%%%%%%%%%%%%%%%%%%%%%%%%%%%%%%%%%%%%%%%%%%%%%%%%%%%%%%%%%%%%%%%%%%%%%%
%%%%%%%%%%%%%%%%%%%%%%%%%%%%%%%%%%%%%%%%%%%%%%%%%%%%%%%%%%%%%%%%%%%%%%%
\begin{equation}
 \Delta F_{\pi}^{\text{an}}(Q^2)
  = \frac{\acal_{2}^{(2)}(\muR)- \left[\asb^{(2)}(\muR)\right]^2}
         {\pi}\,
    {\cal F}_{\pi}^\text{NLO}(Q^2;\muR)
\label{eq:an-amb-NLO}
\end{equation}
%%Eq (7.1) Analytization ambiguity at NLO
and the ratio
$\Delta F_{\pi}^{\text{an}}/
\left[F_{\pi}^{\text{Fact}}\right]_\text{MaxAn}$
(right panel), employing the BMS\ DA and the
$\overline{\strut\text{MS}}$
scheme with different scale settings.
Analogous results for the $\alpha_V$-scheme are also included;
using APT there is no need to introduce external IR regulators.

Summarizing the results in Fig.\ \ref{fig:23+24}, the main observations
are:
\begin{itemize}
\item The NLO analytization ambiguity,
$\Delta F_{\pi}^{\text{an}}(Q^2)$, (left panel) and
 the ratio, $\Delta F_{\pi}^{\text{an}}/F_{\pi}^{\text{Fact}}$,
 (right panel), the latter being computed with the ``Maximally
 Analytic'' procedure within the \MS scheme with the
 $\mu_\text{R}^2=Q^2$ (dashed line) and \BLM (solid line) scale
 settings, is small and almost scaling with $Q^2$ above about
 $10$~GeV$^2$, albeit in the second case there is a sign change around
 $18$~GeV$^2$.
 This is because below this momentum, the term $F_\pi^{(1,\text{FG})}$,
 which is negative, prevails, while above that scale the term
 $F_\pi^{(1,\beta)}$ becomes dominant due to
 $-\ln(Q^2/\mu_\text{R}^2)$---in contrast to the former case in which
 the interplay between these two terms is fixed owing to the absence
 of the log term.
 For that reason, the ``Maximally Analytic'' procedure with the \BLM
 scale setting enhances the form factor at higher $Q^2$ relative to
 the ``Naive'' one.
\item The results with the BLM\ scale setting (dotted line) resemble
 those computed with the $\alpha_V$ scheme (dash-dotted line).
 In both cases,
 $\acal_{2}^{(2)}(\muR)- \left[\asb^{(2)}(\muR)\right]^2$ is large
 and negative (cf.\ Fig.\ \ref{fig-A2400}---right panel), while
 ${\cal F}_\pi^{(1,\text{FG})}$ is also negative.
 Hence, the overall sign of $\Delta F_{\pi}^{\text{an}}(Q^2)$ is plus
 because the $F_\pi^{(1,\beta)}$ is absent.
 However, in the $\alpha_V$ scheme the shift towards smaller values
 of the $\alpha_s$ argument is much less pronounced and consequently
 the enhancement provided by the use of $\acal_{2}^{(2)}(\muR)$
 instead of $\left[\asb^{(2)}(\muR)\right]^2$ is rather weak (see also
 Eq.\ (\ref{eq:FpialphaV})).
\end{itemize}

%%%%%%%%%%%%%%%%%%%%%%%%%%%%%%%%%%%%%%%%%%%%%%%%%%%%%%%%%%%%%%%%%%%%%%%
%                              FIGURE 11                              %
%%%%%%%%%%%%%%%%%%%%%%%%%%%%%%%%%%%%%%%%%%%%%%%%%%%%%%%%%%%%%%%%%%%%%%%
\begin{figure}[t]
\vspace*{0mm}
\centerline{\includegraphics[width=0.9\textwidth]{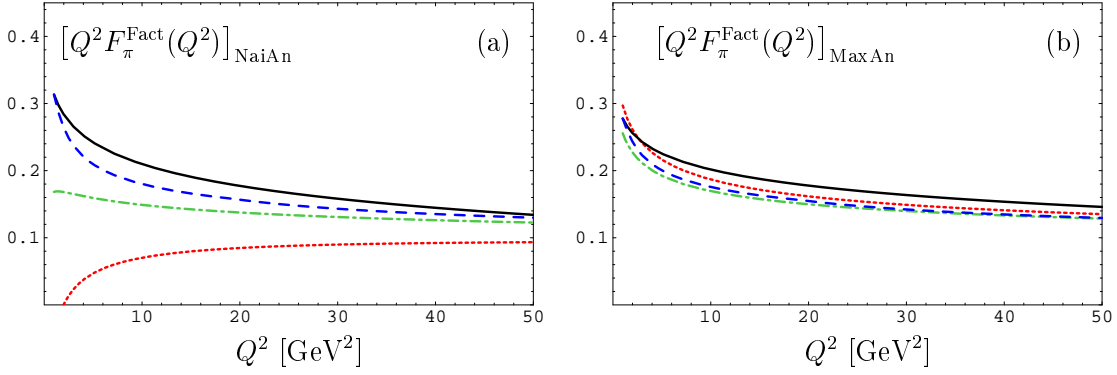}}
\caption{\footnotesize NLO predictions for $Q^2 F_{\pi}^{\text{Fact}}$
         vs.\ $Q^2$, using the ``Naive Analytic'' (a) and ``Maximally
         Analytic'' (b) procedures and employing the BMS\ DA.
         The dashed line corresponds to $\muR=Q^2$, the dotted one
         denotes the result obtained with the standard BLM\ scale
         setting, whereas the solid line shows the result calculated
         with the modified $\overline{\strut\text{BLM}}$ scale setting
         and the cutoff scale $\muO=1$~GeV${}^2$.
         The results obtained with the $\alpha_V$-scheme are displayed
         as a dash-dotted line in both panels.
 \label{fig:APT_msbarav}}
\end{figure}
%%%%%%%%%%%%%%%%%%%%%%%%%%%%%%%%%%%%%%%%%%%%%%%%%%%%%%%%%%%%%%%%%%%%%%%

%%%%%%%%%%%%%%%%%%%%%%%%%%%%%%%%%%%%%%%%%%%%%%%%%%%%%%%%%%%%%%%%%%%%%%%
%                              FIGURE 12                              %
%%%%%%%%%%%%%%%%%%%%%%%%%%%%%%%%%%%%%%%%%%%%%%%%%%%%%%%%%%%%%%%%%%%%%%%
\begin{figure}[b]
 \centerline{\includegraphics[width=0.9\textwidth]{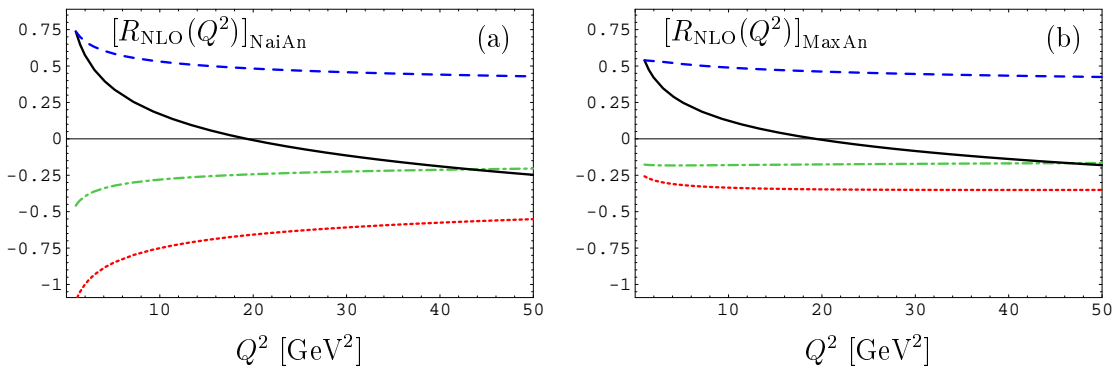}}
 \caption{\footnotesize Results obtained for the ratio
         $R_\text{NLO}(Q^2)$ using the ``Naive Analytic'' (a) and
         ``Maximally Analytic'' (b) procedures and employing the BMS\
         DA.
         Notations are the same as in figure \ref{fig:APT_msbarav}.
          \label{fig:Rat_NLO_MSbarav}}
\end{figure}
%%%%%%%%%%%%%%%%%%%%%%%%%%%%%%%%%%%%%%%%%%%%%%%%%%%%%%%%%%%%%%%%%%%%%%%

Next, we present the results for the factorized pion form factor
derived with APT at the NLO level and adopting either the ``Naive
Analytic'' or the ``Maximally Analytic'' procedure.
From Fig.\ \ref{fig:APT_msbarav}, we see that for both analytization
procedures the results for the $\muR=Q^2$ (dashed line) and
$\overline{\strut\text{BLM}}$ (solid line) scale settings are very
close to each other.
Note that the $\alpha_V$-scheme yields a similar result (dash-dotted
line), but with a much smaller steepness of the curve at low $Q^2$.
On the other hand, the standard BLM\ scale setting (dotted line)
produces even an exact cancellation of the NLO and LO terms at the
momentum value $Q^2\approx 2$~GeV$^2$ (so too behaves the ratio
$R_\text{NLO}(Q^2) = F_\pi^\text{NLO}(Q^2)/F_\pi^\text{LO}(Q^2)$---see
Fig.\ \ref{fig:Rat_NLO_MSbarav}(a)).
The origin of this cancellation is, however, purely accidental and
unphysical: the BLM\ scale at this point is
$\mu_\text{BLM}^2\approx 0.02$~GeV$^2$ with $\alpha_s\approx 0.75$,
rendering the pQCD expansion unreliable.
This deficiency is lifted when applying the ``Maximally Analytic''
procedure---see Fig.\ \ref{fig:APT_msbarav}(b).
Indeed, such is the impact of the ``Maximally Analytic'' condition
that all renormalization-scheme and scale-setting ambiguities are
diminished, with all results for the form factor almost coinciding,
as it is obvious from Fig.\ \ref{fig:APT_msbarav}(b).
Moreover, from Fig.\ \ref{fig:APT_msbarav}b, we can estimate the
effect of varying $\mu_\text{min}^2=\mu_{0}^2$ in the \BLM scale
setting procedure by comparing the \BLM (black solid) and the BLM
(red dotted) curves.
Indeed, $\mu^2_\text{BLM}$ just varies from $0.5$~GeV$^2$ (at
$Q^2=50$~GeV$^2$) to $0.02$~GeV$^2$ (at $Q^2=2$~GeV$^2$), while the
difference between these two curves is no more than $10\%$
(using the ``Maximally Analytic'' condition).
Actually, for $\mu^2_\text{min}$ varying in the range
$[1,\; 0.5]$~GeV$^2$, this difference does not even exceed the level of
$5\%$.

Let us close this discussion with some brief comments on the behavior
of the ratio $R_\text{NLO}(Q^2)$.
The message from Fig.\ \ref{fig:Rat_NLO_MSbarav} is that, except for
the BLM\ scale setting (already discussed), all other scale-settings
are not sensitive to the analytization procedure adopted.
The induced differences are indeed marginal, with $R_\text{NLO}(Q^2)$
being positive, large, and practically scaling with $Q^2$ for the
$\muR=Q^2$ scale setting (dashed line), while this quantity in the
$\alpha_V$ scheme (dash-dotted line) exhibits the same behavior but
with the reverse sign and having approximately half of its magnitude.
The situation for the \BLM scale setting is somewhat transient
between these two options, providing with both analytization procedures
enhancement at the low end of $Q^2$ and reducing the form factor at
$Q^2$ values higher than $Q^2\simeq 20$~GeV$^2$.
This effect is due to the (negative) term $F_\pi^{(1,\text{FG})}$
gaining ground against $F_\pi^{(1,\beta)}$ that becomes smaller
because $\ln(Q^2/\muR)$ is growing.

%%%%%%%%%%%%%%%%%%%%%%%%%%%%%%%%%%%%%%%%%%%%%%%%%%%%%%%%%%%%%%%%%%%%%%%
 \subsection{Non-factorizable Contribution to the Pion Form Factor}
  \label{sect:NFPFF}
%%%%%%%%%%%%%%%%%%%%%%%%%%%%%%%%%%%%%%%%%%%%%%%%%%%%%%%%%%%%%%%%%%%%%%%
So far we have discussed only the factorizable part of the pion form
factor (cf. (\ref{eq:pff})).
But as argued originally in Refs. \cite{NR82,NR83,IL84,Rad90}, and
confirmed latter on in several works, for instance, in
\cite{JK93,AMN95,SSK99,BRS00,SSK00}, the dominant contribution at low
to moderate values of the momentum transfer
$Q^2 \leq 10~\text{GeV}^2$ originates mainly from the soft contribution
that involves no hard-gluon exchanges and is attributed to the Feynman
mechanism.
At present there is no unique way to calculate this contribution from
first principles at the partonic level.
One has to resort to theoretical models, based on assumptions that
attempt to capture the characteristic features of nonperturbative QCD.
In the present investigation we use the Local Duality (LD) approach to
calculate the soft contribution, in  which it is assumed that the pion
form factor is dual to the free quark spectral density
\cite{NR82,Rad95}, i.e.,
\begin{eqnarray}
 F_{\pi}^\text{LD}(Q^2)
 &=& \frac{1}{\pi^2 f_{\pi}^2}
  \int_{0}^{s_0}\!\!\!\int_{0}^{s_0}
   \rho_3(s,s^\prime,Q^2)\ ds\ ds^\prime
 \ =\ 1 - \frac{1 + 6\ s_0/Q^2}{\left(1 + 4\ s_0/Q^2\right)^{3/2}}
 \,,
\label{eq:FFQuarLd}
\end{eqnarray}
%%Eq (7.2) Form factor in quadratic local duality approximation
with the 3-point spectral density
$\rho_{3}(s,s^\prime,t=Q^2)$
given by
\begin{equation}
 \rho_{3}(s,s^\prime,t)
  = \left[t^2 \frac{d^2}{dt^2}
      + \frac{t^3}{3} \frac{d^3}{dt^3}
      \right] \frac{1}{\lambda\left(s,s^\prime,t\right)}
\label{eq:RoSq}
\end{equation}
%Eq (7.3) Integrated 3-point spectral density
where
\begin{equation}
  \lambda\left(s,s^\prime,t\right)
   \equiv
    \sqrt{\left(s + s^\prime + t\right)^2 - 4ss^\prime}\,.
\label{eq:auxlambda}
\end{equation}
%%Eq (7.4) Auxiliary parameter for spectral density
Here the duality interval $s_0$ corresponds to the effective threshold
for the higher excited states and the ``continuum'' in the channels
with the axial-current quantum numbers.
The LD\ prescription for the corresponding correlator \cite{SVZ}
implies the relation
\begin{eqnarray}
 s_0 = 4 \pi^2 f_{\pi}^2\,.
 \label{eq:LDfpi}
\end{eqnarray}
%%Eq (7.5) Duality threshold
A key issue of the soft contribution
is the inclusion of Sudakov-type radiative corrections.
In~\cite{BRS00} only the Sudakov corrections to the quark-photon vertex
were taken into account on the basis of~\cite{ABunpub} leading to a
reduction of the soft contribution by approximately 6\% at low
$Q^2\simeq 2$~GeV${}^2$ and up to 20\% at higher $Q^2$ values.
Just recently, however, it was shown in~\cite{BO04} that taking into
account all radiative corrections to the correlator, the Sudakov
logarithms cancel out.
On the face of this finding, we use in this work Eqs.\
(\ref{eq:FFQuarLd})--(\ref{eq:RoSq}).

%%%%%%%%%%%%%%%%%%%%%%%%%%%%%%%%%%%%%%%%%%%%%%%%%%%%%%%%%%%%%%%%%%%%%%%
%                              FIGURE 13                              %
%%%%%%%%%%%%%%%%%%%%%%%%%%%%%%%%%%%%%%%%%%%%%%%%%%%%%%%%%%%%%%%%%%%%%%%
\begin{figure}[bht]
 \centerline{\includegraphics[width=0.475\textwidth]{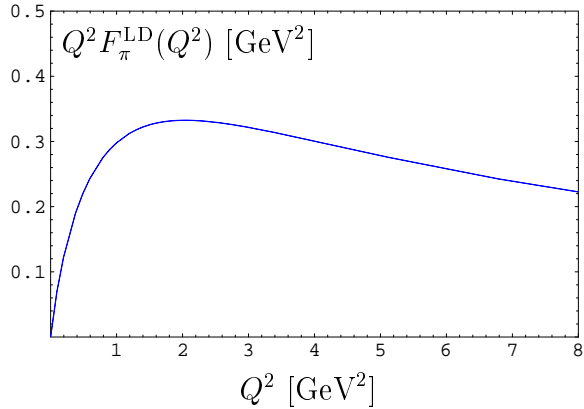}}
  \caption[pion]
  {\footnotesize
   Calculation of the soft part of the pion form factor in the
   Local Duality approach.
   \label{fig-TR}}
\end{figure}
%%%%%%%%%%%%%%%%%%%%%%%%%%%%%%%%%%%%%%%%%%%%%%%%%%%%%%%%%%%%%%%%%%%%%%%
The soft contribution calculated here is consistent with the result
obtained in \cite{SSK00} for the asymptotic pion DA on the basis of the
soft overlap of the pion wave functions, modelling their $k_{\perp}$
dependence in terms of the Brodsky--Huang--Lepage Gaussian ansatz
\cite{BHL81} and using a constituent-like quark mass of $m_q=330$~MeV.
Though the crossover from the soft to the hard regime and the
asymptotic behavior are strongly model dependent, with the mass factor
$\exp\left(-\beta_\text{G}^2m_{q}^2/x\bar{x}\right)$
(where $\beta_\text{G}$ is the width of the Gaussian distribution,
specific for each particular pion DA) playing an important role in
tuning this behavior---see \cite{JK93} for a detailed analysis---the
trend at lower values of $Q^2$ up to about $4$~GeV${}^2$ is
approximately the same.
Similar results were also obtained in \cite{KW93} using a
Bethe-Salpeter equation and a constituent-type quark mass of
$m_q=330$~MeV.
In both approaches mentioned \cite{SSK00,KW93} the quark mass in the
hard part was set equal to zero and the effective QCD coupling was
assumed to saturate at low $Q^2$ with a transition scale from soft to
hard in the range $Q^2\simeq 12 - 18$~GeV${}^2$.

%%%%%%%%%%%%%%%%%%%%%%%%%%%%%%%%%%%%%%%%%%%%%%%%%%%%%%%%%%%%%%%%%%%%%%%
 \subsection{Comparison with experimental data}
  \label{sect:expdata}
%%%%%%%%%%%%%%%%%%%%%%%%%%%%%%%%%%%%%%%%%%%%%%%%%%%%%%%%%%%%%%%%%%%%%%%
It is time to step up one level higher and consider the total form
factor in order to compare our theoretical predictions with the
experimental data.
So far we have considered the factorized hard contribution to the pion
form factor only at higher values of $Q^2$, where pQCD is safe.
However, attempting to compute the total pion form factor in the full
$Q^2$ range, according to Eq.\ (\ref{eq:pff}), we have to combine
this contribution with the soft part.
Recall that we have neglected in the hard-scattering amplitude (i.e.,
in the parton propagators---cf.\ Eqs.\ (\ref{eq:THLOpff}),
(\ref{eq:TH1beta}), and (\ref{eq:TH1C1})) all parton transverse momenta
against the large scale $Q^2$ and integrated out in the pion wave
functions all transverse momenta up to the scale $\muO$.
But below some momentum scale of this order, these contributions in
$T_{\text{H}}$ start to be comparable (especially in the endpoint
region where $x\to 1$) and, a fortiori, the collinear factorization
becomes increasingly unreliable.
To avoid this happening, we have to restrict the evaluation of the hard
form-factor contribution to that $Q^2$ domain compatible with the
collinear approximation.
In technical terms this means that below the scale $s_0$ (the duality
threshold) we have to switch from the parton representation to the
hadron representation according to local duality.\footnote{%
A smooth transition from the partonic to the hadronic regime may go
via an intermediate constituent-quark formation due to QCD dressing.
Because there is no unambiguous way to do this, we prefer to ignore
this regime here (and refer for a discussion of such dressed quarks to
\cite{SSK00}).}

As we have seen in Sec.\ \ref{sect:RenSc}, fixing the renormalization
scale $\muR$ in all considered schemes entails problems related to
the small $Q^2$-behavior of the factorizable term of the pion form
factor: the NLO-term can reach the level of 50\% of the LO part,
casting doubts on the validity of the perturbative expansion per se.
In addition, both terms (LO and NLO) generate a fast growth of the
form factor at small $Q^2$, artificially induced by large values of
the strong coupling and by a $1/Q^2$-factor.
The origin of this failure, as stated above, can be traced to the
violation of the collinear factorization approximation, i.e., the
resurrection of small momenta in $T_\text{H}$ that have initially been
neglected and absorbed into the pion DA.\footnote{Their explicit
inclusion would give logarithmic and power-behaved corrections
amounting to Sudakov-type exponentials containing perturbative
\cite{LS92} and nonperturbative corrections \cite{KS01}.}

Hence, it becomes clear that we must correct the factorization
results in the low-$Q^2$ region in order to ensure that each
contribution lies in the corresponding domain of validity.
To achieve this goal, we need a conceptual framework.

This is provided by gauge invariance that protects the value of
$F_\pi(0)$ by means of the Ward identity relating a three-point
Feynman diagram at zero-momentum transfer to a 2-point diagram.
Consider the $L$-loop approximation in the LD approach.
Then, using the replacements $s_0\to s_0^{L-\text{loop}}$ and
$\rho_{3}(s,s^\prime, Q^2;\muR)
\to \rho_{3}^{L-\text{loop}}(s,s^\prime, Q^2;\muR)$,
Eq.\ (\ref{eq:FFQuarLd}) relates $F_{\pi}^\text{LD}(Q^2)$ to the
integrated 3-point spectral density
$\rho_{3}^{L-\text{loop}}(s,s^\prime, Q^2;\muR)$,
which is now dependent on $\muR$.
Recall that the Ward identity links the 2-point (i.e., axial-axial
current) spectral density $\rho_2^{L-\text{loop}}(s;\muR)$
to the 3-point (vector-axial-axial current) spectral density
$\rho_3^{L-\text{loop}}(s,s',0;\muR)$ in the following way
\begin{eqnarray}
 \label{eq:WI_SD}
  \rho_3^{L-\text{loop}}(s,s',0;\muR)
   &=& \pi\delta(s-s')\rho_2^{L-\text{loop}}(s;\muR) \,.
\end{eqnarray}
%%Eq (7.6) Relation between 2-point and 3-point spectral density
Taking into account the LD expression for the pion decay constant,
\begin{eqnarray}
 f_{\pi}^2
 &=& \frac{1}{\pi}
  \int_{0}^{s_0^{L\text{-loop}}}
   \rho_2^{L\text{-loop}}(s;\muR)\ ds \, ,
\label{eq:fpi_Ld}
\end{eqnarray}
%%Eq (7.7) Relation between f_pi and spectral density
one finds
\begin{eqnarray}
 \label{eq:WI_FF0}
 F_{\pi}^\text{LD}(0;\muR)
   &=& 1 \,.
\end{eqnarray}
%%Eq (7.8) Form factor in local duality approach at Q^2=0
The 2-loop approximation for the spectral density,
$\rho_2^\text{2-loop}(s;\muR)$, can be obtained from the $e^+e^-$
cross-section $R(s)$ \cite{CKT79}, because these quantities in
massless QCD are proportional to each other, so that
\begin{eqnarray}
 \label{eq:2pSD_1L}
%%Eq (7.9) 2-point spectral density in 2-loop approximation
 \rho_2^\text{2-loop}(s;\muR)
  &=& \frac{1}{4\pi}
       \left[1
            + \frac{\alpha_s(\muR)}{\pi}
            + \left(\frac{\alpha_s(\muR)}{\pi}\right)^2
               \left(r_2
                   - \frac{b_0}{4}\ln\frac{s}{\muR}
               \right)
       \right]\,;
 \\
  r_2 &=& \frac{3}{4}\,C_\text{F}
          \left[\frac{1}{12}\,C_\text{A}
              - \frac{1}{8}\,C_\text{F}
              + b_0 \left(\frac{11}{8}-\zeta(3)\right)
          \right]\, ,
\end{eqnarray}
%%Eq (7.10) Definition of $r_2$
where $\zeta(3)$ is the Riemann Zeta function.
Then, Eq.\ (\ref{eq:fpi_Ld}) yields the following nonlinear relation
for the 2-loop effective threshold $s_0^\text{2-loop}$
\begin{eqnarray}
 \label{eq:WI_s0_2L}
  s_0^\text{2-loop}
   \left\{1 + \frac{\alpha_s(\muR)}{\pi}
            + \left(\frac{\alpha_s(\muR)}{\pi}\right)^2
               \left[r_2
                   - \frac{b_0}{4}
                     \left(\ln\frac{s_0^\text{2-loop}}{\muR}
                         - 1
                     \right)
               \right]
   \right\}
 &=& 4\pi^2f_{\pi}^2
\end{eqnarray}
%%Eq (7.11) 2-loop effective threshold
that replaces the standard LD relation, notably, Eq.\ (\ref{eq:LDfpi}).
Note in this context that the effective 2-loop threshold
$s_0^\text{2-loop}$ should be used only in formulas containing the
2-loop spectral density $\rho_3^\text{2-loop}(s,s',Q^2)$.
Were we in the position to write down the 2-loop spectral density
$\rho_3^\text{2-loop}(s,s',Q^2)$ for all $Q^2$ values, then we would
have obtained via Eqs.\ (\ref{eq:FFQuarLd})-(\ref{eq:fpi_Ld}) an
expression for the pion form factor valid at $O(\alpha_s^2)$.
Instead, we use the leading-order LD expression,
$F_{\pi}^\text{LD}(Q^2;\muR)$,
and add perturbative $O(\alpha_s)$- and
$O(\alpha_s^2)$-corrections
explicitly in terms of $F_{\pi}^\text{Fact}(Q^2;\muR)$.
Recalling Eq.\ (\ref{eq:WI_FF0}),
we then have
\begin{eqnarray}
 \label{eq:FFact_Q2=0}
 F_{\pi}^\text{Fact}(Q^2=0;\muR) = 0\,.
\end{eqnarray}
%%Eq (7.12) Limit of factorizable form factor at Q^2=0 using Ward
%%          identity
The next task is to match this low-$Q^2$ value with the large-$Q^2$
result of pQCD, $F_{\pi}^\text{Fact}(Q^2;\muR)$.
The most straightforward way is to adopt
$F_{\pi}^\text{Fact}(Q^2;\muR)$ at large $Q^2$ and correct its singular
($\sim1/Q^2$) behavior at small $Q^2$ by introducing some reasonable
mass scale $M_0^{-1}$ via the replacement\footnote{One may
think of this scale as corresponding to the maximum transverse quark
antiquark separation $b_0 \sim M_0$ still accessible to the hard
form factor via hard-gluon exchange just before the crossover to the
nonperturbative dynamics.}
\begin{eqnarray}
 F_{\pi}^\text{Fact}(Q^2;\muR)
  &\equiv&  \tilde{F}_{\pi}(Q^2;\muR) \frac{M_0^2}{Q^2}
   \ \to\   \tilde{F}_{\pi}(Q^2;\muR) \frac{M_0^2}{M_0^2+Q^2}\,.
\end{eqnarray}
%%Eq (7.13) Removing singular behavior of factorized form factor
However, this expression has the wrong limit at $Q^2=0$, so that one
needs to correct it in order to maintain the Ward identity (WI):
\begin{eqnarray}
  F_{\pi}^\text{Fact-WI}(Q^2;\muR)
  &=& - \tilde{F}_{\pi}(Q^2;\muR)\,\Phi(Q^2/M_0^2)
      +  \tilde{F}_{\pi}(Q^2;\muR) \frac{M_0^2}{M_0^2+Q^2}\,.
\end{eqnarray}
%%Eq (7.14) Modified Form factor to maintain Ward identity
Here the function $\Phi(z)$ is some smooth function with $\Phi(0)=1$
and $z\Phi(z)\to 0$ when $z\to\infty$, introduced to preserve the
high $Q^2$-asymptotics of $F_{\pi}^\text{Fact}(Q^2;\muR)$.
The simplest choice for $\Phi(z)$ is $\Phi(z)=1/(1+z)^2$, yielding
\begin{eqnarray}
  F_{\pi}^\text{Fact-WI}(Q^2;\muR)
  &=& \tilde{F}_{\pi}(Q^2;\muR)
       \frac{M_0^2}{M_0^2+Q^2}
       \left(1 - \frac{M_0^2}{M_0^2+Q^2}
       \right)\nonumber\\
 \label{eq:FMatch}
  &=& F_{\pi}^\text{Fact}(Q^2;\muR)
       \left(\frac{Q^2}{M_0^2+Q^2}\right)^2\,.
\end{eqnarray}
%%Eq (7.15) Simplest choice to fulfill WI
The scale parameter $M_0^2$ should be identified with the threshold
$2s_0^\text{2-loop}$
to read
$M_0^2=2s_0^\text{2-loop}$
because $s_0^\text{2-loop}$ is the ``natural''
scale parameter for the 2-point correlator in the pion case, while the
scale characterizing the 3-point correlator, corresponding to the form
factor, is two times larger \cite{BR91}.

In this way, we finally arrive at
\begin{eqnarray}
 \label{eq:Fpi-Mod}
  F_{\pi}^\text{Fact-WI}(Q^2;\muR)
  &=& \left(\frac{Q^2}{2s_0^\text{2-loop}+Q^2}\right)^2
       F_{\pi}^\text{Fact}(Q^2;\muR)\,.
\end{eqnarray}
%%Eq (7.16) Interpolating expression from low to high Q^2 for
%%          factorizable form factor
We are now in the position to supply an expression for the total pion
form factor valid in the whole $Q^2$ range:
\begin{eqnarray}
  F_{\pi}(Q^{2};\muR)
  =  F_{\pi}^\text{LD}(Q^{2})
  +  F_{\pi}^\text{Fact-WI}(Q^2;\muR)\,.
\label{eq:Q2Pff}
\end{eqnarray}
%%Eq (7.17) Total pion form factor
This expression comprises the NLO prediction for the factorized part
under the proviso of the Ward identity at $Q^2=0$ and the
non-factorizable soft part.
[Parenthetically, note the explicit $\muR$ dependence of this
expression as a consequence of the truncation of the perturbative
series (see Eqs.\ (\ref{eq:Q2pff})--(\ref{eq:Q2pffNLOnondia})).]

\begin{figure}[h]
\centerline{\includegraphics[width=0.475\textwidth]{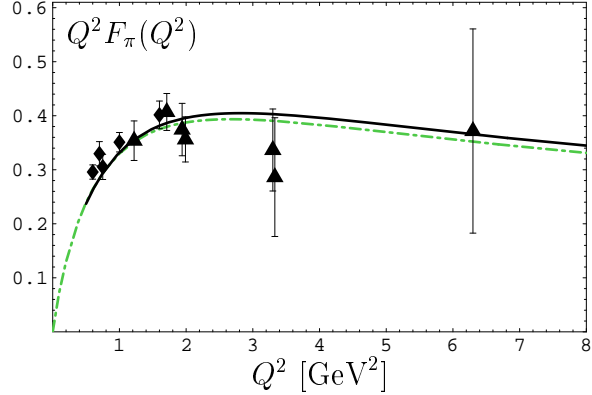}}%
 \caption[pion-AR]{\footnotesize
  Comparison of the pion form factor calculated
  by Radyushkin \protect{\cite{Rad95}} using the LD\ approach and
  interpolation from low to high $Q^2$ (dash-dotted line) with our
  result for the total form factor given by  Eq.\ (\ref{eq:Q2Pff}),
  computed with the asymptotic DA in the \MS scheme with the \BLM scale
  setting and using LO APT (solid line).
  The experimental data are taken from \protect{\cite{JLAB00}}
  (diamonds) and \cite{FFPI73}, \cite{FFPI76} (triangles).
  \label{fig:match_AR}}
\end{figure}

Before continuing with the presentation of our final results, let us
remark that a similar type of matching has been applied by
Radyushkin \cite{Rad95} to describe the pion form factor, providing
the result
\begin{equation}
 F_{\pi}(Q^2)
 =
 \frac{F_{\pi}^\text{LD(0)}(Q^2) +
    \left(\alpha_s/\pi\right)\left[1+Q^2/\left(2s_0\right)\right]^{-1}}
     {1+ \alpha_s/\pi}
 \label{eq:FpiAR}
\end{equation}
%%Eq (7.18) Radyushkin's pion form factor
illustrated in Fig.\ \ref{fig:match_AR} (a) (dash-dotted line).
%%%%%%%%%%%%%%%%%%%%%%%%%%%%%%%%%%%%%%%%%%%%%%%%%%%%%%%%%%%%%%%%%%%%%%%
%                              FIGURE 14                              %
%%%%%%%%%%%%%%%%%%%%%%%%%%%%%%%%%%%%%%%%%%%%%%%%%%%%%%%%%%%%%%%%%%%%%%%
%%%%%%%%%%%%%%%%%%%%%%%%%%%%%%%%%%%%%%%%%%%%%%%%%%%%%%%%%%%%%%%%%%%%%%%
\noindent
In this equation---which follows the Brodsky--Lepage interpolation
formula for the $\pi\gamma$-transition form factor~\cite{BL80mad}---the
first term means the soft form factor calculated with the LD\ approach,
while the second one includes the LO radiative corrections.
It is evident from this figure that Radyushkin's result is very close
to that given by Eq.\ (\ref{eq:Q2Pff}), evaluated with the asymptotic
DA in the \MS scheme with the \BLM scale setting using for the sake of
comparison LO APT (solid line).

\begin{figure}[t]
\centerline{\includegraphics[width=0.475\textwidth]{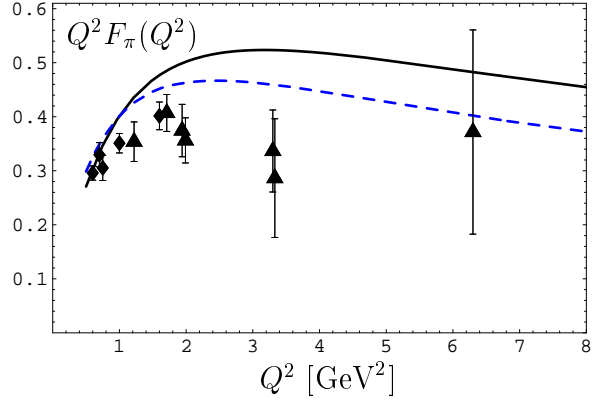}}
 \caption[pion]{\footnotesize
  Theoretical predictions for $Q^2F_\pi(Q^2)$
  obtained with the BMS\ pion DA using standard pQCD within
  the  $\overline{\strut \text{MS}}$ scheme and adopting
  the $\muR=Q^2$ (dashed line) and
  $\overline{\strut\text{BLM}}$ (solid line) scale settings.
  The experimental data are taken from \protect{\cite{JLAB00}}
  (diamonds) and \protect\cite{FFPI73}, \protect\cite{FFPI76}
  (triangles).
 \label{fig:FFSum-Sta}}
\end{figure}

\begin{figure}[b]
\centerline{\includegraphics[width=0.975\textwidth]{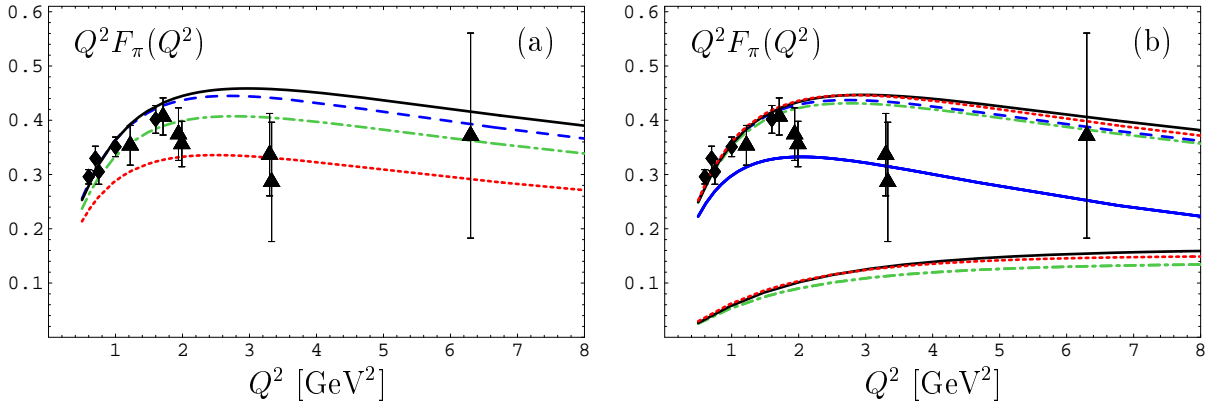}}
 \caption[pion]{\footnotesize
  Theoretical predictions for $Q^2F_\pi(Q^2)$
  using analytic perturbation theory and the BMS\ DA in conjunction
  with the ``Naive Analytic'' (a) and ``Maximally Analytic'' (b)
  analytization procedures.
  Different scale settings within the $\overline{\strut \text{MS}}$
  scheme are used: $\muR=Q^2$ (dashed line),
  BLM\ (dotted line), and $\overline{\strut\text{BLM}}$ (solid line).
  The dash-dotted line represents the prediction obtained with the
  $\alpha_V$-scheme.
  Also included are the prediction for the soft form-factor part
  (solid blue line) and below this, the hard contributions in
  correspondence with the predictions for the total form factor on the
  upper part of the figure.
  The experimental data are taken from \protect{\cite{JLAB00}}
 (diamonds) and \cite{FFPI73}, \cite{FFPI76} (triangles).
 \label{fig:FFSum-Apt}}
\end{figure}

Employing the above considerations we now present predictions for the
total scaled form factor vs.\ $Q^2$ in different renormalization schemes
and perturbation-theory approaches using the BMS\ pion DA.
Figure \ref{fig:FFSum-Sta} shows the results for the standard
perturbation theory within the \MS scheme adopting the $\muR=Q^2$
(dashed line) and the \BLM (solid line) scale settings.
In Fig.\ \ref{fig:FFSum-Apt} we present analogous predictions
calculated with the APT.
In this case, it is possible to include results computed with the BLM\
scale setting and to use the $\alpha_V$ scheme.
We observe from this figure (left panel) that the ``Naive
Analytization'' gives results that bear a rather strong scheme and
scale-setting dependence.
In contrast, applying the ``Maximal Analytic'' procedure, the
arbitrariness in the scheme and scale setting is minimized---Fig.\
\ref{fig:FFSum-Apt}(b) being a graphic proof of that.
Note that this figure shows also separately the soft part of the form
factor, displayed in Fig.\ \ref{fig-TR}, and the hard contributions
corresponding to the various scheme and scale settings discussed above
and presented in Fig.\ \ref{fig:APT_msbarav}(b).
%%%%%%%%%%%%%%%%%%%%%%%%%%%%%%%%%%%%%%%%%%%%%%%%%%%%%%%%%%%%%%%%%%%%%%%
%                              FIGURE 15                              %
%%%%%%%%%%%%%%%%%%%%%%%%%%%%%%%%%%%%%%%%%%%%%%%%%%%%%%%%%%%%%%%%%%%%%%%
%%%%%%%%%%%%%%%%%%%%%%%%%%%%%%%%%%%%%%%%%%%%%%%%%%%%%%%%%%%%%%%%%%%%%%%

%%%%%%%%%%%%%%%%%%%%%%%%%%%%%%%%%%%%%%%%%%%%%%%%%%%%%%%%%%%%%%%%%%%%%%%
%                              FIGURE 16                              %
%%%%%%%%%%%%%%%%%%%%%%%%%%%%%%%%%%%%%%%%%%%%%%%%%%%%%%%%%%%%%%%%%%%%%%%
%%%%%%%%%%%%%%%%%%%%%%%%%%%%%%%%%%%%%%%%%%%%%%%%%%%%%%%%%%%%%%%%%%%%%%%

\begin{figure}[t]
\centerline{\includegraphics[width=0.975\textwidth]{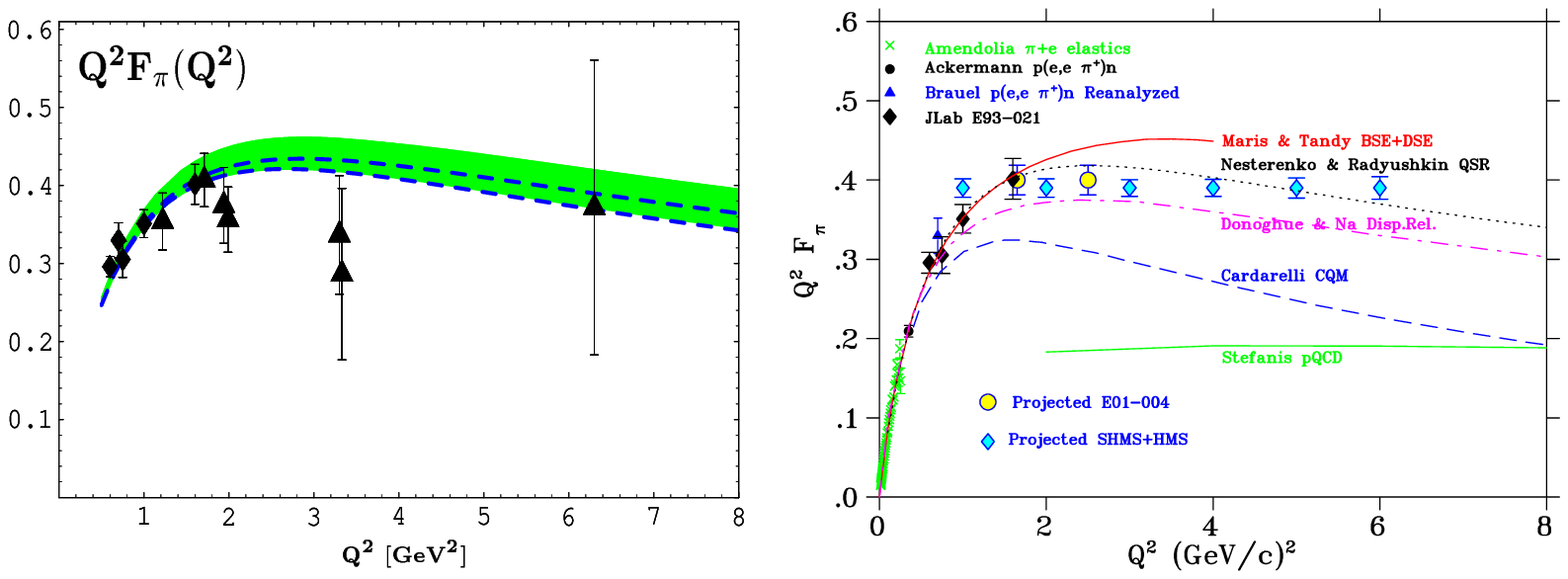}}
 \caption[fig:pidata]
        {\footnotesize (Left) Predictions for the scaled pion form
        factor calculated with the BMS\ bunch (green strip)
        encompassing nonperturbative uncertainties from nonlocal QCD
        sum rules \protect{\cite{BMS01}} and renormalization scheme and
        scale ambiguities at the level of the NLO accuracy, as
        discussed in Fig.\ \protect{\ref{fig:FFSum-Apt}}.
        The dashed lines inside the strip indicate the corresponding
        area of predictions obtained with the asymptotic pion DA.
        Note that this strip contains only perturbative scheme and
        scale ambiguities at the level of the NLO accuracy, calculated
        in APT with the ``Maximally Analytic'' procedure.
        The experimental data are taken from \protect{\cite{JLAB00}}
        (diamonds) and \cite{FFPI73}, \cite{FFPI76} (triangles).
        (Right) Summary of existing and projected experimental data on
        the electromagnetic pion form factor in comparison with the
        results of various theoretical calculations; figure taken from
        \protect{\cite{Blok02}} (see there for explanations).
  \label{fig:pidatasum}}
\end{figure}

The phenomenological upshot of our analysis is summarized in the left
panel of Fig.\ \ref{fig:pidatasum}, where we show predictions for the
whole BMS\ ``bunch'' of pion DAs \cite{BMS01}.
The shaded strip incorporates the nonperturbative uncertainties related
to non local QCD sum rules and also the ambiguities induced by the
scheme and renormalization scale setting---in correspondence to Fig.\
\ref{fig:FFSum-Apt}.
Note that the two broken lines mark the region of predictions
associated with the asymptotic pion DA.
These results can be compared with previous theoretical predictions
and also with further experimental data to be obtained at JLab (see
right part of Fig.\ \ref{fig:pidatasum}, taken from \cite{Blok02}).
The data points extending to $Q^2$ of $6$~GeV$^2$ are expectations
from projected experiments at JLab\ after the planned upgrade of
CEBAF\ to 12~GeV (we refer to \cite{Blok02} for further explanations
and related references).

These striking findings give convincing evidence that the end-point
suppressed structure of the BMS\ type pion DA not only provides
best agreement with the CLEO and CELLO data (cf.\ Fig.\
\ref{fig:Formfactor}), it also allows to describe the pion form-factor
data with at least the same quality as with the asymptotic pion DA---as
it becomes evident from the LHS of Fig.\ \ref{fig:pidatasum}.

%%%%%%%%%%%%%%%%%%%%%%%%%%%%%%%%%%%%%%%%%%%%%%%%%%%%%%%%%%%%%%%%%%%%%%%
%                              FIGURE 17                              %
%%%%%%%%%%%%%%%%%%%%%%%%%%%%%%%%%%%%%%%%%%%%%%%%%%%%%%%%%%%%%%%%%%%%%%%
%%%%%%%%%%%%%%%%%%%%%%%%%%%%%%%%%%%%%%%%%%%%%%%%%%%%%%%%%%%%%%%%%%%%%%%

%\cleardoublepage

%%%%%%%%%%%%%%%%%%%%%%%%%%%%%%%%%%%%%%%%%%%%%%%%%%%%%%%%%%%%%%%%%%%%%%%
%%%%%%%%%%%%%%%%%%%%%%%%%%%%%%%%%%%%%%%%%%%%%%%%%%%%%%%%%%%%%%%%%%%%%%%
\section{Summary and conclusions}
\label{sect:concl}
%%%%%%%%%%%%%%%%%%%%%%%%%%%%%%%%%%%%%%%%%%%%%%%%%%%%%%%%%%%%%%%%%%%%%%%
In summary, the key concepts and merits arising from this analysis are
as follows.
\begin{itemize}
\item We worked out interpolation formulas for the analytic
coupling constant and its analytic second power that take into account
heavy-flavor thresholds and greatly facilitate calculations.
This allowed us to develop a theoretical procedure and apply its
numerical realization in order to compute the evolution of the pion DA
using NLO analytic perturbation theory.
The hard form factor was corrected at low $Q^2$ as to fulfill the
Ward identity and was added to the soft form factor, derived via Local
Duality, without introducing double counting.
\item On the theoretical front, we found that replacing the QCD
effective coupling and its powers by their analytic images---a
procedure we termed ``Maximally Analytic''---not only provides IR
protection to the coupling, it also helps diminishing the
renormalization scheme and scale-setting dependence of the form-factor
predictions already at the NLO level, rendering the calculation of
still higher-order corrections virtually superfluous.
\item From the phenomenological point of view, our most discernible
result is that the BMS pion DA \cite{BMS01} (out of a ``bunch'' of
similar doubly-peaked endpoint-suppressed pion DAs) yields to
predictions for the electromagnetic form factor very close to those
obtained with the asymptotic pion DA.
Hence, concerns that a double-humped pion DA could jeopardize the
sound application of pQCD are unduly.
Conversely, we have shown that a small deviation of the prediction
for the pion form factor from that obtained with the asymptotic pion DA
does not necessarily imply that the underlying pion DA has to be close
to the asymptotic profile.
Much more important is the behavior of the pion DA in the endpoint
region $x\to 0\,, 1$.
\end{itemize}

Looking further into the future is yet more exciting.
With the planned upgrade of the CEBAF\ experiment to 12~GeV, the pion's
electromagnetic form factor can be studied up to
$Q^2\simeq 6$~GeV${}^2$ \cite{Blok02},
providing crucial constraints to verify the various theoretical
predictions discussed here and elsewhere.
The apparently good agreement of our results with the available
experimental data (see Fig.\ \ref{fig:pidatasum}) is encouraging.

\bigskip

\acknowledgments
We wish to thank Bla\v{z}enka Meli\'{c}, Sergey Mikhailov,
Dieter M\"{u}ller, Dan Pirjol, Dmitry Shirkov and Igor Solovtsov
for useful discussions, and Badri Magradze for supplying us with
unpublished numerical data on $\asb^{(2)}(Q^2)$ and
${\cal A}_{2}^{(2)}(Q^2)$ that take into account quark mass thresholds.
We would also like to thank Henk Blok and Garth Huber for providing
information on the current pion form-factor experimental data.
Two of us (A.P.B.\ and K.P.-K.) are indebted to Prof.\ Klaus Goeke
for the warm hospitality at Bochum University,
where the major part of this investigation was carried out.
This work was supported in part by the Deutsche Forschungsgemeinschaft
(Projects 436 KRO 113/6/0-1 and 436 RUS 113/752/0-1),
the Heisenberg--Landau Programme (grants 2002 and 2003),
the COSY Forschungsprojekt J\"{u}lich/Bochum,
the Russian Foundation for Fundamental Research
(grants No.\ 03-02-16816 and 03-02-04022),
the INTAS-CALL 2000 N 587,
and by the Ministry of Science and Technology of the Republic of
Croatia under Contract No.\ 0098002.
The work of W.S. was supported in part by a Feodor-Lynen Fellowship
from the Alexander von Humboldt Foundation and in part by the
U.S.~Department of Energy (D.O.E.) under cooperative research
agreement \#DF-FC02-94ER40818.

\cleardoublepage
\begin{appendix}
\appendix
%%%%%%%%%%%%%%%%%%%%%%%%%%%%%%%%%%%%%%%%%%%%%%%%%%%%%%%%%%%%%%%%%%%%%%%
%%%%%%%%%%%%%%%%%%%%%%%%%%%%%%%%%%%%%%%%%%%%%%%%%%%%%%%%%%%%%%%%%%%%%%%
\section{Hard-scattering amplitude for the pion's electromagnetic form
         factor at NLO}
\label{app:HSAnlo}
%%%%%%%%%%%%%%%%%%%%%%%%%%%%%%%%%%%%%%%%%%%%%%%%%%%%%%%%%%%%%%%%%%%%%%%
In this section we list the NLO results for the hard-scattering
amplitude \cite{FGOC81,DR81,Sar82,RK85,KMR86,BT87,MNP99a}, used in our
analysis.

The LO contribution to $T_\text{H}(x, y, Q^2;\muF)$, expanded as in
\req{eq:TH}, reads
\begin{eqnarray}
   T_\text{H}^{(0)}(x,y,Q^2)
   = \frac{N_\text{T}}{Q^2} \,\frac{1}{\x \y}
        \,,
\label{eq:THLOpff}
\end{eqnarray}
%%Eq (A1) Hard-scattering amplitude in LO
where
\begin{eqnarray}
   N_\text{T} = \frac{2\, \pi \, C_\text{F}}{C_\text{A}}
       = \frac{8\pi}{9}\, ,
\label{eq:NTpff}
\end{eqnarray}
%%Eq (A2) Color factor for LO pion form factor
$C_\text{F}=\left(N_{\rm c}^{2}-1\right)/2N_{\rm c}=4/3$,
$C_\text{A}=N_{\rm c}=3$ are the color factors of
$SU(3)_{\rm c}$, and the notation $\bar{z} \equiv 1-z$ has been used.
The usual color decomposition of the NLO correction---marked by
self-explainable labels---is given by
\begin{eqnarray}
 T_\text{H}^{(1)}\left(Q^2;\muF,\muR\right)
  = C_\text{F}\,T_\text{H}^{(1,\text{F})}\left(Q^2;\muF\right)
  + b_0\,T_\text{H}^{(1,\beta)}\left(Q^2;\muR\right)
  + C_\text{G}\,T_\text{H}^{(1,\text{G})}\left(Q^2\right)\,,
 \label{eq:THNLOpff}
\end{eqnarray}
%%Eq (A3) Color decomposition of hard-scattering amplitude in NLO
where $C_\text{G}=(C_\text{F}-C_\text{A}/2)$
and the first coefficients of the $\beta$
function are
\begin{eqnarray}
    b_0=\frac{11}{3}C_\text{A} - \frac{4}{3}T_\text{R} N_f
    \,,\qquad \qquad
    b_1=\frac{34}{3}C_{\text{A}}^{2}-
    \left(4C_\text{F} + \frac{20}{3}C_\text{A}\right)T_\text{R} N_f
    \,.
\label{eq:beta0&1}
\end{eqnarray}
%%Eq (A4) First two beta-function coefficients: b_0, b_1
Here, $T_\text{R}=1/2$ and $N_f$ denotes the number of flavors, whereas
the expansion of the $\beta$-function in the NLO approximation is given
by
\begin{equation}
 \beta(\alpha_{s}(\mu^2))
  = -\alpha_{s}(\mu^2)
      \left\{b_0\left[\frac{\alpha_{s}(\mu^2)}{4 \pi}\right]
           + b_1\left[\frac{\alpha_{s}(\mu^2)}{4 \pi}\right]^2
      \right\}\,.
 \label{eq:betaf}
\end{equation}
%%Eq (A5) NLO approximation for beta function
With reference to the application of the BLM\ \cite{BLM83} scale
setting in fixing the renormalization point, we single out the
$b_0$-proportional (i.e., $N_f$-dependent) term, given by
\begin{subequations}
\label{eq:TH1}
\begin{eqnarray}
  T_\text{H}^{(1,\beta)}\left(x,y,Q^2;\muR\right)
& = &
     \frac{N_\text{T}}{Q^2}\,
     \frac{1}{\x\y}\,
     \left[\frac{5}{3}
           - \ln (\x \y)
           - \ln \frac{Q^2}{\muR}
     \right]\,,
\label{eq:TH1beta}
\end{eqnarray}
%%Eq (A6a) Hard-scattering amplitude term proportional to beta function
and present the remainder of $T_\text{H}$ as a color decomposition in
the form
\begin{eqnarray}
  T_\text{H}^{(1,\text{F})}(x,y,Q^2;\muF)
& = &
  \frac{N_\text{T}}{Q^2} \,
        \frac{1}{\x \y} \,
         \left[
               -\frac{28}{3}
               + \left( 6 -\frac{1}{x} \right) \ln \x
               + \left( 6 -\frac{1}{y} \right) \ln \y
               + \ln^2 (\x \y)
\right. \nonumber \\[0.15cm]
&&  \left.
               + 2 \ln \frac{Q^2}{\muF}
              \left(3 + \ln (\x\y) \right)
         \right]
\label{eq:TH1F}
\end{eqnarray}
%%Eq (A6b) (1,F) hard-scattering amplitude term
\begin{eqnarray}
  T_\text{H}^{(1,\text{G})}(x,y,Q^2)
& = &
       \frac{N_\text{T}}{Q^2} \,
       \frac{1}{\x \y} \,
         \left[
               -\frac{20}{3}
               -8 \frac{\ln\x}{x}
               -8 \frac{\ln\y}{y}
               -2 \ln \x \ln \y
               -2 \ln x \ln y
\right. \nonumber \\[0.15cm]
&&  \left.
               + 2 \ln x \ln \y + 2 \ln \x \ln y
              -2 (1-x-y) \text{H} (x,y) -2 \,  \text{R} (x,y)
         \right]
\nonumber \\ [0.15cm]
\label{eq:TH1G}
\end{eqnarray}
\end{subequations}
%%Eq (A6c) (1,G) hard-scattering amplitude term
for the color singlet and color non-singlet parts, respectively.
For calculational convenience, we also supply the sum of these terms
(cf.\ Eq.\ (\ref{eq:Q2pff1FG})):
\begin{eqnarray}
 T_\text{H}^{(1,\text{FG})}\left(x,y,Q^2;\muF\right)
 &=& C_\text{F}\,T_\text{H}^{(1,\text{F})}\left(x,y,Q^2;\muF\right)
   + C_\text{G}\,T_\text{H}^{(1,\text{G})}\left(x,y,Q^2\right)
\nonumber\\
 &=& \frac{N_\text{T}}{Q^2}\,
      \frac{1}{\x\y}\,
       \frac{1}{3}\,
        \Biggl\{-34
             + 24\ln(\x\y)
             + 4\ln^2(\x\y)
          \nonumber \\
  &&         + \ln x\ln y
             + \ln\x\ln\y
             - \ln x\ln\y
             - \ln\x\ln y
         \nonumber \\
  &&         + (1-x-y)\text{H}(x,y)
             + \text{R}(x,y)
             + 8\,\left[3 + \ln(\x\y)\right]\,
                \ln\frac{Q^2}{\muF}
         \Biggr\}\, ,
 \label{eq:TH1C1}
\end{eqnarray}
%%Eq (A7) (1,FG) Hard-scattering part
where
\begin{subequations}
\label{eq:HR}
\begin{eqnarray}
 \text{H}(x,y)
  &=& \frac{1}{1-x-y}
       \left[\Li\left(\frac{\y}{x}\right)
           + \Li\left(\frac{\x}{y}\right)
           + \Li\left(\frac{x y}{\x\y}\right)
        \right. \nonumber \\
  & &   \left.\qquad\qquad\
           - \Li\left(\frac{x}{\y}\right)
           - \Li\left(\frac{y}{\x}\right)
           - \Li\left(\frac{\x\y}{x y}\right)
         \right]
\label{eq:H}
\end{eqnarray}
%%Eq (A8a) Function termed H(x,y) in NLO hard-scattering amplitude
with $\Li$ being the di-logarithm function, defined by
$\Li(z)=-\int_{0}^{z}\frac{\ln (1-t)dt}{t}$,
and
\begin{eqnarray}
 \text{R}(x,y)
  &=& \frac{1}{(x-y)^{2}}
       \Bigl[(2xy-x-y)(\ln x+\ln y)
       \nonumber \\
  & &      + (-2xy^{2}-2y^{2}+10xy-2y-4x^{2})
              \frac{\ln\y}{y}
       \nonumber \\
  & &      + (-2yx^{2}-2x^{2}+10xy-2x-4y^{2})
              \frac{\ln\x}{x}
       \nonumber \\
  & &
           - (y\y^{2}+x\x^{2})\text{H}(x,\y)
       \Bigr]\,.
\label{eq:R}
\end{eqnarray}
%%Eq (A8b) Function termed R(x,y) in NLO hard-scattering amplitude
\end{subequations}

%%%%%%%%%%%%%%%%%%%%%%%%%%%%%%%%%%%%%%%%%%%%%%%%%%%%%%%%%%%%%%%%%%%%%%%
%%%%%%%%%%%%%%%%%%%%%%%%%%%%%%%%%%%%%%%%%%%%%%%%%%%%%%%%%%%%%%%%%%%%%%%
\section{Factorization scale dependence in standard pQCD}
\label{app:Fact}
%%%%%%%%%%%%%%%%%%%%%%%%%%%%%%%%%%%%%%%%%%%%%%%%%%%%%%%%%%%%%%%%%%%%%%%
Here we examine the $\muF$-dependence of the hard-scattering
amplitude (see, for example,
\cite{LB79,ER80,MNP01a,MNP01b,Passek01}). We start with the
representation for $T_H(\muF,\muR)$, given by (\ref{eq:TH}), to
get
\begin{eqnarray}
 T_\text{H}(\muF,\muR)
  &=& \alpha_s(\mu_R^2)
       \left[T_\text{H}^{(0)}
           + \frac{\alpha_s(\mu_R^2)}{4\pi}
             T_\text{H}^{(1)}(\muF,\muR)
       \right]
\end{eqnarray}
%%Eq (B1) NLO approximation of Hard Scattering Amplitude
with $T_\text{H}^{(0)}(x,y,Q^2)$ and
$T_\text{H}^{(1)}(x,y,Q^2;\muF,\muR)$
as in (\ref{eq:THLOpff}) and (\ref{eq:THNLOpff}) for the LO and NLO,
respectively.
These functions can be represented as follows
\begin{eqnarray}
 \label{eq:T0_PT}
  T_\text{H}^{(0)}
    &=& N_\text{T} C_0(x,y,Q^2)\,,\quad \text{and~~~} C_0
    =\frac{1}{\bar{x}\bar{y}Q^2}\,,\\
 \label{eq:T1_PT}
  T_\text{H}^{(1)}(\muF,\muR)
    &=& \ln\left(\frac{Q^2}{\muF}\right)
         \left[T_\text{H}^{(0)}(x,s,Q^2)\convo{s}V_0(s,y)
             + V_0(s,x)\convo{s}T_\text{H}^{(0)}(s,y,Q^2)
         \right]\nonumber\\
    &-& b_0 \ln\left(\frac{Q^2}{\muR}\right)
         T_\text{H}^{(0)}(x,y,Q^2)
      + N_\text{T} C_1(x,y,Q^2)\,,
\end{eqnarray}
%%Eq (B2) LO Hard Scattering Amplitude
%%Eq (B3) NLO contribution to Hard Scattering Amplitude
where $C_1(x,y,Q^2)$ absorbs all other $\muF$- and $\muR$-independent
terms from (\ref{eq:THNLOpff}).
Using this structure with respect to the $\muF$-dependence, we can
conclude that
\begin{eqnarray}
\label{eq:PT_Factorization_1L}
 \frac{dT_H}{d\ln\muF}
  &=& - T_H(\muF,\muR)\convo{}V(\alpha_s(\muR))
      - V(\alpha_s(\muR))\convo{}T_H(\muF,\muR)
      + O(\alpha_s^3)\,,
\end{eqnarray}
%%Eq (B4) Evolution equation for Hard Scattering Amplitude
where $V(\alpha_s(\muR))$ is the ERBL evolution kernel
(\ref{eq:kernel}).
Then, the whole derivative of the form factor (\ref{eq:pff-Fact})
is
\begin{eqnarray}
 \frac{d F_\pi(Q^2;\muR)}{d\ln\muF}
  &=& \Phi_\pi(\muF)\convo{}T_H(\muF,\muR)\convo{}
       \left[V(\alpha_s(\muF))- V(\alpha_s(\muR))\right]
        \convo{}\Phi_\pi(\muF)\nonumber\\
  &+& \Phi_\pi(\muF)\convo{}\left[V(\alpha_s(\muF))
   -  V(\alpha_s(\muR))\right]
       \convo{}T_H(\muF,\muR)
        \convo{}\Phi_\pi(\muF)\nonumber\\
  &+& O(\alpha_s^3)\,.
\end{eqnarray}
%%Eq (B5) Evolution equation for Pion FF to 2 loops
Recalling that in the 2-loop approximation of the standard pQCD
\begin{eqnarray}\label{eq:PT_beta}
  \frac{d\alpha_s(\mu^2)}{d\ln\mu^2}
   &=& -4\pi b_0
         \left[\frac{\alpha_s(\mu^2)}{4\pi}\right]^2
          \left[1+O(\alpha_s(\mu^2))\right],
\end{eqnarray}
%%Eq (B6) RGE for strong coupling
we have
\begin{eqnarray}
  V\left(\alpha_s(\muF)\right)-V\left(\alpha_s(\muR)\right)
  &=& \left[\frac{\alpha_s(\muF)-\alpha_s(\muR)}{4\pi}\right]
      V_0 + O(\alpha_s^2)
  \ =\ O(\alpha_s^2)\,,
\end{eqnarray}
%%Eq (B7) Difference of evolution kernels
so that
\begin{eqnarray}
  \frac{d F_\pi(Q^2;\muR)}{d\ln\muF}
  &=& O(\alpha_s^3)\,.
\end{eqnarray}
%%Eq (B8) Violation of factorization-scale independence of pion FF
Hence, we conclude that at the level of the NLO approximation
of the standard pQCD, the violation of the
factorization-scale independence is one order of $\alpha_s$
higher.\footnote{Moreover, the dependence on $\muF$ of the NLO
prediction for the pion form factor was investigated in
\cite{MNP99a}, where it was found that these results vary only
slightly with $\muF$ rendering the ``factorization-scheme
ambiguity'' to be small.}

%%%%%%%%%%%%%%%%%%%%%%%%%%%%%%%%%%%%%%%%%%%%%%%%%%%%%%%%%%%%%%%%%%%%%%%
%%%%%%%%%%%%%%%%%%%%%%%%%%%%%%%%%%%%%%%%%%%%%%%%%%%%%%%%%%%%%%%%%%%%%%%
\section{Two-loop evolution of the pion distribution amplitude in
         standard pQCD}
\label{app:DAnlo}
%%%%%%%%%%%%%%%%%%%%%%%%%%%%%%%%%%%%%%%%%%%%%%%%%%%%%%%%%%%%%%%%%%%%%%%
The pion distribution amplitude $\varphi_\pi(x,\muF)$ satisfies an
evolution equation of the form
\begin{eqnarray}
  \frac{d\, \varphi_\pi(x,\muF)}{d \ln\muF}
   = V(x,u,\alpha_s(\muF))\convo{u}\varphi_\pi(u,\muF)\,,
\label{eq:eveq}
\end{eqnarray}
%%Eq (C1) Evolution equation for pion DA
where $V(x,u,\alpha_s(\muF))$ is the perturbatively calculable NLO
evolution kernel
\begin{eqnarray}
  V(x,u,\alpha_s)
        =  \frac{\alpha_s}{4 \pi} \, V_0(x,u) +
                \frac{\alpha_s^2}{(4 \pi)^2}  V_1(x,u)\,.
\label{eq:kernel}
\end{eqnarray}
%%Eq (C2) Evolution kernel in 2-loop order
If the distribution amplitude $\varphi_\pi(x,\muO)$ is determined at an
initial momentum scale $\muO$ (using some nonperturbative methods),
then the integro-differential evolution equation \req{eq:eveq} can be
integrated using the moment method to give $\varphi_\pi(x,\muF)$ at any
momentum scale $\muF$.
The one-\cite{LB80} and two-loop \cite{DR84,Sarmadi84,MR85} corrections
to the evolution kernel were determined in the \MS-scheme, but
because of the complicated structure of the two-loop corrections, only
the numerical evaluation of the (first few) moments of the evolution
kernel was possible \cite{Katz85,MR86ev}.
However, making use of conformal-symmetry constraints, the complete
analytical form of the NLO solution of the evolution equation
\req{eq:eveq} has been obtained \cite{Mul94,Mul95}.
We note that for $\muF \to \infty$ the solution of Eq. \req{eq:eveq}
takes the asymptotic form
$\varphi_\pi(x,\muF \to \infty)\equiv \varphi_{\text{as}}(x)=6x(1-x)$.

The pion DA can be cast in the form
\begin{eqnarray}
 \varphi_\pi(x, \muF)
  = U(x, s; \muF, \muO)\convo{s}
    \varphi_\pi(s, \muO)\,,
 \label{eq:phiUphi}
\end{eqnarray}
%%Eq (C3) Evolution form of pion DA
where the operator $U(x, s; \muF, \muO)$ describes the evolution
from the scale $\muO$ to the scale $\muF$ and represents the solution
of an evolution equation equivalent to \req{eq:eveq}, given by
\begin{eqnarray}
  \frac{d}{d \ln\muF}\, U(x,s;\muF,\muO)
   = V(x,u,\alpha_s(\muF))\convo{u}U(u,s;\muF,\muO)\,.
 \label{eq:eveqU}
\end{eqnarray}
%%Eq (C4) Solution to evolution equation

It is convenient to express the nonperturbative input DA,
$\varphi_\pi(x,\muO)$, as an expansion over Gegenbauer polynomials
$C_{k}^{3/2}(2 x-1)$ which represent the eigenfunctions of the LO
kernel $V_0$, i.e.,
\begin{eqnarray}
 \varphi_\pi(x,\mu_0^2)
  = 6 x (1-x)
     \left[ 1
          + \sum_{m=2}^{\infty}{}' a_{m}(\muO) \, C_{m}^{3/2}(2x -1)
     \right]\,,
\label{eq:phi0-m}
\end{eqnarray}
%%Eq (C5) Expansion of pion DA in terms of Gegenbauer polynonials
in which ${\sum}'$ denotes the sum over even indices only.
The nonperturbative input is now contained in the $a_m(\muO)$
coefficients.
The Gegenbauer polynomials $C_n^{3/2}(2 x -1)$ satisfy the
orthogonalization condition
\begin{eqnarray}
 \int_0^1 dx \, x (1-x) \, C_n^{3/2}(2 x -1)\,  C_m^{3/2}(2x - 1)
      = N_n \,  \delta_{nm}
\label{eq:ortGegen}
\end{eqnarray}
%%Eq (C6) Orthogonalization condition of Gegenbauer polynomials
with respect to the weight $x(1-x)$, where
\begin{eqnarray}
    N_n  =  \frac{(n+1) (n+2)}{4 (2 n + 3)} \, .
\label{eq:Nn}
\end{eqnarray}
%%Eq (C7) Normalization factor
The moments of the evolution kernel
\begin{subequations}
\begin{eqnarray}
 M_{kn}\left[\alpha_s(\muF)\right]
  &=& C_k^{3/2}(2 x-1)\convo{x}
       V\left(x,y;\alpha_s(\muF)\right)\convo{y}
        \frac{y(1-y)}{N_n}C_n^{3/2}(2 y-1)\,,
\end{eqnarray}
%%Eq (C8) Moments of evolution kernel
or, equivalently, the anomalous dimensions
\begin{eqnarray}
 \gamma_{kn}\left[\alpha_s(\muF)\right]
 = -2  M_{kn}\left[\alpha_s(\muF)\right]\,,
\end{eqnarray}
%%Eq (C9) Anomalous dimensions
\end{subequations}
represent the elements of the triangular matrix ($k \ge n$).
While the LO kernel is diagonal with respect to the Gegenbauer
polynomials $C_n^{3/2}$ (only the
$\gamma_{nn}^{(0)} \equiv \gamma_n^{(0)}$ elements appear), the
structure of the NLO and still higher-order kernels leads to the
appearance of off-diagonal terms in the matrix of the anomalous
dimensions (both types of terms
$\gamma_{nn}^{(1)} \equiv \gamma_n^{(1)}$
as well as
$\gamma_{kn}^{(1)}$, $k>n$ are present).
Accordingly, the solution of the evolution equation \req{eq:eveqU}
takes the general form
\begin{eqnarray}
\lefteqn{U(x,s;\muF,\muO)}
\nonumber \\ &=&
\sum_{n=0}^{\infty} {}' \,
 \, E_n(\muF,\muO) \,
\left[ C_n^{3/2}(2 x-1)
+ \frac{\alpha_s(\muF)}{4 \pi}
 \sum_{k=n+2}^{\infty}{'} \, d_{kn}^{(1)}(\muF,\muO) C_k^{3/2}(2x-1)
+ {\cal O}(\alpha_s^3) \right]
\nonumber \\ && \times \,
\frac{x (1-x)}{N_n} \,
C_n^{3/2}(2 s -1)
\, .
\label{eq:solU}
\end{eqnarray}
%%Eq (C9) Solution of evolution equation
The effect of the diagonal terms $\gamma_{nn} \equiv \gamma_n$ is
completely contained in the factor $E_n(\muF,\muO)$, which is given by
\begin{eqnarray}
 E_n(\muF,\muO)
  = \exp\left[-\int_{\alpha_s(\muO)}^{\alpha_s(\muF)}d \alpha_s
                \frac{\gamma_{n}(\alpha_s)}
                     {2\,\beta(\alpha_s)}
        \right]\,.
 \label{eq:EnGen}
\end{eqnarray}
%%Eq (C10) Evolution effects of diagonal terms
The expansion of the anomalous dimensions in terms of $\alpha_s$ reads
\begin{subequations}
\begin{eqnarray}
  \gamma_n(\alpha_{s}(\mu^2))
   =  \frac{\alpha_{s}(\mu^2)}{4 \pi}\gamma_n^{(0)}
    + \frac{\alpha_{s}^2(\mu^2)}{(4 \pi)^2}\gamma_n^{(1)} + \ldots\,,
 \label{eq:gamman}
\end{eqnarray}
%%Eq (C11a) Expansion of anomalous dimensions
whereas the lowest order anomalous dimensions can be represented in
closed form by
\begin{eqnarray}
 \gamma_n^{(0)}
   = 2 C_\text{F}\left[4 S_1(n+1) %% := + 4 \sum_{i=1}^{n+1}\frac{1}{i}
                 - 3 - \frac{2}{(n+1) (n+2)}
           \right]
 \label{eq:gamma0}
\end{eqnarray}
%%Eq (C11b) Closed form of anomalous dimensions in LO
with $S_1(n+1)=\sum_{i=1}^{n+1} 1/i =\psi(n+2)+\psi(1)$,
while the function $\psi(z)$
is defined as $\psi(z)= d\ln\Gamma(z)/d z$.
Since the anomalous dimensions $\gamma_n$ coincide with the flavor
non-singlet anomalous dimensions, i.e., the moments of the splitting
kernels in deep inelastic scattering, we can use for $\gamma_n^{(1)}$
the results obtained in \cite{FRS77,GALY79}; viz.,
\begin{eqnarray}
 \gamma_0^{(1)} = 0 \,,\quad
  \gamma_2^{(1)} = \frac{830}{81}N_f
                 - \frac{34450}{243} \,, \quad
   \gamma_4^{(1)} = \frac{31132}{2025}N_f
                  - \frac{662846}{3375} \,,
 \label{eq:gamma1NF}
\end{eqnarray}
%%Eq (C11c) Values of first 3 anomalous dimensions
\end{subequations}
where $N_f$ denotes the number of active flavors.\footnote{For
$Q^2=1.7-18.5$~GeV$^2$ this number is 4, whereas for still higher $Q^2$
values, it starts to be equal 5.}
The nondiagonal matrix elements $\gamma_{kn}$ ($k>n$) manifest
themselves in the $d_{kn}^{(1)}$ terms of the eigenfunctions expansion
and were obtained in closed form in \cite{MR86ev,Mul94,Mul95}:
\begin{eqnarray}
 d_{kn}^{(1)}(\muF,\muO)
  = 2\,\frac{N_n}{N_k} \,
    S_{kn}(\muF,\muO) \,
    C_{kn}^{(1)}\,,
\end{eqnarray}
%%Eq (C12) d_kn terms of nondiagonal matrix elements
\begin{subequations}
where
\begin{eqnarray}
 S_{kn}(\muF,\muO)
  &=&
 \frac{\gamma_k^{(0)}-\gamma_n^{(0)}}%
 {\gamma_k^{(0)}-\gamma_n^{(0)}-2b_0}
 \left\{1 - \left[\frac{\alpha_{s}(\muF)}{\alpha_{s}(\muO)}
           \right]^{-1+(\gamma_k^{(0)}-\gamma_n^{(0)})/(2b_0)}
 \right\}
            \label{eq:Skn}
\end{eqnarray}
%%Eq (C13a) S_kn coefficients
and
\begin{eqnarray}
  C_{kn}^{(1)}
   = (2 n + 3)
     \left\{\frac{-\gamma_n^{(0)}-2 b_0+8 C_\text{F} A_{kn}}%
     {2 (k-n)(k+n+3)}
   + \frac{ 2 C_\text{F} \left[ A_{kn} - \psi (k+2) + \psi(1) \right]}
          {(n+1)(n+2)}
     \right\}
  \label{eq:Ckn1}
\end{eqnarray}
%%Eq (C13b) C_kn coefficients
with
\begin{eqnarray}
    A_{kn} = \psi \left(\frac{k+n+4}{2}\right)
              -\psi \left(\frac{k-n}{2}\right)+
              2 \psi(k-n) - \psi(k+2)- \psi(1)\, .
\label{eq:Akn}
\end{eqnarray}
%%Eq (C13c) A_kn definition
\label{eq:SCAkn}
\end{subequations}
We turn now our attention to the finite-order solutions of the
evolution equation \req{eq:eveqU}, i.e., \req{eq:eveq}.
Denoting the formal solution of the LO equation, which contains only
the $V_0$ kernel, by $U^{\text{LO}}(x,s;\muF,\muO)$, the
corresponding function $E_n$, defined in Eq. \req{eq:EnGen}, becomes
\begin{eqnarray}
 E_n^{\text{LO}}(\muF,\muO)
  = \left[\frac{\alpha_{s}(\muF)}{\alpha_{s}(\muO)}
    \right]^{\gamma_n^{(0)}/(2b_0)}\, .
\label{eq:EnLO}
\end{eqnarray}
%%Eq (C14) Evolution factor via anomalous dimensions
Analogously, the solution of the NLO equation, containing both kernels
$V_0$ and $V_1$, will be represented by
$U^{\text{NLO}}(x,s;\muF,\muO)$.
This expression contains contributions coming from both the diagonal
($E_n$) and the nondiagonal ($d_{kn}^{(1)}$) parts.
One finds in the literature two representations for the
$E_n^{\text{NLO}}(\muF,\muO)$ function.
The form which retains the manifest renormalization-group property
$\left[E_n^{\text{NLO}}(\mu_1^2,\mu_2^2)
       E_n^{\text{NLO}}(\mu_2^2,\mu_3^2)
=E_n^{\text{NLO}}(\mu_1^2,\mu_3^2)\right]$
reads \cite{MR86ev}
\begin{eqnarray}
 \label{eq:EnNLO}
  E_n^{\text{NLO}}(\muF,\muO)
  &=& \frac{e_n(\muF)}{e_n(\muO)}\,;
  \qquad
  \omega(n)\
   \equiv\
    \frac{\gamma_n^{(1)}b_0-\gamma_n^{(0)}b_1}
           {2b_0b_1}\,;
  \\
  e_n(\muF)
   &\equiv& \Big[\alpha_s(\muF)\Big]^{\gamma_n^{(0)}/(2b_0)}
       \Big[1+\frac{b_1}{4\pi b_0}\ \alpha_s(\muF)\Big]^{\omega(n)}\, .
\end{eqnarray}
%%Eq (C15) Evolution factor in NLO
%%Eq (C16) Terms of evolution factor in NLO
Alternatively, one can recast $E_n^{\text{NLO}}$ in the form
\cite{MNP99a,Mul98}
\begin{eqnarray}
{\widehat{E}}_n^{\text{NLO}}(\muF,\muO) =
       \left[\frac{\alpha_{s}(\muF)}{\alpha_{s}(\muO)}
       \right]^{\gamma_n^{(0)}/(2b_0)}
     \left\{1 +\frac{b_1}{4\pi b_0}\ \alpha_s(\muF)
              \left[1-\frac{\alpha_{s}(\muO)}{\alpha_{s}(\muF)} \right]
              \omega(n)\,
     \right\}\,,
\label{eq:EnNLO1}
\end{eqnarray}
%%Eq (C17) Alternative form for evolution factor in NLO
which corresponds to a resummation of the leading logarithms
associated with the diagonal terms, while the subleading ones are
expanded with respect to $\alpha_s$.
In this work we employ \req{eq:EnNLO}.\footnote{The expression
\req{eq:EnNLO} is obtained by expanding $\gamma_n$ and $\beta$ to NLO
in Eq. \req{eq:EnGen} and integrating over $\alpha_s$.
To obtain the form \req{eq:EnNLO1}, one expands the integrand in
\req{eq:EnGen} over $\alpha_s$ and performs subsequently the
integration.}

Finally, we systematize below some previous results by recasting them
in a form that is more suitable for practical purposes.
As mentioned in previous sections, the coefficients $a_n(\muF)$
encapsulate nonperturbative information about the binding dynamics
inside the pion and correspond to matrix elements of local operators
according to the OPE, determined at some low-energy scale,
characteristic of the nonperturbative dynamics employed
\cite{PPRWG99,Pra01,ADT00,BMS01,Dor02}.
To obtain these coefficients at a higher scale, say, $\muF$, one has to
apply LO or NLO ERBL evolution.
Specifically the coefficients which correspond to the LO evolution
equation are given by
\begin{eqnarray}
 a_n^{\text{LO}}(\muF)=a_n^{\text{D,LO}}(\muF)
 = a_n(\muO) \, E_n^{\text{LO}}(\muF,\muO)\,,
 \label{eq:anLO}
\end{eqnarray}
%%Eq (C18) Coefficients for LO evolution
while those corresponding to the NLO evolution equation can be written
in the form
\begin{subequations}
\label{eq:anNLOall}
\begin{eqnarray}
a_n^{\text{NLO}}(\muF)=
a_n^\text{D,NLO}(\muF)
+ \frac{\alpha_s(\muF)}{4 \pi}
a_n^\text{ND,NLO}(\muF)\, ,
\label{eq:anNLO}
\end{eqnarray}
%%Eq (C19a) Coefficients for NLO evolution
where
\begin{eqnarray}
a_n^\text{D,NLO}(\muF)&=&
a_n(\muO) \, E_n^{\text{NLO}}(\muF,\muO)\, ,
\label{eq:anNLO0}
\\
a_n^\text{ND,NLO}(\muF)&=&
\sum_{k=0}^{n-2}
a_k(\muO) \, E_k^{\text{NLO}}(\muF,\muO)
d_{nk}^{(1)}(\muF,\muO)\, .
\label{eq:anNLO1}
\end{eqnarray}
%%Eq (C19b) Diagonal coefficients for NLO evolution
%%Eq (C19c) Non-Diagonal coefficients for NLO evolution
\end{subequations}
We note that using instead of \req{eq:EnNLO} the expression
\req{eq:EnNLO1}, would introduce only minor numerical corrections
in the $a_n^{\text{NLO}}$ coefficients of the order $\alpha_s^2$
(amounting to, for example, a $1\%$ relative deviation in
$a_n^{\text{NLO}}$ at 10 GeV$^2$).

%%%%%%%%%%%%%%%%%%%%%%%%%%%%%%%%%%%%%%%%%%%%%%%%%%%%%%%%%%%%%%%%%%%%%%%
%%%%%%%%%%%%%%%%%%%%%%%%%%%%%%%%%%%%%%%%%%%%%%%%%%%%%%%%%%%%%%%%%%%%%%%
\section{NLO Evolution by taking into account heavy-quark thresholds}
\label{app:ThresholdEvo}
%%%%%%%%%%%%%%%%%%%%%%%%%%%%%%%%%%%%%%%%%%%%%%%%%%%%%%%%%%%%%%%%%%%%%%%
We describe here the modification of the evolution formulas,
presented in Appendix~\ref{app:DAnlo}, due to the inclusion of
heavy-flavor thresholds at $M_4=1.3$ GeV and $M_5=4.3$ GeV (with
$M_1=M_2=M_3=0$).
First of all, for calculational convenience, we limit our study to pion
DAs that include at the initial scale $\muO$ only two Gegenbauer
coefficients (i.e., eigenfunctions)
\begin{eqnarray}
 \varphi_\pi(x,\muO)
  = 6 x (1-x)
     \left[1
         + a_{2}^0\, C_{2}^{3/2}(2 x -1)
         + a_{4}^0\, C_{4}^{3/2}(2 x -1)
     \right]
\end{eqnarray}
%%Eq (D1) Pion DA in terms of first 2 Gegenbauer coefficients
and rewrite expressions (\ref{eq:anNLOall}) in terms of these
coefficients $\{a_{2}^0, a_{4}^0\}$ in the more compact form
\begin{subequations}
\begin{eqnarray}
 a_2^{\text{NLO}}(\muF)
 = \tilde{E}_{2}(\muF,\muO)\, a_{2}^0
   + \tilde{D}_{20}(\muF,\muO)\,;
 \qquad \qquad \qquad \qquad \qquad \qquad \qquad \qquad \quad\ \\
 a_4^{\text{NLO}}(\muF)
 = \tilde{E}_{4}(\muF,\muO)\, a_{4}^0
   + \tilde{D}_{42}(\muF,\muO)\, a_{2}^0
   + \tilde{D}_{40}(\muF,\muO)\,;
 \qquad \qquad \qquad \qquad \qquad \\
 a_{n>4}^{\text{NLO}}(\muF)
 = \tilde{D}_{n4}(\muF,\muO)\, \tilde{E}_{4}(\muF,\muO)\, a_{4}^0
   + \tilde{D}_{n2}(\muF,\muO)\, \tilde{E}_{2}(\muF,\muO)\, a_{2}^0
   + \tilde{D}_{n0}(\muF,\muO)\,,
\end{eqnarray}
%%Eq (D2a) Gegenbauer coefficient a_2^NLO
%%Eq (D2b) Gegenbauer coefficient a_4^NLO
%%Eq (D2c) Gegenbauer coefficient a_n>4^NLO
where
\begin{eqnarray}
 \tilde{D}_{nk}(\muF,\muO)
  &\equiv& \frac{\alpha_s(\muF)}{4\pi}\,
           d_{nk}^{(1)}(\muF,\muO)\,.
\end{eqnarray}
%%Eq (D2d) Coefficients D_nk(tilded)
\end{subequations}
We start the evolution from the initial scale $\muO=1$~GeV$^2$, which
corresponds to $N_f=3$.
When $\muF\in [M_4^2,M_5^2]$, we need to change our evolution formulas
by adopting the value $N_f=4$.
Finally, when $\muF\geq M_5^2$, we need to use the value $N_f=5$.
Besides changing the number of active flavors $N_f$, we need also to
match the initial conditions of evolution in each considered region.
This generates the following evolution functions (omitting in the
following expressions the superscript NLO):
\begin{subequations}
\begin{eqnarray}
\label{eq:ThrDiaEvo}
 \tilde{E}_n(\muF,\muO)
  &=& \left\{\begin{array}{lcl}
            E_n(\muF,\muO;N_f=3)\,,&~~& \muF\leq M_4^2\,, \\
            E_n(\muF,M_4^2;N_f=4)\tilde{E}_n(M_4^2,\muO)\,,
                                  &~~& \muF\in(M_4^2,M_5^2]\,, \\
            E_n(\muF,M_5^2;N_f=5)\tilde{E}_n(M_5^2,\muO)\,,
                                  &~~& \muF > M_5^2\,.
             \end{array}
      \right.
\end{eqnarray}
%%Eq (D3a) Evolution equations with heavy-quark mass thresholds
We define in this way the diagonal part of the evolution equation
from a fixed initial scale $\muO$ using
\begin{eqnarray}
\label{eq:ThrDiaEvoArb}
 \tilde{E}_n(\muF,q^2)
  &\equiv& \tilde{E}_n(\muF,\muO)\tilde{E}^{-1}_n(q^2,\muO)\,.
\end{eqnarray}
%%Eq (D3b) Diagonal part of evolution equation
\end{subequations}
In this way we are able to derive the non-diagonal evolution
functions (for the sake of brevity, we omit the explicit indication of
the corresponding $\mu_{\rm F}^2$ regions); namely,
\begin{eqnarray}
 \label{eq:ThrEvo20}
 \tilde{D}_{20}(\muF,\muO)
  &=& \left\{\begin{array}{l}
            D_{20}(\muF,\muO;N_f=3)\,, \\
            D_{20}(\muF,M_4^2;N_f=4)
             +\tilde{E}_2(\muF,M_4^2)\tilde{D}_{20}(M_4^2,\muO)\,,\\
            D_{20}(\muF,M_5^2;N_f=5)
             +\tilde{E}_2(\muF,M_5^2)\tilde{D}_{20}(M_5^2,\muO)\,;
             \end{array}\right.\\
 \label{eq:ThrEvo40}
 \tilde{D}_{40}(\muF,\muO)
  &=& \left\{\begin{array}{l}
            D_{40}(\muF,\muO;N_f=3)\,,\\
            D_{40}(\muF,M_4^2;N_f=4)
             +\tilde{E}_4(\muF,M_4^2)\tilde{D}_{40}(M_4^2,\muO)\,,\\
            D_{40}(\muF,M_5^2;N_f=5)
             +\tilde{E}_4(\muF,M_5^2)\tilde{D}_{40}(M_5^2,\muO)\,;
             \end{array}\right.\\
 \label{eq:ThrEvo42}
 \tilde{D}_{42}(\muF,\muO)
  &=& \left\{\begin{array}{l}
            D_{42}(\muF,\muO;N_f=3)\tilde{E}_2(\muF,\muO) \,,\\
            D_{42}(\muF,M_4^2;N_f=4)\tilde{E}_2(\muF,\muO)
             +\tilde{E}_4(\muF,M_4^2)\tilde{D}_{42}(M_4^2,\muO)\,,\\
            D_{42}(\muF,M_5^2;N_f=5)\tilde{E}_2(\muF,\muO)
             +\tilde{E}_4(\muF,M_5^2)\tilde{D}_{42}(M_5^2,\muO)\,.
             \end{array}\right.
\end{eqnarray}
%%Eq (D4) Coefficients D_20 (tilded) for non-diagonal evolution
%%Eq (D5) Coefficients D_40 (tilded) for non-diagonal evolution
%%Eq (D6) Coefficients D_42 (tilded) for non-diagonal evolution
For $n>4$ and $k=0, 2, 4$, we have
\begin{eqnarray}
 \label{eq:ThrEvo_nk}
 \tilde{D}_{nk}(\muF,\muO)
  &=& \left\{\begin{array}{l}
            D_{nk}(\muF,\muO;N_f=3)\,,\\
            \tilde{E}_{n}(\muF,M_4^2)\tilde{E}_{k}^{-1}(\muF,M_4^2)
             D_{nk}(\muF,M_4^2;N_f=4)+\tilde{D}_{nk}(M_4^2,\muO)\,,\\
            \tilde{E}_{n}(\muF,M_5^2)\tilde{E}_{k}^{-1}(\muF,M_5^2)
             D_{nk}(\muF,M_5^2;N_f=5)+\tilde{D}_{nk}(M_5^2,\muO)\,.
             \end{array}\right.
\end{eqnarray}
%%Eq (D7) Coefficients D_nk (tilded) for non-diagonal evolution
Using these expressions, we can revert to our previous formulas
(\ref{eq:anNLOall}) and write
\begin{subequations}
\label{eq:anNLOallThr}
\begin{eqnarray}
 a_n^{\text{NLO}}(\muF)
 &=& a_n^\text{D,NLO}(\muF)
  + \frac{\alpha_s(\muF)}{4 \pi}\,
     a_n^\text{ND,NLO}(\muF)\,,
 \label{eq:anNLOThr}
\end{eqnarray}
%%Eq (D8a) Coefficients a_n^NLO
where
\begin{eqnarray}
 a_n^\text{D,NLO}(\muF)
  &=& a_n(\muO) \, \tilde{E}_n(\muF,\muO)
   \,,\label{eq:anNLO0Thr}\\
 a_n^\text{ND,NLO}(\muF)
  &=& \sum_{k=0}^{n-2} a_k(\muO)\,\tilde{E}_k(\muF,\muO)
       \tilde{d}_{nk}^{(1)}(\muF,\muO)
\label{eq:anNLO1Thr}
\end{eqnarray}
%%Eq (D8b) Coefficients a_n^NLO, diagonal
%%Eq (D8c) Coefficients a_n^NLO, non-diagonal
with
\begin{eqnarray}
 \tilde{d}_{nk}^{(1)}(\muF,\muO)
  &\equiv&
   \left[\frac{4\pi}{\alpha_s(\muF)}\right]
    \tilde{D}_{nk}(\muF,\muO)\,.
\label{eq:dnkNLO}
\end{eqnarray}
%%Eq (D8d) Coefficients d_nk (tilded)
\end{subequations}

%%%%%%%%%%%%%%%%%%%%%%%%%%%%%%%%%%%%%%%%%%%%%%%%%%%%%%%%%%%%%%%%%%%%%%%
%%%%%%%%%%%%%%%%%%%%%%%%%%%%%%%%%%%%%%%%%%%%%%%%%%%%%%%%%%%%%%%%%%%%%%%
\section{Evolution of the pion distribution amplitude in Analytic
         Perturbation Theory}
\label{app:DAnloSS}
%%%%%%%%%%%%%%%%%%%%%%%%%%%%%%%%%%%%%%%%%%%%%%%%%%%%%%%%%%%%%%%%%%%%%%%
The pion DA satisfies an evolution equation of the form
\begin{eqnarray}
 \frac{d \varphi_\pi(x,\muF)}{d \ln\muF}
  = V(x,u,\asb(\muF))\convo{u}
    \varphi_\pi(u,\muF)\, ,
 \label{eq:eveq_SS}
\end{eqnarray}
%%Eq (E1) Evolution equation for pion DA
with $V(x,u,\alpha)$ having the same functional dependence on $\alpha$
as in (\ref{eq:kernel}).
Let us rewrite this equation for the coefficients
of expansion (\ref{eq:phi0-m}) in terms of Gegenbauer polynomials,
using the notations proposed in~\cite{MR86ev} and linking them to
those in \cite{Mul95}:
\begin{eqnarray}
 \label{eq:EVO_SS-a_n}
 \frac{d a_{n}(\muF)}{d\ln\muF}
  &=& \frac{-\asb(\muF)}{8\pi}
       \left[\gamma_n^{(0)}
           + \frac{\asb(\muF)}{4\pi}\gamma_n^{(1)}
       \right]a_{n}(\muF)
   + \frac{1}{2}\left[\frac{\asb(\muF)}{4\pi}\right]^2
      {\sum\limits_{0\leq j<n}}'
        M_{j,n}a_{j}(\muF)\,,\nonumber\\
 M_{j,n}
   &=& 2\,\frac{N_j}{N_n}\,
        C_{nj}^{(1)}\,
         \big[\gamma_n^{(0)}-\gamma_j^{(0)}\big]\,.
 \label{eq:Exact_Mkn}
\end{eqnarray}
%%Eq (E2) Evolution equation for coefficients a_n with respect to mu_F
First, we define the diagonal evolution operator
$\hat{E}^\text{NLO}_n(Q^2,\mu_0^2)=\hat{e}_n(L_Q)/\hat{e}_n(L_{\mu_0})$
with
\begin{eqnarray}
 \frac{d \hat{e}_n\left(L_\mu\right)}{d L_\mu}
   &=& \frac{-\asb^{(2,\text{fit})}(\mu^2)}{8\pi}
       \left[\gamma_n^{(0)}
           + \frac{\asb^{(2,\text{fit})}(\mu^2)}{4\pi}\gamma_n^{(1)}
       \right]\hat{e}_n\left(L_\mu\right)\,,
\end{eqnarray}
%%Eq (E3) Diagonal evolution operator
where $L_\mu\equiv\ln(\mu^2/\Lambda_{21}^2)$ and
$\asb^{(2,\text{fit})}(\mu^2)$ is given by (\ref{eq:asb_2-App-Global}),
i.e.,
\begin{eqnarray}
  \asb^{(2,\text{fit})}(\mu^2)
  &=& \frac{4\pi}{b_0(3)}\,
       \bar{a}_s\left[\ell\left(L_\mu,c_{21}^\text{fit}
                           \right)
                 \right]
  \ \equiv\ \frac{4\pi}{b_0(3)}\,
  \bar{A}_s^{(2,\text{fit})}\left(L_\mu\right)\,.
\end{eqnarray}
%%Eq (E4) Analytic 2-loop cpupling fitted
Then we have
\begin{eqnarray}
 \label{eq:Exp_Evo_SS_NLO}
 \hat{e}_n^\text{NLO}\left(L_\mu\right)
   &=& \exp\left(-\int_{L_0}^{L_\mu}
                   \left\{\frac{\gamma_n^{(0)}}{2b_0(3)}\
                         \bar{A}_s^{(2,\text{fit})}(L)
                       + \frac{\gamma_n^{(1)}}{2b_0(3)^2}\
                       \left[\bar{A}_s^{(2,\text{fit})}(L)\right]^2
                   \right\} d L
           \right)\,.
\end{eqnarray}
%%Eq (E5) Evolution coefficient
In principle, the lower limit of integration, $L_0$, can be chosen
to be an arbitrary positive number, but it is more convenient to set it
equal to the average value of $L[\muF]$ under actual consideration.
Then, we can represent our solution in a modified form---as compared to
(\ref{eq:anNLOall}); namely,
\begin{eqnarray}
  a^\text{An}_{n}(\muF)
  = \hat{e}_n^\text{NLO}\left(L_{\mu_\text{F}}\right)
    \left[\frac{a_{n}(\mu_0^2)}%
               {\hat{e}_n^\text{NLO}\left(L_{\mu_0}\right)}
        + \frac{\asb(\muF)}{4\pi}
          {\sum\limits_{0\leq j<n}}'
            \hat{d}_{j,n}(\muF,\mu_0^2)
             \frac{a_{j}(\mu_0^2)}
                  {\hat{e}_{j}^\text{NLO}\left(L_{\mu_0}\right)}
     \right]\,.
 \label{eq:EvoSS}
\end{eqnarray}
%%Eq (E6) Solution to evolution for Shirkov-Solovtsov coupling function
The advantage of introducing the same factor
$\hat{e}_n^\text{NLO}\left[L_{\mu_\text{F}}\right]$
for the whole function $a^\text{An}_{n}(\muF)$ is that it ensures exact
cancellation of the diagonal terms in Eq. (\ref{eq:EVO_SS-a_n}).
What is left over after this cancellation
provides an equation for $\hat{d}_{j,n}(\muF,\mu_0^2)$:
\begin{eqnarray}
  \frac{d\!\left[\asb(\muF)\,
  \hat{d}_{j,n}(\muF,\mu_0^2)\right]}{d L_{\mu_\text{F}}}
 &=& \frac{\asb^2(\muF)}{8\pi}
\left[M_{j,n}\,\frac{\hat{e}_j^\text{NLO}\left(L_{\mu_\text{F}}\right)}
{\hat{e}_n^\text{NLO}\left(L_{\mu_\text{F}}\right)}
     \right.
 \nonumber \\
 &&   \left.\qquad\qquad
    + \frac{\asb(\muF)}{4\pi}
            {\!\!\sum\limits_{j\leq m<n}\!}'
              \hat{d}_{j,m}(\muF,\mu_0^2)M_{m,n}\,
               \frac{\hat{e}_m^\text{NLO}\left(L_{\mu_\text{F}}\right)}
                    {\hat{e}_n^\text{NLO}\left(L_{\mu_\text{F}}\right)}
      \right].
\label{eq:evol_djn}
\end{eqnarray}
%%Eq (E7) Remainder after cancellation of diagonal term
In the NLO approximation this expression becomes
\begin{eqnarray}
 \frac{d}{d L_{\mu_\text{F}}}
 \left[\bar{A}_s^{(2,\text{fit})}\left(L_{\mu_\text{F}}\right)\,
                      \hat{d}_{j,n}(\muF,\mu_0^2)\right]
  \approx
   \frac{M_{j,n}}{2b_0(3)}
     \left[\bar{A}_s^{(2,\text{fit})}
     \left(L_{\mu_\text{F}}\right)\right]^2
      \frac{\hat{e}_j^\text{NLO}\left(L_{\mu_\text{F}}\right)}
           {\hat{e}_n^\text{NLO}\left(L_{\mu_\text{F}}\right)}
\label{eq:evol_NLO_A2}
\end{eqnarray}
%%Eq (E8) NLO approximation of remainder
and its solution is given by
\begin{eqnarray}
 \hat{d}_{j,n}(\muF,\mu_0^2)
  \approx \frac{M_{j,n}}{2b_0(3)\, \bar{A}_s^{(2,\text{fit})}
          \left(L_{\mu_\text{F}}\right)}
           \int_{L_{\mu_0}}^{L_{\mu_\text{F}}}
            \left[\bar{A}_s^{(2,\text{fit})}(L)\right]^2
             \frac{\hat{e}_j^\text{NLO}(L)}{\hat{e}_n^\text{NLO}(L)}
              d L\; .
\end{eqnarray}
%%Eq (E9) Solution for evolution coefficients d_j,n (tilded)
Since we do not take into account the nondiagonal part of the evolution
equation (see for more details in Sec.\ \ref{sect:PFFNLO:AR}, just
after Eq.\ (\ref{eq:Q2pffNLOnondia})), we can use an approximate form
of Eq.\ (\ref{eq:EvoSS}), viz.,
\begin{eqnarray}
 a_n^{\text{An;\,D,NLO}}(\muF)
 = a_n(\muO)\,
   \frac{\hat{e}^\text{NLO}_n\left(L_{\mu_\text{F}}\right)}
        {\hat{e}^\text{NLO}_n\left(L_{\mu_0}\right)}\,,
 \label{eq:SS_NLO}
\end{eqnarray}
%%Eq (E10) Approximate form evolution of analytic coefficients
where the functions $\hat{e}^\text{NLO}_n\left(L_{\mu_\text{F}}\right)$
are defined in a two-step numerical procedure:
\begin{enumerate}
 \item We determine first by numerical integration of Eq.\
 (\ref{eq:Exp_Evo_SS_NLO}) the functions
 $\hat{e}^\text{num}_{n}(L)$ for $L\in[0,10]$ and $n=2, 4$.
 \item We construct then interpolating functions
 $\hat{e}^\text{NLO}_{n}(L)$ for all functions determined in step one.
\end{enumerate}
In order to obtain the term
$\tilde{\cal F}_{\pi}^\text{LO}(\muF,\muR)$
in the approximate formula (\ref{eq:Q2pffApprox}) for the factorized
form factor, we also need the LO-part of the evolution, namely,
\begin{eqnarray}
 \label{eq:Exp_Evo_SS_LO}
 \hat{e}_n^\text{LO}\left(L_\mu\right)
  &=& \exp\left[-\int_{L_{\mu_0}}^{L_\mu}
                \frac{\gamma_n^{(0)}}{2b_0(3)}\
                \bar{A}_s^{(2,\text{fit})}(L)
                d L
           \right]\,;\\
 a_n^{\text{An;\,LO}}(\muF)
  &=& a_n(\muO)\,
   \frac{\hat{e}^\text{LO}_n\left(L_{\mu_\text{F}}\right)}
        {\hat{e}^\text{LO}_n\left(L_{\mu_0}\right)}
 \label{eq:SS_LO}\, .
\end{eqnarray}
%%Eq (E11) LO evolution factor (fitted)
%%Eq (E12) LO analytic coefficients

\end{appendix}

%finishes textpart
%\bibliographystyle{apsrev}
%\bibliography{pion,lambda}
%\end{document}

\newcommand{\noopsort}[1]{} \newcommand{\printfirst}[2]{#1}
  \newcommand{\singleletter}[1]{#1} \newcommand{\switchargs}[2]{#2#1}

\end{document}